\PassOptionsToPackage{unicode}{hyperref}
\PassOptionsToPackage{hyphens}{url}
\PassOptionsToPackage{dvipsnames,svgnames,x11names}{xcolor}
\documentclass[]{article}

\usepackage{iftex}
\ifPDFTeX
  \usepackage[T1]{fontenc}
  \usepackage[utf8]{inputenc}
  \usepackage{textcomp}
\else
  \usepackage{unicode-math}
  \defaultfontfeatures{Scale=MatchLowercase}
  \defaultfontfeatures[\rmfamily]{Ligatures=TeX,Scale=1}
\fi
\usepackage{lmodern}

\IfFileExists{upquote.sty}{\usepackage{upquote}}{}
\IfFileExists{microtype.sty}{
  \usepackage[]{microtype}
  \UseMicrotypeSet[protrusion]{basicmath}
}{}

\usepackage[top=30mm,left=30mm,heightrounded]{geometry}
\makeatletter
\@ifundefined{KOMAClassName}{
  \IfFileExists{parskip.sty}{\usepackage{parskip}}{
    \setlength{\parindent}{0pt}
    \setlength{\parskip}{6pt plus 2pt minus 1pt}
  }
}{
  \KOMAoptions{parskip=half}
}
\makeatother
\usepackage{orcidlink}
\usepackage{authblk}
\usepackage{appendix}

\makeatletter
\ifx\paragraph\undefined\else
  \let\oldparagraph\paragraph
  \renewcommand{\paragraph}{
    \@ifstar\xxxParagraphStar\xxxParagraphNoStar}
  \newcommand{\xxxParagraphStar}[1]{\oldparagraph*{#1}\mbox{}}
  \newcommand{\xxxParagraphNoStar}[1]{\oldparagraph{#1}\mbox{}}
\fi
\ifx\subparagraph\undefined\else
  \let\oldsubparagraph\subparagraph
  \renewcommand{\subparagraph}{
    \@ifstar\xxxSubParagraphStar\xxxSubParagraphNoStar}
  \newcommand{\xxxSubParagraphStar}[1]{\oldsubparagraph*{#1}\mbox{}}
  \newcommand{\xxxSubParagraphNoStar}[1]{\oldsubparagraph{#1}\mbox{}}
\fi
\makeatother

\providecommand{\tightlist}{\setlength{\itemsep}{0pt}\setlength{\parskip}{0pt}}

\usepackage{amsmath}

\usepackage{longtable,booktabs,array,calc,multirow,wrapfig,float,colortbl,
            pdflscape,tabu,threeparttable,threeparttablex,makecell,tabularx,adjustbox}

\usepackage{etoolbox}
\makeatletter
\patchcmd\longtable{\par}{\if@noskipsec\mbox{}\fi\par}{}{}
\makeatother

\IfFileExists{footnotehyper.sty}{\usepackage{footnotehyper}}{\usepackage{footnote}}
\makesavenoteenv{longtable}

\usepackage{graphicx}
\makeatletter
\def\maxwidth{\ifdim\Gin@nat@width>\linewidth\linewidth\else\Gin@nat@width\fi}
\def\maxheight{\ifdim\Gin@nat@height>\textheight\textheight\else\Gin@nat@height\fi}
\makeatother
\setkeys{Gin}{width=\maxwidth,height=\maxheight,keepaspectratio}
\def\fps@figure{htbp}

\usepackage{xcolor}
\makeatletter
\@ifpackageloaded{tcolorbox}{}{\usepackage[skins,breakable]{tcolorbox}}
\@ifpackageloaded{fontawesome5}{}{\usepackage{fontawesome5}}
\definecolor{quarto-callout-color}{HTML}{909090}
\definecolor{quarto-callout-note-color}{HTML}{0758E5}
\definecolor{quarto-callout-important-color}{HTML}{CC1914}
\definecolor{quarto-callout-warning-color}{HTML}{EB9113}
\definecolor{quarto-callout-tip-color}{HTML}{00A047}
\definecolor{quarto-callout-caution-color}{HTML}{FC5300}
\definecolor{quarto-callout-color-frame}{HTML}{acacac}
\definecolor{quarto-callout-note-color-frame}{HTML}{4582ec}
\definecolor{quarto-callout-important-color-frame}{HTML}{d9534f}
\definecolor{quarto-callout-warning-color-frame}{HTML}{f0ad4e}
\definecolor{quarto-callout-tip-color-frame}{HTML}{02b875}
\definecolor{quarto-callout-caution-color-frame}{HTML}{fd7e14}
\makeatother

\usepackage{caption}
\usepackage{subcaption}

\makeatletter
\@ifundefined{c@chapter}{
  \newfloat{codelisting}{h}{lop}
}{
  \newfloat{codelisting}{h}{lop}[chapter]
}
\floatname{codelisting}{Listing}

\makeatother

\usepackage{calc}
\newlength{\cslhangindent}
\newlength{\csllabelwidth}
\newenvironment{CSLReferences}[2]
  {\begin{list}{}{\setlength{\itemindent}{0pt}
                  \setlength{\leftmargin}{0pt}
                  \setlength{\parsep}{0pt}
                  \ifodd #1
                    \setlength{\leftmargin}{\cslhangindent}
                    \setlength{\itemindent}{-1\cslhangindent}
                  \fi
                  \setlength{\itemsep}{#2\baselineskip}}}
  {\end{list}}

\NewDocumentCommand\citeproctext{}{}

\makeatletter
\let\@cite@ofmt\@firstofone
\def\@biblabel#1{}
\def\@cite#1#2{{#1\if@tempswa , #2\fi}}
\makeatother

\usepackage{bookmark}
\IfFileExists{xurl.sty}{\usepackage{xurl}}{}
\hypersetup{
  pdftitle={Deceptive uses of Artificial Intelligence in elections strengthen support for AI ban},
  colorlinks=true,
  linkcolor=blue,
  filecolor=Maroon,
  citecolor=Blue,
  urlcolor=Blue,
  pdfcreator={LaTeX via pandoc}
}

\ifLuaTeX
  \usepackage{selnolig}
\fi

\AtBeginDocument{%

}

\usepackage{xcolor}
\usepackage{hyperref}
\hypersetup{
  colorlinks=true,
  linkcolor=blue,
  citecolor=blue,
  urlcolor=blue,
  pdftitle={Artificial Intelligence in Election Campaigns: Perceptions, Penalties, and Implications},
  pdfcreator={Andreas Jungherr}
}

\usepackage{authblk}
\usepackage{orcidlink}

\title{Artificial Intelligence in Election Campaigns: Perceptions, Penalties, and Implications}
\author[1]{Andreas Jungherr \orcidlink{0000-0003-2598-2453}}
\author[2]{Adrian Rauchfleisch \orcidlink{0000-0003-1232-083X}}
\author[3]{Alexander Wuttke \orcidlink{0000-0002-9579-5357}}
\affil[1]{University of Bamberg}
\affil[2]{National Taiwan University}
\affil[3]{LMU Munich}
\date{\today}

\begin{document}
\maketitle
\begin{abstract}
As political parties around the world experiment with Artificial
Intelligence (AI) in election campaigns, concerns about deception and
manipulation are rising. This article examines how the public reacts to
different uses of AI in elections and the potential consequences for
party evaluations and regulatory preferences. Across three preregistered
studies with over 7,600 American respondents, we identify three
categories of AI use -- campaign operations, voter outreach, and
deception. While people generally dislike AI in campaigns, they are
especially critical of deceptive uses, which they perceive as norm
violations. However, parties engaging in AI-enabled deception face no
significant drop in favorability, neither with supporters nor opponents.
Instead, deceptive AI use increases public support for stricter AI
regulation, including calls for an outright ban on AI development. These
findings reveal a misalignment between public disapproval of deceptive
AI and the political incentives of parties, underscoring the need for
targeted regulatory oversight. Rather than banning AI in elections
altogether, regulation should distinguish between harmful and beneficial
applications to avoid stifling democratic innovation.
\end{abstract}

Artificial Intelligence (AI) is starting to feature in election
campaigns around the world (Barredo-Ibáñez et al. 2021; Raj 2024; Sifry 2024; Swenson, Merica, and Burke 2024). From targeting
voters to generating messages, AI offers political actors powerful new
tools (Foos 2024; Kruschinski et al. 2025; Tomić, Damnjanović, and Tomić
2023). Yet these technological advances are often met with sharp
criticism from journalists and experts, who frequently highlight AI's
perceived potential to fuel electoral deception and manipulation
(Sanders and Schneier 2024; Verma and Vynck 2024). This tension between
the opportunities AI offers to political actors and the public's growing
concerns about its misuse raises urgent questions for democracies;
particularly regarding the role of public opinion in constraining the
behavior of political elites, shaping campaign tactics, and informing
appropriate regulatory responses. This article contributes to that
debate by providing evidence on how citizens perceive the use of AI in
electoral campaigns and whether they sanction political actors seen as
violating electoral norms.

Across three preregistered studies with more than 7,600 American
respondents, we find that while the public strongly disapproves of
AI-enabled deception in election campaigns, this disapproval does not
translate into favorability penalties for those responsible. Instead, it
leads to heightened support for strict AI regulation, including calls
for a general ban on AI development. This reveals a troubling
misalignment between public opinion and political incentives. This
dynamic has significant implications for the governance of AI and the
future of democratic campaigning.

We introduce a new framework for understanding how people assess the use
of AI in elections. We distinguish between three core types of AI use:
improving campaign operations, enhancing voter outreach, and engaging in
deception. We argue that public reactions vary based on the perceived
norm violation of each use, which in turn affects not only attitudes
toward specific campaigns but also broader beliefs about democracy,
personal control, and technology regulation.

Our findings are based on three studies: (1) a nationally representative
survey of public attitudes toward various campaign-related uses of AI
(Perceptions: n = 1,199), (2) a survey experiment testing how different
AI use types affect political attitudes and regulatory preferences
(Reactions: n = 1,985), and (3) a partisan-based experiment assessing
whether deceptive AI use leads to electoral penalties (Penalties: n =
4,451). Together, these studies provide a systematic account of how the
public views AI in elections and what those views mean for democratic
institutions and AI policy.

The implications of these findings extend beyond campaign strategy. If
public backlash against AI-enabled deception leads to broad, sweeping
regulation with parties facing little to no political cost for using
such tactics, democracies risk a dual threat: political actors may have
no incentive to refrain from deceptive or inflammatory uses of AI, and
the resulting public pressure could drive overly restrictive governance
that stifles beneficial innovation. This combination could erode the
quality of political competition while curbing the democratic potential
of AI. A nuanced regulatory approach is therefore essential; one that
targets harmful practices without discouraging legitimate and
constructive applications of AI in democratic politics.

\section{Theoretical Framework: Public Reactions to AI Use in
Elections}\label{theoretical-framework-public-reactions-to-ai-use-in-elections}

We propose a framework to explain how the public reacts to different
uses of AI in election campaigns and how these reactions affect
attitudes toward democracy and technology governance. Prior research
shows that attitudes toward AI are not uniform. Instead, they seem to
depend on how and to what purpose AI is used (Raviv 2025; Zhang and
Dafoe 2019). Similarly, we do not expect public opinion on the uses of
AI in election campaigns to be uniform. We expect people's reactions to
depend on how AI is used and whether these uses violate norms of
appropriate political conduct. These perceptions can influence judgments
across multiple domains, including evaluations of campaigning practices,
democratic legitimacy, personal autonomy, and support for regulation.

Our framework consists of four interconnected steps:

\begin{itemize}
\tightlist
\item
  categorizing different types of AI use in campaigns;
\item
  identifying perceived norm violation as key mechanism;
\item
  outlining the key domains in which public attitudes are affected; and
\item
  anticipating the downstream consequences of norm-violating uses.
\end{itemize}

\subsection{Step 1: Types of AI Use in Election
Campaigns}\label{step-1-types-of-ai-use-in-election-campaigns}

A consistently growing set of academic studies and journalistic accounts
from all over the world show that the use of AI in election campaigns is
highly varied and extends far beyond the often-discussed threat of
deepfakes (Barredo-Ibáñez et al. 2021; Dommett 2023; Foos 2024;
Garimella and Chauchard 2024; Kamal and Kaur 2025; Kruschinski et al.
2025; Raj 2024; Sifry 2024; Swenson, Merica, and Burke
2024; Tomić, Damnjanović, and Tomić 2023). But with journalistic
discourse heavily featuring often exaggerated, if not downright
imagined, AI-risks (Sanders and Schneier 2024; Verma and Vynck 2024), we
risk overlooking more common and consequential applications. This skewed
attention can obscure how campaigns actually adopt AI, typically
selecting tools based on their anticipated contribution to specific
campaign objectives and organizational needs (Earl and Kimport 2011;
Hindman 2005; Jungherr, Rivero, and Gayo-Avello 2020).

To capture this functional diversity, we introduce an original
categorization scheme that distinguishes three primary types of AI use
in election campaigns: campaign operations, voter outreach, and
deception. These categories emerged inductively through a review of
documented AI applications in journalistic reporting and academic
literature. The resulting framework reflects the distinct ways in which
campaigns employ AI tools to pursue strategic goals as well as manage
everyday operational demands.

\subsubsection{AI in Campaign
Operations.}\label{ai-in-campaign-operations.}

Campaigns use AI to streamline internal processes and improve
organizational efficiency. Applications include automated content
generation, chatbot-based communication with volunteers or supporters,
and algorithmic segmentation of donor and walk lists (Carrasquillo 2024;
Chow 2024; Foos 2024; Sifry 2024; Swenson 2023). These uses tend to
receive limited public attention but play a significant role in shaping
how campaigns allocate resources, communicate internally, and adapt to
fast-changing circumstances.

Past research on digital campaigning shows that technological adoption
is often driven less by public-facing spectacle than by improvements to
everyday processes (Hindman 2005; Karpf 2012; Kreiss 2011, 2012; Nielsen
2011). In this regard, AI continues the evolution of campaign
infrastructure, offering new tools for scaling operations with limited
staff and budgets.

\subsubsection{AI in Voter Outreach.}\label{ai-in-voter-outreach.}

AI is also used to enhance voter outreach by identifying likely
supporters, testing and optimizing message content, and delivering
personalized appeals across digital platforms (Burke and Sunderman 2024;
Chatterjee 2024; Hackenburg and Margetts 2024). These practices build on
a well-established tradition of data-driven campaigning, including
microtargeting and experimental message testing (Dommett, Kefford, and
Kruschinski 2024; Green and Gerber {[}2004{]} 2023; Hersh 2015; McCarthy
2020; Nickerson and Rogers 2014; Turrow et al. 2012).

In this context, AI enables campaigns to refine their persuasive
strategies and tailor messages to specific audiences more efficiently
than before. While some outreach practices may raise concerns about
manipulation or privacy, they are broadly consistent with conventional
campaign goals and practices, that are now augmented by machine learning
and generative tools.

\subsubsection{AI in Deception.}\label{ai-in-deception.}

AI can also be used to intentionally mislead, impersonate, or obscure
key information. Deceptive practices include the generation of synthetic
audio or video that falsely portrays a candidate or opponent, the
impersonation of political figures using deepfakes, and the deployment
of AI-generated content in coordinated astroturfing campaigns (e.g., via
bots or automated email outreach to journalists and voters) (Barari,
Lucas, and Munger 2025; Bond 2024; Coltin 2024; Linvill and Warren 2024;
Ternovski, Kalla, and Aronow 2022).

Such uses build on familiar political tactics such as negative
campaigning and strategic misinformation (Austen-Smith 1992; Bucciol and
Zarri 2013; Haselmayer 2019; Jay 2010; Lau and Rovner 2009; Nai 2020).
AI is feared to enhance these tactics by facilitating likeness,
scalability, and speed, which may raise their potential impact. A Pew
Research Center study from Summer 2024 found that Republicans and
Democrats alike are equally worried about the impact of AI on the 2024
presidential campaign (Gracia 2024). At the same time, early fears about
widespread societal harm from AI-enabled deception, especially
deepfakes, appear to be overstated, and evidence on their actual
influence remains scarce (Simon, Altay, and Mercier 2023).

We treat deception as a distinct category of AI use because it reflects
an intent to influence public perception through misrepresentation,
rather than merely an unintended side effect of technological
deployment. As with operational and outreach applications, deceptive
uses represent deliberate choices about how campaigns seek to achieve
specific communication goals.

By distinguishing between operations, voter outreach, and deception, our
framework offers a comprehensive account of how campaigns use AI,
capturing applications that range from routine and arguably constructive
to controversial and potentially harmful.

\subsection{Step 2: Norm Violation as Key
Mechanism}\label{step-2-norm-violation-as-key-mechanism}

In their choices of tactics, technologies, and communication strategies
campaigns are constrained by norms of acceptable political behavior and
by the extent to which the public and press are willing to enforce these
norms and sanction violations. This also applies to the use of AI in
election campaigns.

Norms can be understood as shared, often implicit, expectations that
define what is considered appropriate within a group or society
(Cialdini and Trost 1998). In interpersonal contexts, norms guide
behavior and shape expectations. Violations of these norms often provoke
negative reactions, including social sanction or reduced credibility
(Burgoon and Le Poire 1993; van Kleef et al. 2015). This mechanism also
extends to how people assess the actions of organizations and
institutional actors (Dahl, Frankenberger, and Manchanda 2003).

In electoral politics, actors who violate established norms often face
public backlash. Such norm violations, such as excessive negativity or
incivility, can reduce support for the offending party or candidate,
both in terms of attitude and vote choice (Ansolabehere et al. 1994;
Fridkin and Kenney 2011; Muddiman 2017). We argue that the same
evaluative logic applies to the use of AI: people are likely to judge
parties' AI uses based on their perceived adherence to, or deviation
from, normative expectations of legitimate political competition.

Crucially, not all uses of AI are equally likely to be seen as norm
violations. Deceptive uses, such as generating fake content or
impersonating political figures, clearly conflict with general social
norms. In many domains, deception leads to negative affective and
behavioral responses toward the deceiver (Croson, Boles, and Murnighan
2003; Ohtsubo et al. 2010; Tyler, Feldman, and Reichert 2006). We
therefore expect that when people learn a party has used AI deceptively,
they will react with disapproval, perceiving the act as a violation of
democratic norms.

The case of voter outreach is more ambiguous. On the one hand, critics
have long argued that targeting voters with individually tailored
messages, especially based on psychological profiling, threatens
informational autonomy and undermines democratic deliberation (Bayer
2020; Bennett and Manheim 2006). Survey evidence also shows that people
often feel uneasy about behavioral targeting by both commercial and
political actors (McCarthy 2020; Turrow et al. 2012). On the other hand,
such techniques may now be sufficiently normalized that people no longer
view them as norm violations, even if they remain uncomfortable with
them. Moreover, targeted outreach may help parties reach disengaged or
hard-to-reach citizens in fragmented media environments, potentially
contributing to democratic inclusion (Dommett, Kefford, and Kruschinski
2024; Jungherr, Rivero, and Gayo-Avello 2020). As such, public reactions
to AI-enabled targeting may depend on whether it is framed as
manipulative or as a means of broadening participation.

Finally, the use of AI to improve internal campaign operations, such as
scheduling, donor segmentation, or content drafting, is unlikely to be
viewed as normatively problematic. On the contrary, people may see these
uses as legitimate or even beneficial. Lowering the cost of campaign
operations could expand political participation by reducing barriers to
entry and enhancing representational diversity, potentially
strengthening democratic competition.

In sum, we expect perceived norm violation to be a key mechanism shaping
public responses to campaign uses of AI. The intensity and direction of
those responses, however, are likely to vary by context and function:
strongly negative in the case of deception, more ambivalent in the case
of voter outreach, and potentially positive in the case of operational
uses.

\subsection{Step 3: Related Attitudes}\label{step-3-related-attitudes}

Public experiences with AI use in politics are likely to shape more than
just opinions about election campaigns. Because campaigns attract
intense media coverage and public scrutiny, they often serve as
high-profile contexts in which emerging technologies are introduced,
contested, and symbolically evaluated. As such, campaign-related AI uses
frequently become exemplars (Murphy 2002), cases that people refer to
when forming judgments about AI's broader role in politics and society,
especially in areas where personal experience or direct evidence is
limited.

Historical cases illustrate this dynamic. For example, the campaigns of
Howard Dean and Barack Obama became emblematic of the democratic
potential of digital technologies, contributing to hopeful narratives
about online mobilization and participatory empowerment (Kreiss 2016;
Shirky 2008). In contrast, accounts of Cambridge Analytica's role in the
Brexit and Trump campaigns have become touchstones in narratives about
the dangers of data-driven surveillance and political manipulation,
fueling public demands for stronger oversight of digital platforms (West
and Allen 2020; Weiss-Blatt 2021).

Likewise, contemporary uses of AI in campaigns, whether celebrated or
condemned, may shape public attitudes across a range of domains. At the
most immediate level, individuals form judgments about specific campaign
applications of AI: whether they are appropriate, effective, or
concerning. These evaluations can then extend to more general political
beliefs, such as perceptions of party integrity, election fairness, or
the overall quality of democracy. Over time, campaign-related exemplars
may even influence how people perceive AI's role in their personal
lives, particularly in relation to autonomy and control. For example,
encountering deceptive AI uses in politics may heighten feelings of
disempowerment or suspicion toward algorithmic systems more broadly.

Finally, these perceptions can also shape public attitudes toward AI
governance. If political uses of AI are seen as norm-violating or
harmful, they may drive support for more restrictive or precautionary
regulation, not just within the political sphere but across all domains
of AI development and deployment. In this way, campaign experiences with
AI can have ripple effects, shaping how societies understand, accept,
and regulate emerging technologies well beyond the electoral arena.

\subsection{Step 4: Anticipating Downstream
Effects}\label{step-4-anticipating-downstream-effects}

Building on the previous steps, we can now specify our expectations
regarding how different uses of AI in election campaigns affect public
attitudes. When campaign uses of AI are perceived as norm violations we
expect a cluster of negative reactions. These include:

\begin{itemize}
\tightlist
\item
  Lower favorability toward the implicated party;
\item
  Diminished trust in democratic institutions, including the fairness of
  elections;
\item
  Reduced sense of personal autonomy and control; and
\item
  Stronger support for regulatory oversight, including more restrictive
  or precautionary AI policies.
\end{itemize}

These reactions are not limited to evaluations of campaign tactics; they
extend to broader judgments about democracy, technology, and the
legitimacy of emerging systems of governance. In this sense,
norm-violating AI uses in politics can act as focal points that shape
public opinion across multiple domains of political and personal life.

However, because these responses occur in a political context, they are
unlikely to be uniform across the electorate. A robust body of research
shows that political partisans are prone to motivated reasoning: they
tend to discount or rationalize information that reflects poorly on
their preferred party, while reacting strongly to norm violations by
political opponents (Jost, Hennes, and Lavine 2013; Kahan 2016; Williams
2023). We therefore expect heterogeneous effects across partisan lines.
While norm-violating AI use may provoke strong disapproval in general,
this disapproval is likely to be asymmetrical, more intense among
opponents of the implicated party and more muted or absent among its
supporters.

This dynamic helps explain a core tension explored in this study:
although people express strong negative views about deceptive AI use,
these attitudes do not always translate into electoral penalties for the
responsible parties. Instead, motivated reasoning may buffer in-group
parties from reputational harm, even as public concern drives broader
demands for regulatory intervention.

\section{Materials and Methods}\label{materials-and-methods}

\begin{figure}

\centering{

\includegraphics[keepaspectratio]{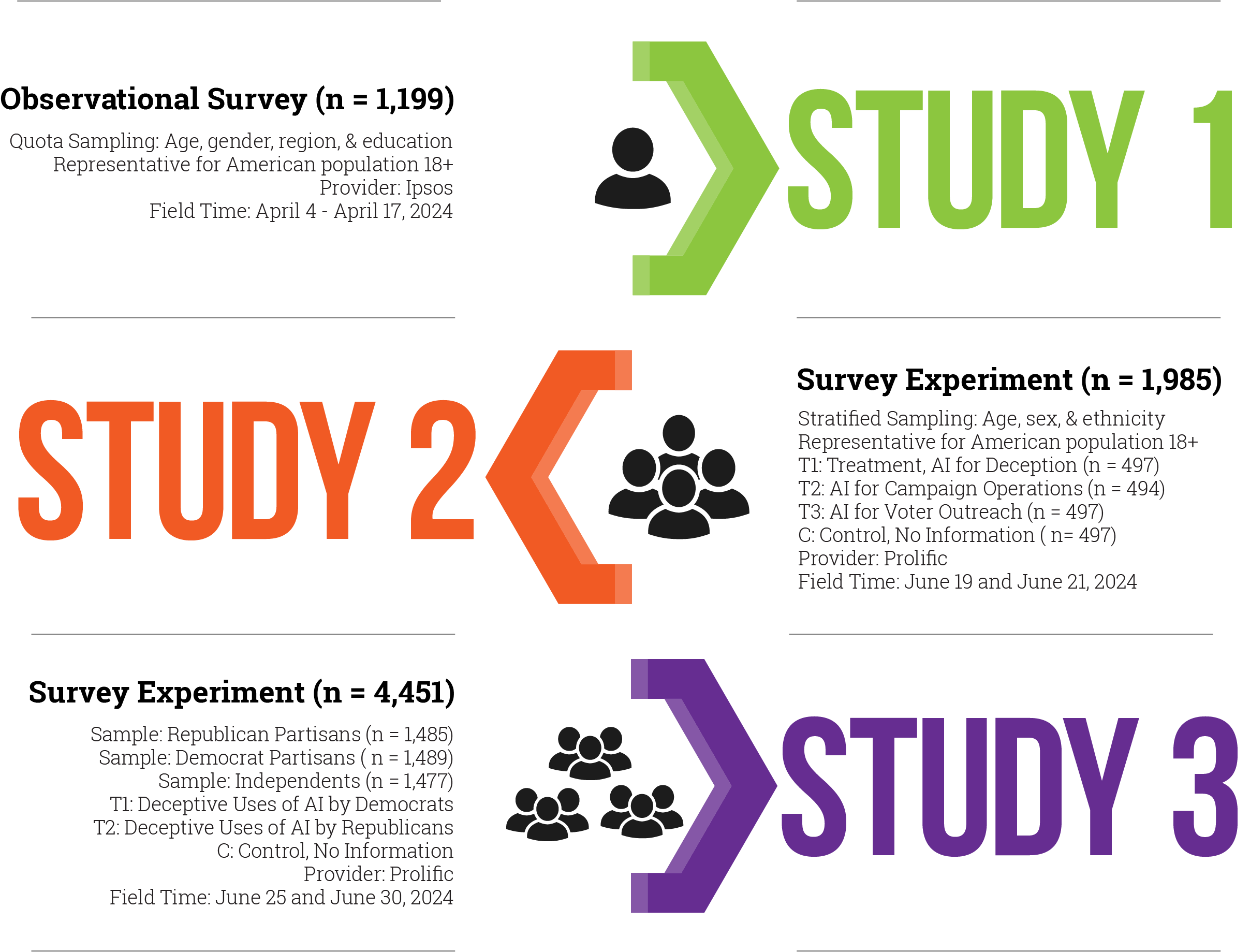}

}

\caption{\label{fig-research_design}Research Design.}

\end{figure}

To test our framework for the explanation of public reactions to AI uses
in election campaigns, we ran three preregistered surveys, including two
survey experiments (for research design see
Figure~\ref{fig-research_design} and Online Appendix). All studies were
approved by the Institutional Review Board at the home institution of
one of the authors. Each study focuses on a distinct aspect of the
framework: public attitudes toward different AI uses (Study 1:
Perceptions), causal effects of exposure to those uses (Study 2:
Reactions), and whether parties are penalized for using deceptive AI
(Study 3: Penalties). With this study, we present first systematic
evidence on how Americans think about different uses of AI in election
campaigns.

\subsection{Study 1 (Perceptions): Observational survey (n =
1,199)}\label{study-1-perceptions-observational-survey-n-1199}

In Study 1, we queried people for their opinions on specific uses of AI
in elections and their views on the benefits and risks of AI in other
areas. We ran a preregistered survey (n=1,199) among members of an
online panel that the market and public opinion research company Ipsos
provided. We used quotas on age, gender, region, and education to
realize a sample representative of the US electorate (See Online
Appendix 1 for details). Before running the survey, we registered our
research design, analysis plan, and hypotheses about
outcomes.\footnote{Preregistration Study 1:
  \url{https://osf.io/3nrb4/?view_only=1d82e100d6084edd81d9c4af46f31a30}.}
We did not deviate from the registered procedure. We provided
respondents with short descriptions of various campaigning tasks for
which parties and candidates use AI. These tasks fall into three broad
categories: (I) campaign operations, (II) voter outreach, (III)
deception. For each category, we identified five example tasks that
practitioners and journalists have documented and discussed. We showed
each respondent three randomly drawn tasks for each category and asked
them whether this use of AI in elections: (1) worried them, (2) felt
like a norm violation, (3) was likely to make politics more interesting
to voters, and (4) increase participation. We combined items ``This use
of AI makes politics more interesting to voters'', and ``This use of AI
increases voter engagement'' into one index (Rise in Voter Involvement)
to capture AI's likely impact on voter interest and mobilization. As two
distinct independent variables, we also measured AI risk ($\alpha$ = 0.78)
and benefit perceptions ($\alpha$ = 0.84) with three items each. For each
dependent variable, we estimated a multilevel model with varying
intercepts for cases (15) and participants. Furthermore, as specified in
the preregistration, we used data imputation to fill in the missing
responses for AI benefits and AI risks (both models, with and without
data imputation, show the same results; see Online Appendix 5.1). See
Online Appendix 1.1 for an overview of variables, question wordings,
operationalizations, key diagnostics of item measurements, and the data
imputation procedure for Study 1.

\subsection{Study 2 (Reactions): Survey experiment (n =
1,985)}\label{study-reactions-survey-experiment-n-1985}

In our second study, we test the causal effects of learning about
different types of AI use in elections. We ran a preregistered survey
experiment with members of an online panel provided by \emph{Prolific}
(n=1,985).\footnote{Prior research shows that Prolific provides
  excellent data quality. Peer et al. (2022), for example, show that
  Prolific provides even higher data quality than widely used panels
  such as Qualtrics and Dynata. For Study 1, which focuses on general
  attitudes toward different AI use categories, we relied on a
  high-quality Ipsos panel. In Studies 2 and 3, which emphasize
  experimental effects, we used Prolific with quota sampling, as
  detailed in the Online Appendix. Given the sampling approach and the
  large sample sizes (which are especially important for statistical
  power in experimental designs), we are confident in the
  generalizability of our findings.} We used quotas to realize a sample
resembling the US electorate (See Online Appendix 1.2 for details). The
survey was fielded between June 19 and June 21, 2024. We divided
respondents into three treatment and one control group (C, n = 497).
Treatment 1 (T1, n = 497) contained information about campaigns' uses of
AI for deception. Treatment 2 (T2, n = 494) contained information about
campaigns' uses of AI for improving campaign operations. Treatment 3
(T3, n = 497) contained information about campaigns' uses of AI for
voter outreach. The dependent variables in this study cover four broad
areas: campaign use (worry, norm violations, and positive impact on
politics), democracy (fairness of elections, favorability of parties in
general and specific parties), lifeworld (personal loss of control), and
regulation (of AI use in elections, AI in general, priority of
regulation over innovation, and support for an AI moratorium). We
registered our research design, analysis plan, and hypotheses about
outcomes before the survey.\footnote{Preregistration Study 2:
  \url{https://osf.io/wsrkv/?view_only=6d55d846ae8d4ba886c3e3ce8076d845}.}
We did not deviate from the registered procedure. For detailed
information on treatments, variables, question wordings,
operationalizations, key diagnostics of item measurements, and
manipulation checks for Study 2, see Table 11 in the Online Appendix.

\subsection{Study 3 (Penalties): Survey experiment (n =
4,451)}\label{study-3-penalties-survey-experiment-n-4451}

In Study 3, we test whether parties face a penalty for deceptive AI uses
attributed to them and whether partisans' group-protective cognitions
lead to heterogeneous effects of being informed about deceptive uses.
For this preregistered study,\footnote{Preregistration Study 3:
  \url{https://osf.io/vugp8/?view_only=5e4387422dc94458bb355e6e2e5fba3d}}
we recruited three samples containing only respondents identifying as
partisans for (1) Democrats (n=1,489), (2) Republicans (n=1,485), or as
(3) Independent (n=1,477). Prolific prescreened partisans. No attempt to
be representative was made. The survey was fielded between June 25 and
June 30, 2024. Respondents in these three samples were exposed to either
of two treatments or were assigned a pure control group that did not
receive any information. Treatments contained information about
deceptive uses of AI by candidates from the Democratic Party (T1) or the
Republican Party (T2). This allows us to identify whether
group-protective cognition leads partisans to discount information about
uses of AI by parties they support, compared to adjusting related
attitudes when being informed about deceptive uses by parties they
oppose, and how this compares to reactions by Independents. For the
dependent variables, we focused on three broad areas: campaign use
(worry, norm violations, and positive impact on politics), democracy
(fairness of elections, favorability of specific parties), and
regulation (priority of regulation over innovation and support for an AI
moratorium). We registered our research design, analysis plan, and
hypotheses about outcomes before the survey. We did not deviate from the
registered procedure. For detailed information on treatments, variables,
question wordings, operationalizations, key diagnostics of item
measurements, and manipulation checks for Study 3, see Online Appendix
1.3.

\section[Results]{\texorpdfstring{Results\footnote{For an overview of
  all preregistered hypotheses with all estimates and statistical tests,
  see Appendix 2, and for the complete tables of the models, see
  Appendix 5.}}{Results}}\label{resultsmore}

\subsection{Study 1, Perceptions: Public Disapproval of AI in Campaigns
Varies by
Use}\label{study-1-perceptions-public-disapproval-of-ai-in-campaigns-varies-by-use}

We asked a representative sample of Americans (n=1,199) for their
opinions on specific uses of AI (see Online Appendix 1.1 for details).
In our preregistered study, we provided respondents with a selection
from fifteen short descriptions of various campaigning tasks for which
AI might be used. These tasks fall into three functional categories:

\begin{itemize}
\tightlist
\item
  Campaign operations: including automated content generation,
  chatbot-based communication, and the segmentation of donor and walk
  lists.
\item
  Voter outreach: including identifying persuadable voters, optimizing
  message appeal (either at scale or individually), and generating
  personalized ads.
\item
  Deception: including the undeclared use of AI to create misleading
  media (e.g., deepfakes), impersonate candidates, or conduct
  coordinated astroturfing via bots and large language models.
\end{itemize}

Each category included five representative tasks. For each task they
were shown, respondents were asked how much the use worried them,
whether they perceived it as violating campaign norms, and whether they
believed it could increase voter engagement.

\begin{longtable}[]{@{}
  >{\raggedright\arraybackslash}p{(\linewidth - 6\tabcolsep) * \real{0.4286}}
  >{\raggedleft\arraybackslash}p{(\linewidth - 6\tabcolsep) * \real{0.2000}}
  >{\raggedleft\arraybackslash}p{(\linewidth - 6\tabcolsep) * \real{0.1905}}
  >{\raggedleft\arraybackslash}p{(\linewidth - 6\tabcolsep) * \real{0.1810}}@{}}
\caption{Share responses that agree with assessment (Study
1)}\label{tbl-response_shares_main}\tabularnewline
\toprule\noalign{}
\begin{minipage}[b]{\linewidth}\raggedright
Campaign Task
\end{minipage} & \begin{minipage}[b]{\linewidth}\raggedleft
Worry (in \%)
\end{minipage} & \begin{minipage}[b]{\linewidth}\raggedleft
Norm Violation (in \%)
\end{minipage} & \begin{minipage}[b]{\linewidth}\raggedleft
Rise Voter Involvement (in \%)
\end{minipage} \\
\midrule\noalign{}
\endfirsthead
\toprule\noalign{}
\begin{minipage}[b]{\linewidth}\raggedright
Campaign Task
\end{minipage} & \begin{minipage}[b]{\linewidth}\raggedleft
Worry (in \%)
\end{minipage} & \begin{minipage}[b]{\linewidth}\raggedleft
Norm Violation (in \%)
\end{minipage} & \begin{minipage}[b]{\linewidth}\raggedleft
Rise Voter Involvement (in \%)
\end{minipage} \\
\midrule\noalign{}
\endhead
\bottomrule\noalign{}
\endlastfoot
Deception: Astroturfing (interactive) & 68.57 & 64.49 & 33.88 \\
Deception: Astroturfing (social media) & 76.37 & 69.86 & 32.18 \\
Deception: Deceptive Robocalls & 68.78 & 64.14 & 28.69 \\
Deception: Deepfakes (negative campaigning) & 71.58 & 68.26 & 29.46 \\
Deception: Deepfakes (self-promotional) & 71.80 & 68.76 & 36.23 \\
& & & \\
Outreach: Ad Optimization & 63.54 & 57.78 & 39.02 \\
Outreach: Data Driven Targeting & 56.11 & 55.90 & 34.16 \\
Outreach: Fundraising & 58.41 & 55.09 & 39.38 \\
Outreach: Message Testing \& Opinion Research & 60.83 & 54.92 & 35.63 \\
Outreach: Outreach Optimization & 60.91 & 58.85 & 32.30 \\
& & & \\
Operations: Automating Interactions & 67.72 & 63.56 & 35.84 \\
Operations: Deepfakes (benign) & 56.43 & 57.11 & 40.18 \\
Operations: Resource Allocation & 49.80 & 47.83 & 41.70 \\
Operations: Transcription & 53.83 & 53.42 & 33.13 \\
Operations: Writing & 62.91 & 55.10 & 29.50 \\
\end{longtable}

Table~\ref{tbl-response_shares_main} summarizes the proportion of respondents
who rated each item above the midpoint on a 7-point scale (responses
\textgreater{} 4), excluding missing values. The results show clear
distinctions across categories. Respondents were most concerned about
deceptive uses of AI, which they saw as more likely to violate
democratic norms and less likely to boost voter involvement. While all
AI uses met with some degree of concern, certain campaign operations,
particularly low-profile, non-communicative tasks like resource
allocation and transcription, were viewed most favorably.
Outreach-related uses were generally seen as more problematic than these
operational tasks, but less troubling than deceptive practices. However,
operational tasks involving AI-generated interaction or writing also
drew substantial concern, underscoring that reactions vary within
categories as well as across them.

\begin{figure}

\centering{

\includegraphics[keepaspectratio]{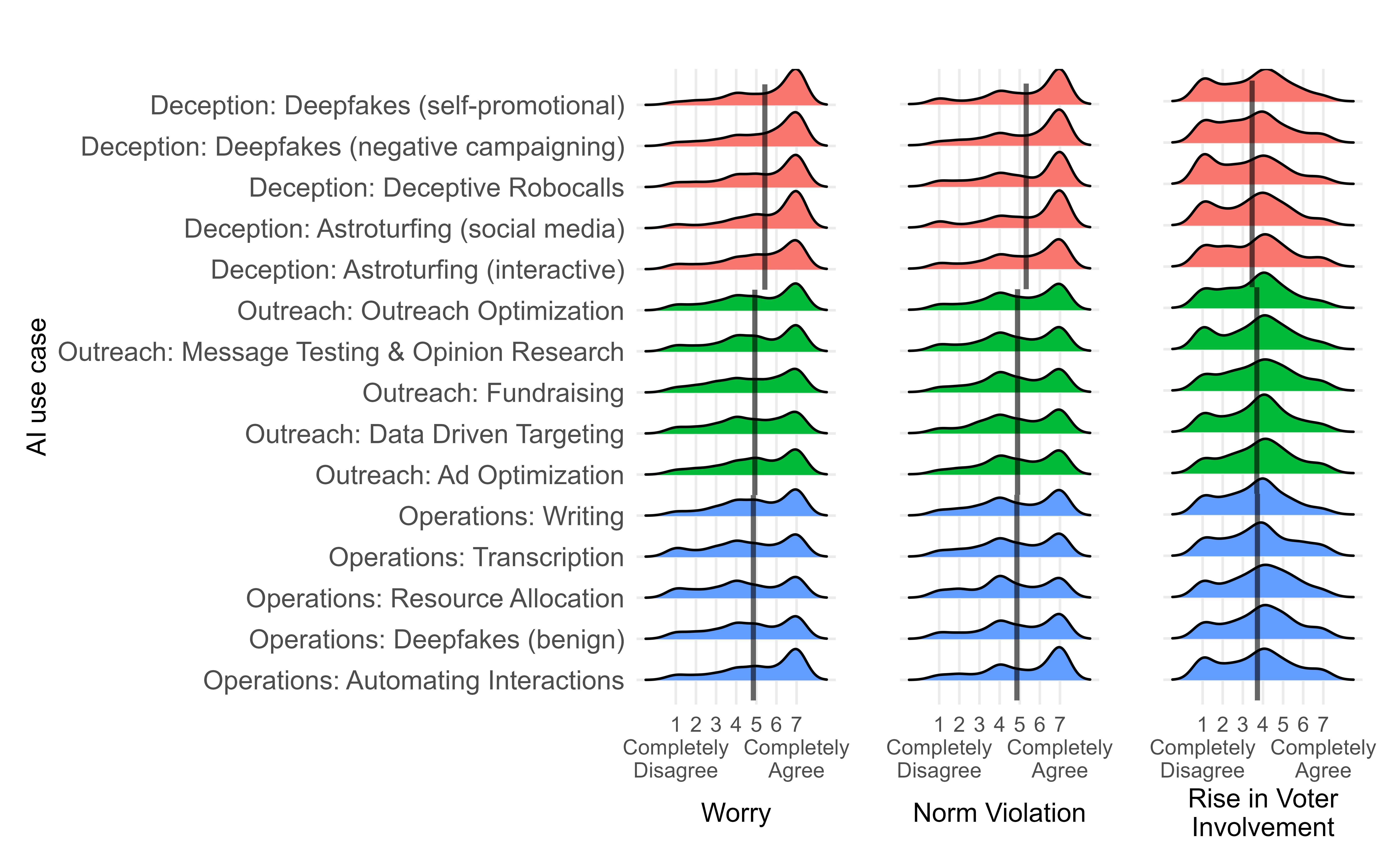}

}

\caption{\label{fig-attitudes_ai_elections}Attitudes toward AI uses in
elections by type. The vertical line indicates the mean per category.
The deception category differs significantly (p \textless{} .001) from
the other two for all three outcome variables.}

\end{figure}%

Figure~\ref{fig-attitudes_ai_elections} visualizes the distribution of
responses across the fifteen use cases by category. The ridgeline plots
show that, although public opinion on AI in politics is generally
cautious, deceptive uses consistently meet the strongest disapproval.
Across all three measures (i.e.~worry, norm violation, and perceived
potential to increase engagement) AI for deception is judged most
negatively. These plots also highlight the internal variation within
each category, reinforcing the importance of distinguishing between
specific applications of AI in campaign contexts.

\begin{figure}

\centering{

\includegraphics[keepaspectratio]{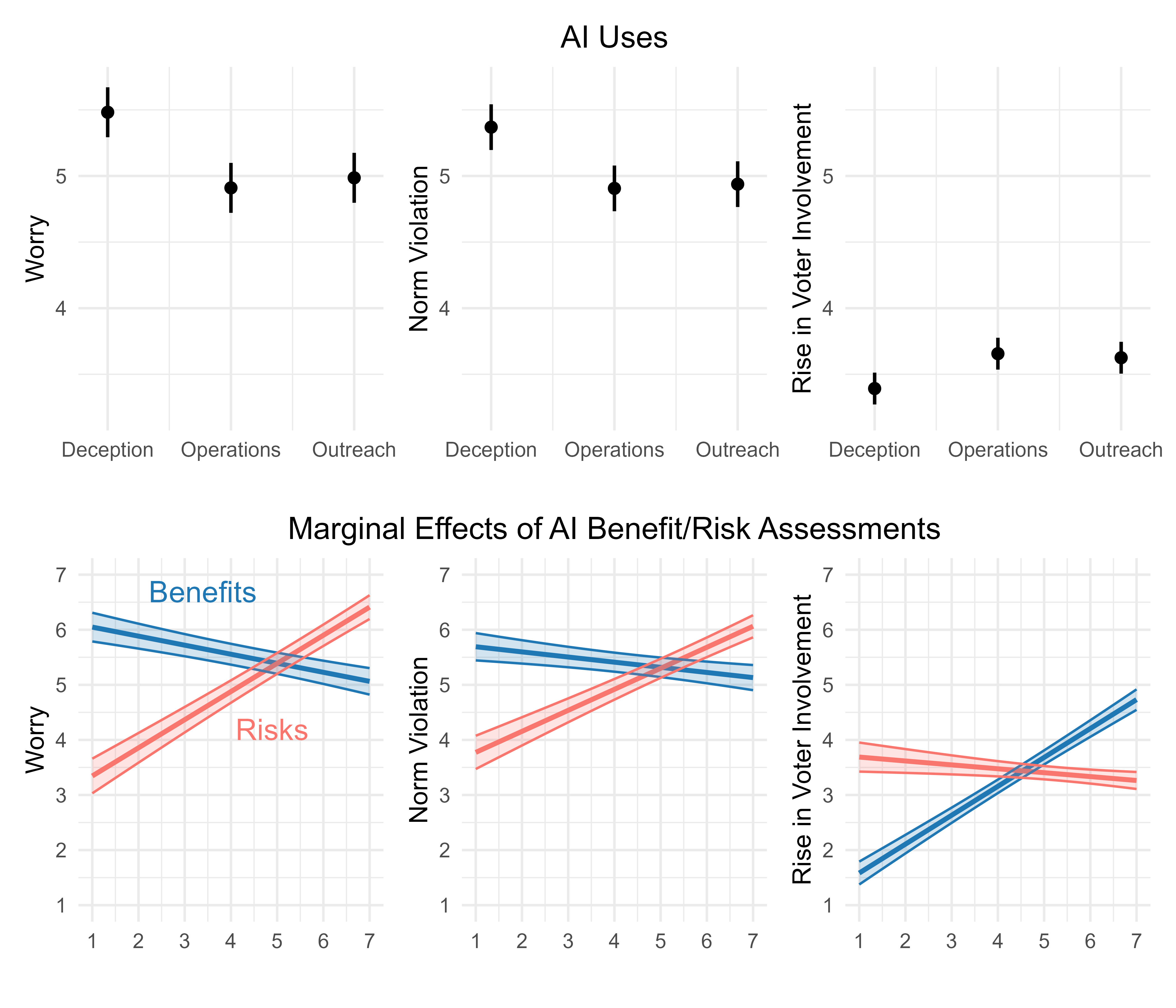}

}

\caption{\label{fig-ai_campaing_attitudes_explained}Attitudes toward AI
uses in elections, regressions (for campaign tasks, deception is used as
a reference group). Estimates with 95\%-CIs. The deception category
differs significantly (p \textless{} .001) from the other two for all
three outcome variables. Also, both AI benefit and risk perception are
significant for all three outcome variables.}

\end{figure}%

To further examine these differences, we estimated regression models
predicting levels of worry, perceived norm violation, and belief in
increased voter involvement (n = 1,199). As shown in the top row of
Figure~\ref{fig-ai_campaing_attitudes_explained}, even after controlling
for individual characteristics, deceptive uses of AI consistently met
with greater concern and were more likely to be viewed as norm
violations, while also being least likely to be seen as enhancing
democratic participation. Estimations for operational and outreach uses
overlap, suggesting that respondents perceive these uses in broadly
similar ways but distinct from how they view deceptive applications.

\subsection{Study 2: Reactions to AI
Uses}\label{study-2-reactions-to-ai-uses}

In a preregistered follow-up experiment (n = 1,985), we examined the
causal effects of learning about different types of AI use in election
campaigns (see Online Appendix 1.2 for details). Respondents were
randomly assigned to one of three treatment groups or a control group (n
= 497). Each treatment provided information about a different category
of campaign AI use:

\begin{itemize}
\tightlist
\item
  Deception Treatment (n = 497): presented examples of AI-enabled
  deception by campaigns.
\item
  Operations Treatment (n = 494): presented examples of AI used to
  improve internal campaign processes.
\item
  Outreach Treatment (n = 497): presented examples of AI used to enhance
  voter targeting and communication.
\end{itemize}

Since Study 1 showed that deceptive uses of AI were perceived most
negatively, we preregistered the deception condition as the reference
group.

\begin{figure}

\centering{

\includegraphics[keepaspectratio]{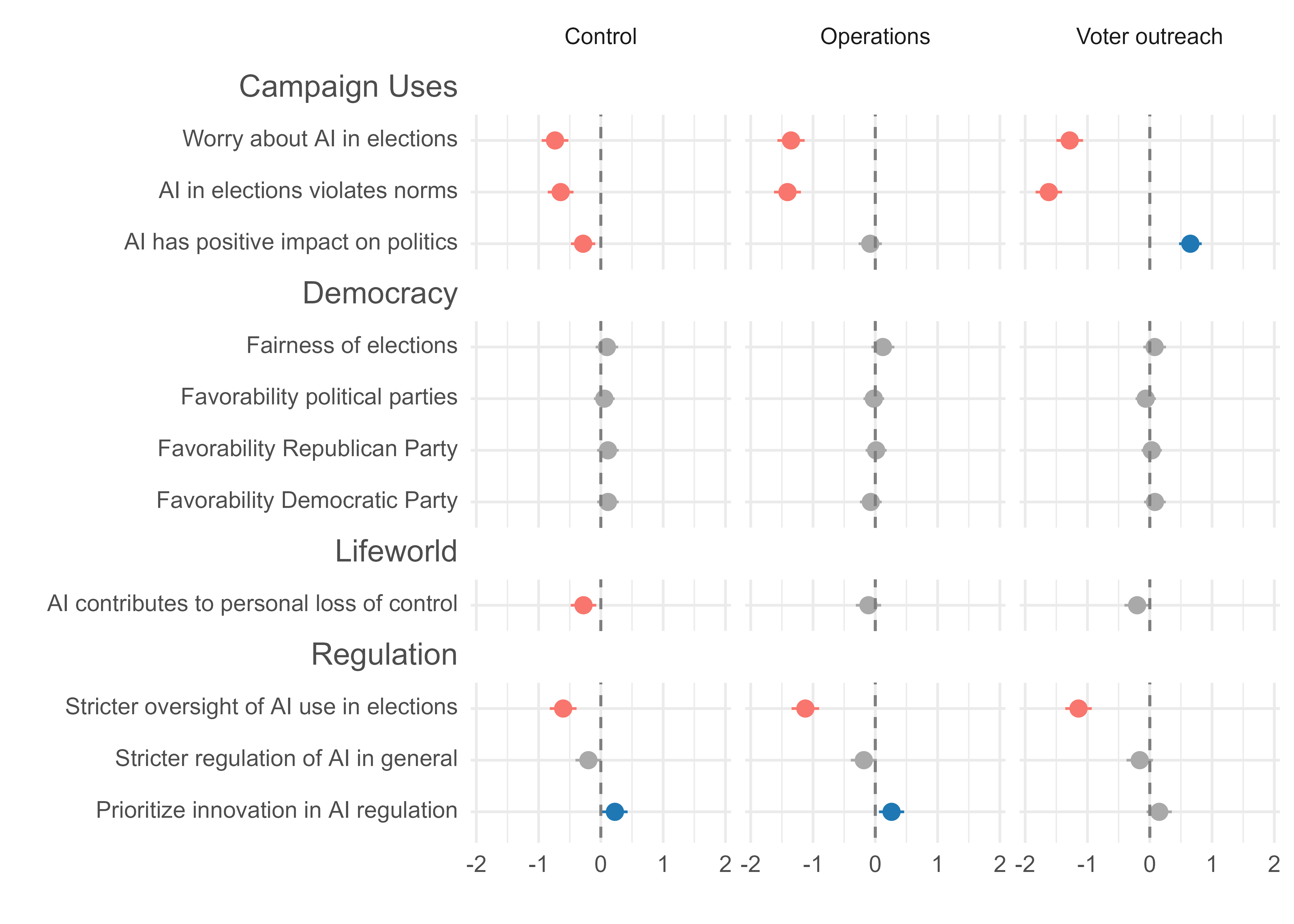}

}

\caption{\label{fig-effects_ai_elections_s2}Effects of Information about
Different Uses of AI in Elections (Reference Category: Deception).
Estimates with 95\%-CIs. Grey estimates are not significant.}

\end{figure}%

The results indicate that perceived norm violation plays a central role
in shaping public reactions, but only for specific outcome dimensions
(see Figure~\ref{fig-effects_ai_elections_s2}). Respondents who received
the deception treatment expressed significantly higher levels of worry
and were more likely to view the behavior as a violation of campaign
norms, compared to all other groups.

However, the effects did not extend to broader democratic attitudes.
Exposure to AI-enabled deception did not significantly affect
perceptions of electoral fairness or party favorability. In other words,
while deceptive AI use elicits disapproval, it does not appear to carry
an electoral penalty, at least not in terms of reduced support for the
party involved or diminished trust in the election process. Accordingly,
we tested in a follow-up study directly whether parties suffer a
favorability penalty when deceptive AI use is attributed to them.

\subsection{Study 3: Parties face no favorability penalty for deceptive
AI use}\label{sec-study2}

This preregistered study used three samples composed exclusively of
self-identified partisans: Democrats (n = 1,489), Republicans (n =
1,485), and Independents (n = 1,477) (see Online Appendix 1.3 for
details). Within each group, respondents were randomly assigned to one
of two treatment conditions or a control group. The treatments provided
information about deceptive AI use by either a Democratic or a
Republican candidate.

\begin{figure}

\centering{

\includegraphics[keepaspectratio]{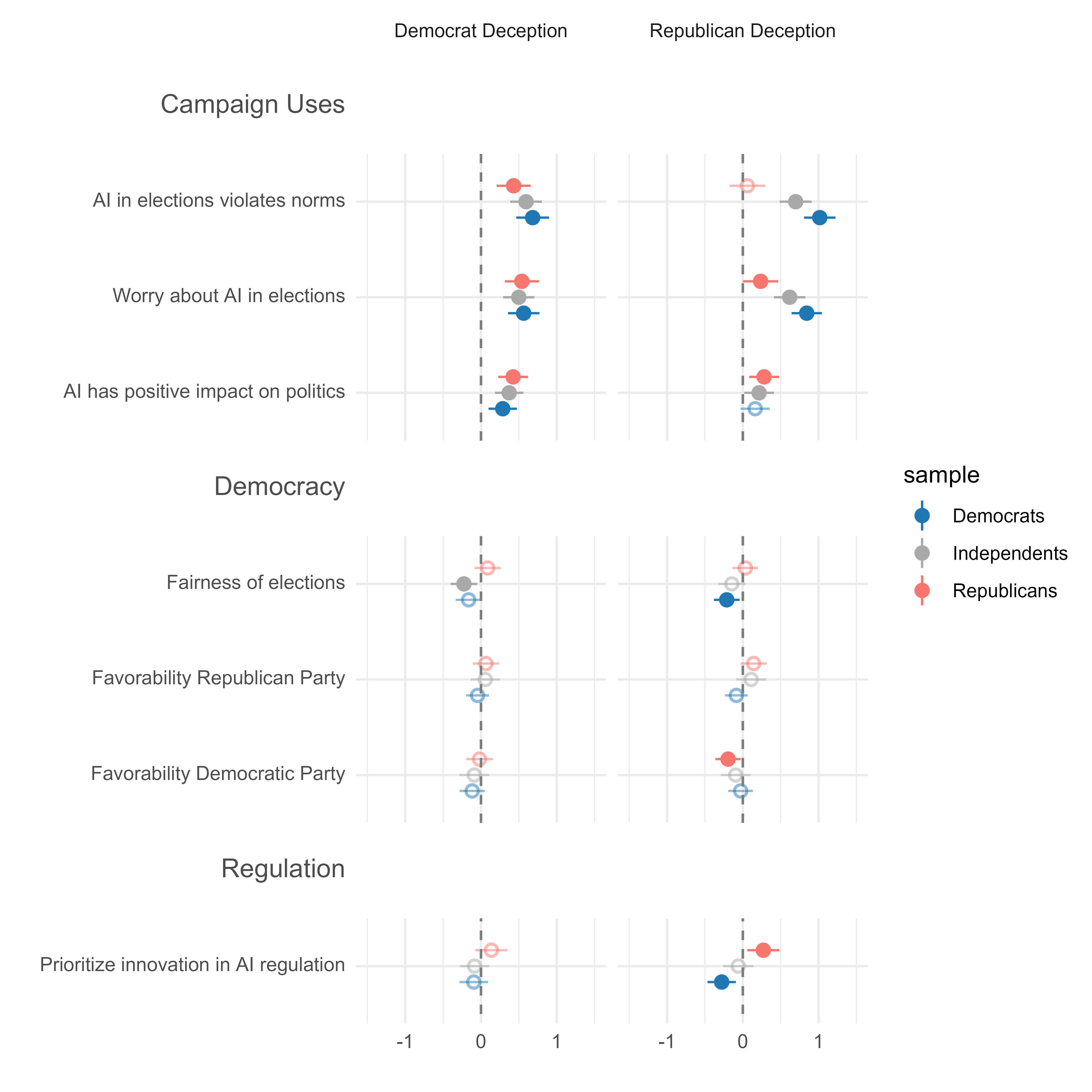}

}

\caption{\label{fig-effects_ai_elections_3}Effects of Information About
Alleged Deceptive Uses of AI for Partisans and Independents (Reference
Category: Control). Estimates with 95\%-CIs. Hollow estimates are not
significant.}

\end{figure}%

The results provide no evidence that deceptive AI use reduces support
for the implicated party (see Figure~\ref{fig-effects_ai_elections_3}).
While both Democratic and Republican respondents expressed a greater
sense of norm violation when exposed to information about their own
party's alleged misconduct, they did not lower their favorability
ratings compared to the control group. In other words, partisans
disapprove of deceptive AI use in principle but are unwilling to
penalize their own party for it in practice.

This pattern extends to independents, who also did not significantly
change their favorability ratings in response to either treatment
condition. Equivalence tests show that no substantial shifts in party
favorability occurred among any group (see Online Appendix 4).

Together with the findings from Study 2, this result highlights a
critical disconnect: although norm-violating uses of AI generate
disapproval, they do not translate into political consequences for the
parties responsible. This suggests that the public's normative
expectations and attitudinal reactions are not sufficient to constrain
campaign behavior, especially in an environment shaped by partisan
loyalty and motivated reasoning.

\subsection{Studies 2 and 3: AI-Enabled Deception Increases Support for
Restrictive AI
Governance}\label{studies-2-and-3-ai-enabled-deception-increases-support-for-restrictive-ai-governance}

Beyond attitudes directly related to campaigning, Studies 2 and 3 also
examined how exposure to AI-enabled deception influences broader views,
particularly assessments of personal autonomy and support for AI
regulation. These outcomes extend the analysis beyond electoral dynamics
to public perceptions of AI's role in society and the legitimacy of its
continued development.

As shown in Figure~\ref{fig-effects_ai_elections_s2}, respondents in
Study 2 who were informed about deceptive campaign uses of AI were
significantly more likely to report a sense of diminished personal
control. This suggests that people use high-salience political examples,
such as deceptive AI in election campaigns, as heuristics for evaluating
the impact of AI on their daily lives and civic agency.

\begin{figure}

\centering{

\includegraphics[keepaspectratio]{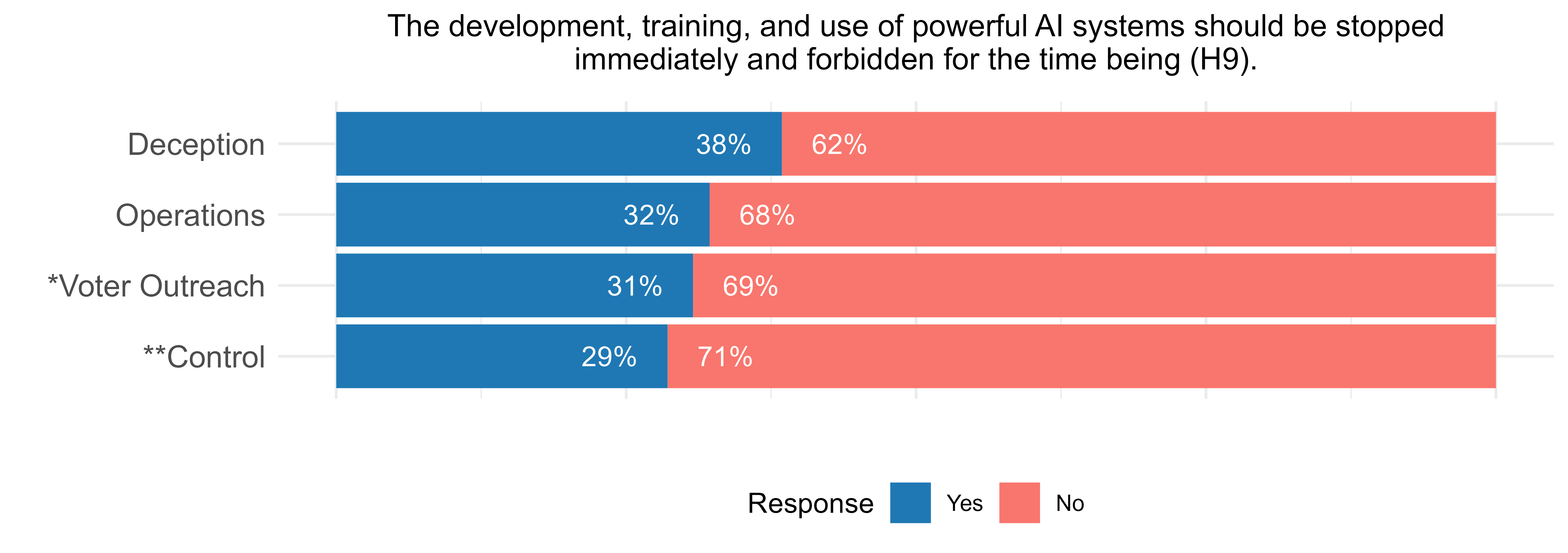}

}

\caption{\label{fig-effects_ai_elections_s2_gov}Effects of Information
about Different Uses of AI in Elections on Governance Preferences
(Reference Category: Deception), p \textless{} 0.05 *), p \textless{}
0.01 (**), p \textless{} 0.001 (***).}

\end{figure}

The effect on regulatory attitudes was even more pronounced. Support for
a complete stop to AI development and use was significantly higher among
those exposed to information about AI deception in elections. While 29\%
of respondents in the control group supported such a ban, 38\% of those
in the deception treatment group did so. These respondents also
expressed stronger support for stricter oversight of AI use in elections
and favored prioritizing safety over innovation in the regulation of
emerging technologies (see Figure~\ref{fig-effects_ai_elections_s2_gov})

\begin{figure}

\centering{

\includegraphics[keepaspectratio]{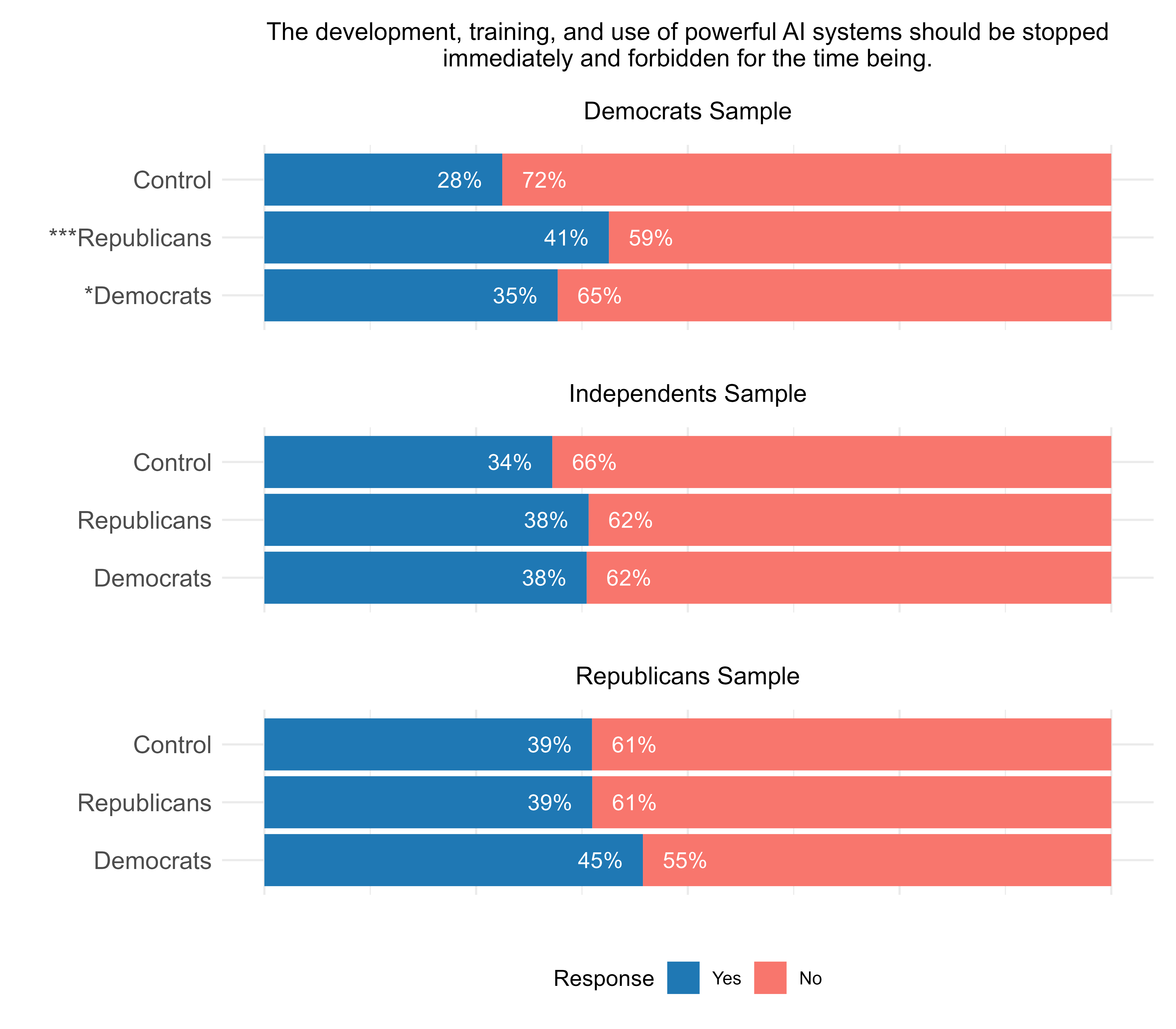}

}

\caption{\label{fig-effects_ai_elections_3_gov}Effects of Information
About Alleged Deceptive Uses of AI on Governance Preferences for
Partisans and Independents (Reference Category: Control), p \textless{}
0.05 (*), p \textless{} 0.01 (**), p \textless{} 0.001 (***).}

\end{figure}

The findings from Study 3 corroborate these results. Again, information
about AI deception increased support for halting AI development (see
Figure~\ref{fig-effects_ai_elections_3_gov}). While baseline support for
an AI ban varied by partisanship, 39\% among Republicans and 28\% among
Democrats in the control group, Democrats significantly increased their
support when exposed to deceptive AI use, regardless of whether the
behavior was attributed to their own party or the opposing one.

These results highlight a critical asymmetry in the political
consequences of AI deception. Although parties do not face direct
electoral penalties for deceptive AI use (as shown in Study 3), the
practice generates negative externalities: it erodes public trust,
undermines perceived autonomy, and increases support for more
restrictive, and potentially hostile, approaches to AI governance.
Notably, we find that exposure to deceptive AI increases public support
for an outright halt to AI development. While this preference for
safety-first regulation is rational in light of perceived threats, it
may carry unintended consequences. Calls for sweeping bans or highly
precautionary regimes can hinder experimentation, limit adaptive
learning, and delay the development of beneficial applications. Public
concern, when expressed through blunt regulatory preferences, may
inadvertently produce policy that slows innovation without significantly
reducing risk. In this way, political uses of AI serve as symbolic
exemplars of the technology's broader societal risks, helping to shape
the trajectory of regulation and public acceptance far beyond the
electoral domain.

\section{Discussion}\label{discussion}

This study provides a systematic account of how people perceive and
react to the use of AI in election campaigns. Across three preregistered
studies, we find that the public broadly disapproves of AI in politics,
especially when it is used deceptively. But this disapproval does not
translate into meaningful electoral penalties for the parties involved.
Instead, public exposure to deceptive AI practices in campaigns leads to
downstream effects that extend beyond politics: heightened support for
AI regulation, increased feelings of personal disempowerment, and
stronger preference for safety over innovation in AI governance.

These findings support our theoretical framework. First, the public
differentiates clearly between types of AI use: campaign operations and
voter outreach are viewed more neutrally, while deception is
consistently seen as a norm violation. Second, AI uses that are seen as
norm violation also met with negative reactions. Third, these reactions
are not confined to campaign evaluations; they spill over into broader
domains, including personal autonomy, and regulatory preference.
Finally, while people disapprove of deceptive uses, they do not penalize
the parties responsible, suggesting a misalignment between public values
and political incentives.

This misalignment is critical. Political actors may gain competitive
advantages from using AI deceptively while facing little or no electoral
cost. Yet their actions carry externalities: they fuel support for
overly restrictive AI policies. In effect, parties can offload the
social costs of AI-enabled deception onto the broader political and
technological system. This is especially concerning as AI becomes an
increasingly prominent feature of democratic processes and the public
arena (Jungherr 2023; Jungherr and Schroeder 2023).

Importantly, not all uses of AI in elections carry such risks. As our
findings show, the public distinguishes between harmful and potentially
helpful applications. AI tools used to improve operations or personalize
voter outreach are viewed with more nuance. This finding emphasizes the
need for research to account for the full breadth of electoral uses of
AI, instead of focusing narrowly on AI-enabled deception (Foos 2024;
Kruschinski et al. 2025; Tomić, Damnjanović, and Tomić 2023). This
differentiation also suggests a way forward: regulatory responses to AI
uses in election campaigns must be targeted and proportionate. Blanket
bans or one-size-fits-all restrictions risk deterring beneficial
innovation while doing little to prevent bad actors from exploiting
loopholes.

Our results also underscore the symbolic power of electoral AI use.
Because election campaigns are high-visibility events, they shape how
people think about AI more generally. Deceptive uses in politics can
become exemplars of technological risk, leading to reactions that affect
AI governance in domains far beyond elections. This amplifies the
responsibility of political actors, regulators, and professional
observers like journalists and academics alike.

Beyond its implications for political practice and regulatory policy,
this study also offers perspectives for future research. The framework
presented here (i.e.~the differentiation of AI use types, the centrality
of perceived norm violations, and the mapping of downstream effects)
provides a generalizable structure for cross-national and longitudinal
work. While our data come from the U.S. electorate, future studies can
apply this framework to other political systems. In countries with
multi-party dynamics, stronger privacy norms, different media ecologies,
or less polarized electorates, both public reactions and campaign
behavior may differ in important ways (Dommett, Kefford, and Kruschinski
2024). For instance, outreach-related AI uses may trigger stronger
reactions in contexts where behavioral targeting is culturally or
legally contested, like Europe. Or deceptive practices might lead to
less of a backlash in more permissible campaign contexts, like Asian
democracies. More broadly, our findings underscore the need to more
prominently incorporate public opinion on AI and regulatory preferences
into wider debates on AI governance and policy design (Bengio et al.
2024; Cohen et al. 2024).

In addition, building on prior research into varying patterns of
technological adoption and the incentives driving them (Bimber,
Flanagin, and Stohl 2012; Earl and Kimport 2011; Jungherr, Schroeder,
and Stier 2019) future studies should investigate how political actors
evaluate the risks and benefits of adopting AI. How do resource-rich
versus resource-constrained campaigns decide whether to use AI tools?
Are some types of parties (e.g., system challengers or populists) more
willing to engage in deceptive uses? Qualitative studies of campaign
staff and consultants could shed light on these organizational dynamics
and further inform regulatory design.

Methodologically, the survey and experimental tools developed here can
be adapted for comparative studies or used to track change over time. As
AI tools and public familiarity evolve, longitudinal designs will be
essential to capturing shifting baselines of acceptance or concern.

Despite the strengths of our design, limitations remain. Our treatments
were short textual descriptions. This may underestimate the real-world
effects of repeated exposure to emotionally charged, or visually
engaging content as well as follow-up coverage in the media and
conversations in social circles. Also, the strong partisan divides in
the U.S. may limit generalizability to less polarized democracies.
Comparative and qualitative work is therefore essential to
contextualizing these findings and extending them to diverse democratic
environments.

How AI is used in election campaigns and which of these uses are
highlighted in public and academic discourse matters not only for
political outcomes, but also for democratic norms, public trust, and the
broader trajectory of technological governance. If deceptive
applications come to dominate the public imagination of AI in politics,
we risk not only undermining confidence in the integrity of electoral
competition but also eroding trust in AI innovation more broadly. To
counter this, public, academic, and regulatory efforts must capture the
full range of electoral AI use cases rather than focusing solely on
evocative but elusive threats. How citizens come to view both elections
and AI will depend on this broader, more nuanced understanding.

\section{Data Availability}\label{data-availability}

Preregistrations, data, and analysis scripts are available at the
project's OSF repository:

\textbf{Study 1}

\begin{itemize}
\tightlist
\item
  Prereg:
  \url{https://osf.io/3nrb4/?view_only=1d82e100d6084edd81d9c4af46f31a30}
\item
  Data and code:
  \url{https://osf.io/gheqz/?view_only=600ff099f37a457681a4b676c6457111}
\end{itemize}

\textbf{Study 2}

\begin{itemize}
\tightlist
\item
  Prereg:
  \url{https://osf.io/wsrkv/?view_only=6d55d846ae8d4ba886c3e3ce8076d845}
\item
  Data and code:
  \url{https://osf.io/8s7ye/?view_only=73e6bc5e3bb9434393efa1f8da4fe81b}
\end{itemize}

\textbf{Study 3}

\begin{itemize}
\tightlist
\item
  Prereg:
  \url{https://osf.io/vugp8/?view_only=5e4387422dc94458bb355e6e2e5fba3d}
\item
  Data and code:
  \url{https://osf.io/r3qa4/?view_only=22ab144a2ec9461a858810cce2abb259}
\end{itemize}

\section{Acknowledgements}\label{acknowledgements}

A. Jungherr, A. Rauchfleisch, and A. Wuttke contributed equally to the
project. A. Jungherr is supported by a grant from the
VolkswagenStiftung. A. Rauchfleisch is supported by a grant from the
National Science and Technology Council, Taiwan. A. Jungherr, A.
Rauchfleisch, and A. Wuttke have no competing interests. Supplementary
Information is available for this paper. Correspondence and requests for
materials should be addressed to A. Jungherr.

\section*{References}\label{references}
\addcontentsline{toc}{section}{References}

\phantomsection\label{refs}

\newpage

\appendix
\appendixpage

\section{Material and Methods}\label{material-and-methods}

\subsection{Study 1}\label{study-1}

In Study 1, we queried people for their opinions on specific uses of AI
in elections and their views on the benefits and risks of AI in other
areas. We ran a preregistered survey (n=1,199) among members of an
online panel that the market and public opinion research company
\emph{Ipsos} provided.

We used quotas on age, gender, region, and education to realize a sample
representative of the US electorate. As Table~\ref{tbl-quotas_study1}
shows, the sampling was largely successful. The average interview length
was 15 minutes. The survey was fielded between April 4 and April 17,
2024. The fieldwork was conducted in compliance with the standards ISO
9001:2015 and ISO 20252:2019, as were all study-related processes.
Before running the survey, we registered our research design, analysis
plan, and hypotheses about outcomes. We did not deviate from the
registered procedure (Preregistration:
\url{https://osf.io/3nrb4/?view_only=1d82e100d6084edd81d9c4af46f31a30}).

\begin{longtable}[]{@{}
  >{\raggedright\arraybackslash}p{(\columnwidth - 6\tabcolsep) * \real{0.1087}}
  >{\raggedright\arraybackslash}p{(\columnwidth - 6\tabcolsep) * \real{0.3478}}
  >{\raggedright\arraybackslash}p{(\columnwidth - 6\tabcolsep) * \real{0.2935}}
  >{\raggedleft\arraybackslash}p{(\columnwidth - 6\tabcolsep) * \real{0.2500}}@{}}
\caption{Comparison between official population census USA and realized
sample, Study 1}\label{tbl-quotas_study1}\tabularnewline
\toprule\noalign{}
\begin{minipage}[b]{\linewidth}\raggedright
Type
\end{minipage} & \begin{minipage}[b]{\linewidth}\raggedright
Category
\end{minipage} & \begin{minipage}[b]{\linewidth}\raggedright
Official Statistics (\%)
\end{minipage} & \begin{minipage}[b]{\linewidth}\raggedleft
Realized distribution (\%)
\end{minipage} \\
\midrule\noalign{}
\endfirsthead
\toprule\noalign{}
\begin{minipage}[b]{\linewidth}\raggedright
Type
\end{minipage} & \begin{minipage}[b]{\linewidth}\raggedright
Category
\end{minipage} & \begin{minipage}[b]{\linewidth}\raggedright
Official Statistics (\%)
\end{minipage} & \begin{minipage}[b]{\linewidth}\raggedleft
Realized distribution (\%)
\end{minipage} \\
\midrule\noalign{}
\endhead
\bottomrule\noalign{}
\endlastfoot
Gender & Male & 49.1 & 45.3 \\
Gender & Female & 50.9 & 54.2 \\
Gender & Diverse & & 0.3 \\
Gender & Other & & 0.3 \\
Age & 18-29 Years & 20.4 & 18.5 \\
Age & 30-44 Years & 25.8 & 26.0 \\
Age & 45-59 Years & 23.4 & 23.8 \\
Age & 60-75 Years & 30.4 & 31.7 \\
Region & New England Division & 4.7 & 4.6 \\
Region & Middle Atlantic Division & 12.8 & 13.9 \\
Region & East North Central Division & 14.1 & 14.9 \\
Region & West North Division & 6.4 & 6.8 \\
Region & South Atlantic Division & 20.4 & 21.2 \\
Region & East South Central Division & 5.8 & 6.1 \\
Region & West South Central Division & 12.1 & 12.4 \\
Region & Mountain Division & 7.6 & 6.8 \\
Region & Pacific Division & 16.0 & 13.3 \\
Education & Low (no college) & 37.7 & 38.4 \\
Education & Medium (some college) & 29.3 & 26.8 \\
Education & High (college plus) & 33.0 & 34.8 \\
\end{longtable}

As specified in the preregistration, we used three attention checks to
identify and exclude inattentive respondents. The first was an
open-ended question, the second was hidden in an item grid, and the
third was a simple single-choice question. The three checks were
distributed throughout the entire survey. Ipsos excluded respondents if
they failed two out of three attention checks. In the final dataset
provided by Ipsos, only one additional respondent remained with two
failed attention checks and was thus excluded from the analysis.
Table~\ref{tbl-attention1} gives an overview of the number of flagged
and excluded participants.

\begin{longtable}[]{@{}ll@{}}
\caption{Number of excluded respondents, Study
1}\label{tbl-attention1}\tabularnewline
\toprule\noalign{}
\textbf{Check} & \textbf{Number} \\
\midrule\noalign{}
\endfirsthead
\toprule\noalign{}
\textbf{Check} & \textbf{Number} \\
\midrule\noalign{}
\endhead
\bottomrule\noalign{}
\endlastfoot
Check 1 (open-ended question) & 11 \\
Check 2 (item grid) & 49 \\
Check 3 (single-choice question) & 49 \\
Excluded Respondents (2 out of 3) & 52 \\
\end{longtable}

Ipsos gave respondents the option of answering the English or Spanish
version of the questionnaire. This takes into account that a growing
population within the US is predominantly Spanish-speaking. Only two
non-excluded respondents chose the Spanish version of the questionnaire.

We provided respondents with short descriptions of various campaigning
tasks for which parties and candidates use AI. These tasks fall into
three broad categories:

\begin{itemize}
\tightlist
\item
  campaign operations;
\item
  voter outreach;
\item
  deception.
\end{itemize}

For each category we identified five example tasks that practitioners
and journalists have documented and discussed. See
Table~\ref{tbl-tasks_items}.

\begin{longtable}[]{@{}
  >{\raggedright\arraybackslash}p{(\columnwidth - 4\tabcolsep) * \real{0.3333}}
  >{\raggedright\arraybackslash}p{(\columnwidth - 4\tabcolsep) * \real{0.3333}}
  >{\raggedright\arraybackslash}p{(\columnwidth - 4\tabcolsep) * \real{0.3333}}@{}}
\caption{AI Campaign Tasks, Categories, Tasks, and Item
Wordings}\label{tbl-tasks_items}\tabularnewline
\toprule\noalign{}
\begin{minipage}[b]{\linewidth}\raggedright
Category
\end{minipage} & \begin{minipage}[b]{\linewidth}\raggedright
Task
\end{minipage} & \begin{minipage}[b]{\linewidth}\raggedright
Item
\end{minipage} \\
\midrule\noalign{}
\endfirsthead
\toprule\noalign{}
\begin{minipage}[b]{\linewidth}\raggedright
Category
\end{minipage} & \begin{minipage}[b]{\linewidth}\raggedright
Task
\end{minipage} & \begin{minipage}[b]{\linewidth}\raggedright
Item
\end{minipage} \\
\midrule\noalign{}
\endhead
\bottomrule\noalign{}
\endlastfoot
Operations & Writing & Many campaigns are turning to AI to help
\textbf{craft emails, speeches, and policy documents}. This technology
offers a cost-saving advantage. \\
& Transcription & Campaigns are turning to AI for the
\textbf{transcription of meetings, speeches, and broadcasts}. This saves
valuable time and resources. \\
& Resource Allocation & Campaigns employ AI to \textbf{optimize walk
lists for door-to-door voter outreach}. These AI-generated lists help
volunteers visit as many homes as possible in a limited time. \\
& Deepfakes (benign) & Some campaigns use AI to \textbf{make funny and
creative pictures of their candidates}, drawing from fantasy and pop
culture. This can make candidates feel more like regular people and
connect with voters better. \\
& Automating Interactions & Campaigns are now using AI to \textbf{create
digital characters that look and sound very real}. These artificial
characters can talk and interact with people on their own. They can
answer questions or even start conversations with visitors on websites
or in online ads, always presenting the campaign's topics and
positions. \\
Voter Outreach & Message Testing \& Opinion Research & Campaigns are
using AI to \textbf{simulate virtual focus groups, testing how different
messages resonate with a wide range of audiences}. This helps campaigns
shape their approach and understand what voters care about. \\
& Data Driven Targeting & Campaigns are using AI to \textbf{pinpoint the
best contacts for voter outreach}. By predicting how individuals might
respond to campaign efforts, AI helps campaigns concentrate on those who
are more likely to be receptive or motivated to vote. \\
& Fundraising & Campaigns are harnessing AI to \textbf{enhance their
fundraising} efforts. By identifying supporters most likely to donate
and optimizing outreach materials, like emails or call scripts, AI can
significantly boost a campaign's financial resources. \\
& Ad Optimization & Campaigns use AI to \textbf{craft tailored digital
ads}. Depending on an individual's interests, concerns, or
characteristics, AI can produce optimized variations of campaign ads. It
can also swiftly adapt ads in response to current events, making them
timely and relevant. \\
& Outreach Optimization & Campaigns employ AI to \textbf{craft tailored
texts for emails and call center scripts}. By anticipating how
individuals might respond to outreach, AI helps fine-tune messages to
inspire actions like voting, volunteering, or donating. \\
Deception & Deceptive Robocalls & Campaigns leverage AI technology for
\textbf{automated, lifelike robocalls} pretending the caller is the
candidate or a volunteer. Within these calls, AI systems can
independently engage with individuals, initiating conversations or
responding to queries. \\
& Deepfakes (self-promotional) & Campaigns are using AI to
\textbf{produce synthetic images, videos, or voice recordings}, commonly
known as deepfakes. These can \textbf{convincingly portray candidates in
a positive light}, often impressing even discerning viewers. \\
& Deepfakes (negative campaigning) & Campaigns are using AI to
\textbf{produce synthetic images, videos, or voice recordings}, commonly
known as deepfakes. These can \textbf{convincingly portray opposing
candidates in a negative light}, often deceiving even discerning
viewers. \\
& Astroturfing (social media) & Some campaigns employ AI to
\textbf{create fake social media posts}, seemingly from supporters of
their candidate. The aim is to sway sentiment in digital communication
environments in their favor. \\
& Astroturfing (interactive) & Some campaigns utilize AI to
\textbf{craft emails and social media messages aimed at journalists and
news editors}, pretending to be genuine supporters to simulate strong
public backing. \\
\end{longtable}

We showed each respondent three randomly drawn tasks for each category
and asked them whether this use of AI in elections

\begin{enumerate}
\def\labelenumi{(\arabic{enumi})}
\tightlist
\item
  worried them,
\item
  felt like a norm violation
\item
  was likely to make politics more interesting to voters, and
\item
  increase participation.
\end{enumerate}

See Table \ref{tab:study1_measurement} for an overview of variables,
question wordings, operationalizations, key diagnostics of item
measurements for Study 1. For the full questionnaire and answer options,
see preregistration
\url{https://osf.io/3nrb4/?view_only=1d82e100d6084edd81d9c4af46f31a30}.

\begin{landscape}
    
    \begin{longtable}{p{3cm}p{6cm}p{5cm}p{3cm}p{2cm}c}
    \caption{Table for measurements Study 1.}
    \label{tab:study1_measurement} \\
    \toprule
    Variable & Question Wording  & Operationalization & $\alpha$ / Spearman-Brown & M (SD) & n\\
    \midrule
    \endfirsthead
    
    \toprule
    Variable & Question Wording  & Answer Options & $\alpha$ / Spearman-Brown & M (SD) & n\\
    \midrule
    \endhead

    \endfoot
    
    \bottomrule
    \endlastfoot

Worry & This use of AI worries me a lot.  &  1 (Completely Disagree) - 7 (Completely Agree) & & 5.07 (1.9) & 7194\\
Norm Violation & This AI use is not how political campaigns should act.  & 1 (Completely Disagree) - 7 (Completely Agree) & & 5.02 (1.92) & 7194\\
\addlinespace
Rise in Voter Involvement & & 1 (Completely Disagree) - 7 (Completely Agree) & 2 items, SB = 0.82-0.90\footnotemark & 3.63 (1.73) & 7194\\
 & This use of AI makes politics more interesting to voters.  & & & 3.53 (1.84) & 7194\\
 & This use of AI increases voter engagement. & &&3.72 (1.82)   &  7194\\
\addlinespace
AI Benefits & & 1 (Completely Disagree) - 7 (Completely Agree) & 3 items, $\alpha$ = 0.84 & 4.53 (1.52) & 920\\
 & AI will drive significant economic expansion in the U.S.   & &  & 4.48 (1.77) & 987\\
 & AI will help governments to more efficiently plan for the future and manage crises.   & & & 4.34 (1.78) & 1024\\
 & AI will provide the U.S. military with advanced  defense capabilities, ensuring national  security.   && & 4.71 (1.71) & 1001\\
 \addlinespace
 \footnotetext{Spearman–Brown values were calculated for each case separately.} \\
AI Risks  & & 1 (Completely Disagree) - 7 (Completely Agree) & 3 items, $\alpha$ = 0.78 & 5.17 (1.38) & 986\\
 & AI is likely to cause widespread job displacement and unemployment.   & & & 5.12 (1.65) & 1081\\
 & Unchecked AI development could pose existential threats to humanity.  & & & 5.47 (1.62) & 1090\\
 & AI in military applications can lead to unintended escalations or conflicts due to a lack of human judgment.  & & & 5.24 (1.59) & 1030\\
 \addlinespace
 Gender & & 1 = male   & &45.37\% & 1199\\
Education (High)  & & 1 = Master's degree or higher && 13.51\% & 1199\\

\end{longtable}
\end{landscape}

We combined items \emph{This use of AI makes politics more interesting
to voters}, and \emph{This use of AI increases voter engagement} into
one index -- \emph{Rise in Voter Involvement} -- to capture AI's likely
impact on voter interest and mobilization. For Spearman-Brown values for
the \emph{Rise in Voter Involvement} index for each campaign task, see
Table \ref{tab:study1_sb_involve}.

\begin{table}
    \caption{Spearman-Brown values for Rise in Voter Involvement index by campaign task.}
    \label{tab:study1_sb_involve}
    \centering
    \begin{tabular}[t]{lrr}
    \toprule
    Case & Spearman-Brown & n\\
    \midrule
    Operations: Writing & 0.86 & 461\\
    Operations: Transcription & 0.90 & 483\\
    Operations: Resource Allocation & 0.84 & 506\\
    Operations: Deepfakes (benign) & 0.90 & 443\\
    Operations: Automating Interactions & 0.90 & 505\\
    \addlinespace
    Outreach: Message Testing \& Opinion Research & 0.91 & 508\\
    Outreach: Data Driven Targeting & 0.88 & 483\\
    Outreach: Fundraising & 0.86 & 452\\
    Outreach: Ad Optimization & 0.88 & 469\\
    Outreach: Outreach Optimization & 0.87 & 486\\
    \addlinespace
    Deception: Deceptive Robocalls & 0.88 & 474\\
    Deception: Deepfakes (self-promotional) & 0.88 & 461\\
    Deception: Deepfakes (negative campaigning) & 0.87 & 482\\
    Deception: Astroturfing (social media) & 0.82 & 491\\
    Deception: Astroturfing (interactive) & 0.88 & 490\\
    \bottomrule
    \end{tabular}
    \end{table}

For key diagnostics of responses on item level, see Tables
\ref{tab:study1_worry}, \ref{tab:study1_norm}, and
\ref{tab:study1_involve}. The column \emph{Mean (SD)} reports means and
standard deviations calculated on the raw responses. The column
\emph{Weighted Mean (Weighted SD)} reports means calculated on the
respondents adjusted by weights provided by \emph{Ipsos} to match our
realized sample more exactly to the US population.

\begin{table}
    \caption{Outcome: Worry}
    \label{tab:study1_worry}
    \centering
    \begin{tabular}[t]{lllr}
    \toprule
    Case & Mean (SD) & Weighted Mean (Weighted SD) & n\\
    \midrule
    Operations: Automating Interactions & 5.24 (1.88) & 5.22 (1.88) & 505\\
    Operations: Deepfakes (benign) & 4.76 (1.94) & 4.74 (1.95) & 443\\
    Operations: Resource Allocation & 4.52 (2.03) & 4.51 (2.02) & 506\\
    Operations: Transcription & 4.66 (2.01) & 4.64 (2.01) & 483\\
    Operations: Writing & 5.08 (1.81) & 5.07 (1.81) & 461\\
    \addlinespace
    Outreach: Ad Optimization & 5.01 (1.85) & 5.01 (1.85) & 469\\
    Outreach: Data Driven Targeting & 4.78 (1.9) & 4.76 (1.9) & 483\\
    Outreach: Fundraising & 4.87 (1.9) & 4.85 (1.91) & 452\\
    Outreach: Message Testing \& Opinion Research & 4.97 (1.91) & 4.95 (1.91) & 508\\
    Outreach: Outreach Optimization & 5 (1.92) & 4.99 (1.91) & 486\\
    \addlinespace
    Deception: Astroturfing (interactive) & 5.23 (1.85) & 5.21 (1.86) & 490\\
    Deception: Astroturfing (social media) & 5.6 (1.74) & 5.58 (1.75) & 491\\
    Deception: Deceptive Robocalls & 5.31 (1.86) & 5.31 (1.86) & 474\\
    Deception: Deepfakes (negative campaigning) & 5.45 (1.8) & 5.42 (1.81) & 482\\
    Deception: Deepfakes (self-promotional) & 5.53 (1.76) & 5.51 (1.77) & 461\\
    \bottomrule
    \end{tabular}
    \end{table}
    
    \begin{table}
    \caption{Outcome: Norm Violation}
    \label{tab:study1_norm}
    \centering
    \begin{tabular}[t]{lllr}
    \toprule
    Case & Mean (SD) & Weighted Mean (Weighted SD) & n\\
    \midrule
    Operations: Automating Interactions & 5.23 (1.92) & 5.21 (1.92) & 505\\
    Operations: Deepfakes (benign) & 4.89 (1.92) & 4.87 (1.93) & 443\\
    Operations: Resource Allocation & 4.56 (1.93) & 4.55 (1.92) & 506\\
    Operations: Transcription & 4.75 (1.92) & 4.74 (1.92) & 483\\
    Operations: Writing & 4.87 (1.92) & 4.85 (1.93) & 461\\
    \addlinespace
    Outreach: Ad Optimization & 4.93 (1.82) & 4.93 (1.81) & 469\\
    Outreach: Data Driven Targeting & 4.88 (1.84) & 4.87 (1.83) & 483\\
    Outreach: Fundraising & 4.85 (1.84) & 4.83 (1.84) & 452\\
    Outreach: Message Testing \& Opinion Research & 4.81 (1.93) & 4.79 (1.91) & 508\\
    Outreach: Outreach Optimization & 4.98 (1.89) & 4.97 (1.89) & 486\\
    \addlinespace
    Deception: Astroturfing (interactive) & 5.17 (1.9) & 5.15 (1.9) & 490\\
    Deception: Astroturfing (social media) & 5.42 (1.94) & 5.39 (1.94) & 491\\
    Deception: Deceptive Robocalls & 5.23 (1.98) & 5.22 (1.97) & 474\\
    Deception: Deepfakes (negative campaigning) & 5.41 (1.87) & 5.38 (1.88) & 482\\
    Deception: Deepfakes (self-promotional) & 5.39 (1.9) & 5.38 (1.9) & 461\\
    \bottomrule
    \end{tabular}
    \end{table}
    
    \begin{table}
    \caption{Rise in Voter Involvement}
    \label{tab:study1_involve}
    \centering
    \begin{tabular}[t]{lllr}
    \toprule
    Case & Mean (SD) & Weighted Mean (Weighted SD) & n\\
    \midrule
    Operations: Automating Interactions & 3.64 (1.82) & 3.68 (1.83) & 505\\
    Operations: Deepfakes (benign) & 3.86 (1.71) & 3.88 (1.7) & 443\\
    Operations: Resource Allocation & 3.84 (1.68) & 3.87 (1.68) & 506\\
    Operations: Transcription & 3.73 (1.77) & 3.75 (1.76) & 483\\
    Operations: Writing & 3.55 (1.63) & 3.57 (1.63) & 461\\
    \addlinespace
    Outreach: Ad Optimization & 3.83 (1.67) & 3.85 (1.68) & 469\\
    Outreach: Data Driven Targeting & 3.67 (1.68) & 3.68 (1.68) & 483\\
    Outreach: Fundraising & 3.8 (1.71) & 3.82 (1.71) & 452\\
    Outreach: Message Testing \& Opinion Research & 3.62 (1.73) & 3.65 (1.72) & 508\\
    Outreach: Outreach Optimization & 3.57 (1.67) & 3.58 (1.66) & 486\\
    \addlinespace
    Deception: Astroturfing (interactive) & 3.56 (1.76) & 3.57 (1.76) & 490\\
    Deception: Astroturfing (social media) & 3.47 (1.71) & 3.49 (1.71) & 491\\
    Deception: Deceptive Robocalls & 3.25 (1.78) & 3.27 (1.78) & 474\\
    Deception: Deepfakes (negative campaigning) & 3.48 (1.75) & 3.51 (1.76) & 482\\
    Deception: Deepfakes (self-promotional) & 3.55 (1.73) & 3.57 (1.74) & 461\\
    \bottomrule
    \end{tabular}
    \end{table}

As specified in the preregistration under \emph{Missing data}, we used
data imputation to fill in the missing responses for \emph{AI Benefits}
and \emph{AI Risks}. Of the 1199 respondents, 134 have a missing
response for at least one of the items. We checked the background of the
respondents with missing data. Older people, women, people with less
prior experience with AI tools, and people with lower education were
likelier to not respond to one of the six items about risks and
benefits. The data is thus not missing completely at random. However, as
we do not expect differences within the different strata regarding
attitudes on average between responding and non-responding participants,
it is justified to assume the data is missing at
random\textsuperscript{26,27}. Furthermore, our items are not sensitive
in any case, which could indicate differences within strata between
responding and non-responding participants.

For data imputation, we followed the procedure recommended in the
literature\textsuperscript{27}. Using the R package
\emph{mice}\textsuperscript{28}, we created 100 datasets with imputed
data for the missing values using predictive mean
matching\textsuperscript{27}. We used all six risks and benefits items
(if available): the use of AI tools in professional life, the use of AI
tools in private life, age, gender (male), education (high), political
orientation, party ID leaning, and geographic region as predictors for
predictive mean matching. After imputing the data, we created mean
indices for benefits and risks within each dataset (as we did in our
incomplete data). We then estimated multilevel models for worry, norm
violation, and rising political involvement based on each imputed
dataset. In the final step, we then pooled the results of the
models\textsuperscript{29,30} with the \emph{mice} package and estimated
the marginal pooled effects with the \emph{marginaleffects package} in
R\textsuperscript{31}.

Overall, the results for all three outcome variables were consistent
between the pooled models with imputed data (see Table
\ref{tab:study1_wo_missings}) and those with missing data (see Table
\ref{tab:study1_imputed}). Our main text reports the results from the
pooled models with complete data.

\subsection{Study 2}\label{study-2}

In our second study, we test the causal effects of learning about
different types of AI use in elections. We ran a preregistered survey
experiment with members of an online panel provided by \emph{Prolific}
(n=1,985). We used quotas to realize a sample resembling the US
electorate. As Table~\ref{tbl-quotas_study2} shows, the sampling was
somewhat successful. The survey was fielded between June 19 and June 21,
2024.

\begin{longtable}[]{@{}
  >{\raggedright\arraybackslash}p{(\columnwidth - 6\tabcolsep) * \real{0.1220}}
  >{\raggedright\arraybackslash}p{(\columnwidth - 6\tabcolsep) * \real{0.2683}}
  >{\raggedright\arraybackslash}p{(\columnwidth - 6\tabcolsep) * \real{0.3293}}
  >{\raggedleft\arraybackslash}p{(\columnwidth - 6\tabcolsep) * \real{0.2805}}@{}}
\caption{Comparison between official population census USA and realized
sample, Study 2}\label{tbl-quotas_study2}\tabularnewline
\toprule\noalign{}
\begin{minipage}[b]{\linewidth}\raggedright
Type
\end{minipage} & \begin{minipage}[b]{\linewidth}\raggedright
Category
\end{minipage} & \begin{minipage}[b]{\linewidth}\raggedright
Official Statistics (\%)
\end{minipage} & \begin{minipage}[b]{\linewidth}\raggedleft
Realized Sample (\%)
\end{minipage} \\
\midrule\noalign{}
\endfirsthead
\toprule\noalign{}
\begin{minipage}[b]{\linewidth}\raggedright
Type
\end{minipage} & \begin{minipage}[b]{\linewidth}\raggedright
Category
\end{minipage} & \begin{minipage}[b]{\linewidth}\raggedright
Official Statistics (\%)
\end{minipage} & \begin{minipage}[b]{\linewidth}\raggedleft
Realized Sample (\%)
\end{minipage} \\
\midrule\noalign{}
\endhead
\bottomrule\noalign{}
\endlastfoot
Gender & Male & 49.1 & 50.5 \\
Gender & Female & 50.9 & 48.7 \\
Gender & Other & & 0.8 \\
Age & 18-29 Years & 20.4 & 17.8 \\
Age & 30-44 Years & 25.8 & 26.9 \\
Age & 45-59 Years & 23.4 & 28.2 \\
Age & 60-75 Years & 30.4 & 24.9 \\
Age & 76+ Years & 30.4 & 2.2 \\
Education & Low (no college) & 37.7 & 34.1 \\
Education & Medium (some college) & 29.3 & 40.2 \\
Education & High (college plus) & 33.0 & 18.3 \\
Education & Other & & 7.4 \\
\end{longtable}

We divided respondents into three treatment and one control group (C, n
= 497). Treatment 1 (T1, n = 497) contained information about campaigns'
uses of AI for deception. Treatment 2 (T2, n = 494) contained
information about campaigns' uses of AI for improving campaign
operations. Treatment 3 (T3, n = 497) contained information about
campaigns' uses of AI for voter outreach. We registered our research
design, analysis plan, and hypotheses about outcomes before the survey.
We did not deviate from the registered procedure (Preregistration:
\url{https://osf.io/wsrkv/?view_only=6d55d846ae8d4ba886c3e3ce8076d845}).

As Table~\ref{tbl-rando_study2} shows randomization between the
treatment groups worked out.

\begin{longtable}[]{@{}
  >{\raggedright\arraybackslash}p{(\columnwidth - 10\tabcolsep) * \real{0.0862}}
  >{\raggedright\arraybackslash}p{(\columnwidth - 10\tabcolsep) * \real{0.1897}}
  >{\raggedleft\arraybackslash}p{(\columnwidth - 10\tabcolsep) * \real{0.1810}}
  >{\raggedleft\arraybackslash}p{(\columnwidth - 10\tabcolsep) * \real{0.2241}}
  >{\raggedleft\arraybackslash}p{(\columnwidth - 10\tabcolsep) * \real{0.1897}}
  >{\raggedleft\arraybackslash}p{(\columnwidth - 10\tabcolsep) * \real{0.1293}}@{}}
\caption{Randomization, Study 2}\label{tbl-rando_study2}\tabularnewline
\toprule\noalign{}
\begin{minipage}[b]{\linewidth}\raggedright
Type
\end{minipage} & \begin{minipage}[b]{\linewidth}\raggedright
Category
\end{minipage} & \begin{minipage}[b]{\linewidth}\raggedleft
T1: Deception (\%)
\end{minipage} & \begin{minipage}[b]{\linewidth}\raggedleft
T2: Voter Outreach (\%)
\end{minipage} & \begin{minipage}[b]{\linewidth}\raggedleft
T3: Operations (\%)
\end{minipage} & \begin{minipage}[b]{\linewidth}\raggedleft
Control (\%)
\end{minipage} \\
\midrule\noalign{}
\endfirsthead
\toprule\noalign{}
\begin{minipage}[b]{\linewidth}\raggedright
Type
\end{minipage} & \begin{minipage}[b]{\linewidth}\raggedright
Category
\end{minipage} & \begin{minipage}[b]{\linewidth}\raggedleft
T1: Deception (\%)
\end{minipage} & \begin{minipage}[b]{\linewidth}\raggedleft
T2: Voter Outreach (\%)
\end{minipage} & \begin{minipage}[b]{\linewidth}\raggedleft
T3: Operations (\%)
\end{minipage} & \begin{minipage}[b]{\linewidth}\raggedleft
Control (\%)
\end{minipage} \\
\midrule\noalign{}
\endhead
\bottomrule\noalign{}
\endlastfoot
Gender & Male & 56.1 & 47.8 & 49.3 & 48.7 \\
Gender & Female & 43.5 & 51.6 & 49.5 & 50.3 \\
Gender & Other & 0.4 & 0.6 & 1.2 & 1.0 \\
Age & 18-29 Years & 14.7 & 19.2 & 19.1 & 18.1 \\
Age & 30-44 Years & 30.0 & 25.9 & 26.8 & 24.9 \\
Age & 45-59 Years & 29.6 & 26.9 & 28.6 & 27.6 \\
Age & 60-75 Years & 24.9 & 25.5 & 22.9 & 26.4 \\
Age & 76+ Years & 0.8 & 2.4 & 2.6 & 3.0 \\
Edu. & Low (no college) & 37.2 & 32.4 & 33.6 & 33.0 \\
Edu. & Medium (some college) & 37.6 & 40.3 & 41.6 & 41.2 \\
Edu. & High (college plus) & 17.3 & 19.0 & 18.7 & 18.3 \\
Edu. & Other & 7.8 & 8.3 & 6.0 & 7.4 \\
\end{longtable}

\begin{tcolorbox}[enhanced jigsaw, rightrule=.15mm, left=2mm, colbacktitle=quarto-callout-tip-color!10!white, breakable, bottomrule=.15mm, leftrule=.75mm, opacitybacktitle=0.6, bottomtitle=1mm, title=\textcolor{quarto-callout-tip-color}{\faLightbulb}\hspace{0.5em}{Treatment 1: Deception}, toprule=.15mm, coltitle=black, toptitle=1mm, colframe=quarto-callout-tip-color-frame, titlerule=0mm, arc=.35mm, opacityback=0, colback=white]

Candidates from all parties, including Republicans and Democrats, and
candidates from various third parties use AI in their campaigns.

They use AI technology to produce videos depicting fictional scenarios
involving their opposing candidates. Picture a scenario where a video
portrays an opposing candidate making controversial statements or
engaging in questionable conduct -- all generated using AI.

These resulting videos are frequently captivating and occasionally gain
substantial traction, especially among demographics typically tricky to
engage with for political parties. However, it's essential to note that
these videos are pure fiction and do not reflect actual events or
actions.

\end{tcolorbox}

\begin{tcolorbox}[enhanced jigsaw, rightrule=.15mm, left=2mm, colbacktitle=quarto-callout-tip-color!10!white, breakable, bottomrule=.15mm, leftrule=.75mm, opacitybacktitle=0.6, bottomtitle=1mm, title=\textcolor{quarto-callout-tip-color}{\faLightbulb}\hspace{0.5em}{Treatment 2: Campaign Operations}, toprule=.15mm, coltitle=black, toptitle=1mm, colframe=quarto-callout-tip-color-frame, titlerule=0mm, arc=.35mm, opacityback=0, colback=white]

Candidates from all parties, including Republicans and Democrats, and
candidates from various third parties use AI in their campaigns.

For example, they use AI to automatically generate emails, speeches, and
policy documents.

By leveraging AI, campaigns can conserve valuable resources through the
automation of repetitive tasks and help with the allocation of funds and
volunteer hours. This enhanced efficiency aids campaigns in pursuing
their objectives effectively and is particularly beneficial for
financially constrained campaigns.

\end{tcolorbox}

\begin{tcolorbox}[enhanced jigsaw, rightrule=.15mm, left=2mm, colbacktitle=quarto-callout-tip-color!10!white, breakable, bottomrule=.15mm, leftrule=.75mm, opacitybacktitle=0.6, bottomtitle=1mm, title=\textcolor{quarto-callout-tip-color}{\faLightbulb}\hspace{0.5em}{Treatment 3: Voter Outreach}, toprule=.15mm, coltitle=black, toptitle=1mm, colframe=quarto-callout-tip-color-frame, titlerule=0mm, arc=.35mm, opacityback=0, colback=white]

Candidates from all parties, including Republicans and Democrats, and
candidates from various third parties use AI in their campaigns.

They use AI technology to create customized voter outreach strategies.
By meticulously analyzing consumer data, online activities, and voting
histories, AI has the capacity to create and distribute personalized
campaign messages that align with each voter's specific interests. For
instance, a tech-savvy urban dweller might receive information about the
party's innovation initiatives, while a young parent could receive
insights on education reform.

Campaigns rely on AI-enabled outreach to distinguish themselves in a sea
of generic political communications and to effectively connect with
voters on subjects that resonate with them.

\end{tcolorbox}

For an overview of variables, question wordings, operationalizations,
and key diagnostics of item measurements for Study 2, see Table
\ref{tab:study2_measurement}. For the complete questionnaire and answer
options, see preregistration
(\url{https://osf.io/wsrkv/?view_only=6d55d846ae8d4ba886c3e3ce8076d845}).

\begin{landscape}
    
    \begin{longtable}{p{3cm}p{6cm}p{5cm}p{3cm}p{2cm}c}
    \caption{Table for measurements Study 2.}
    \label{tab:study2_measurement} \\
    \toprule
    Variable & Question Wording  & Operationalization & $\alpha$ / Spearman-Brown & M (SD) & n\\
    \midrule
    \endfirsthead
    
    \toprule
    Variable & Question Wording  & Answer Options & $\alpha$ / Spearman-Brown & M (SD) & n\\
    \midrule
    \endhead
    
    \multicolumn{6}{r}{{ }} \\
    \endfoot
    
    \bottomrule
    \endlastfoot

    Worry & The use of AI in campaigns worries me a lot. &  1 (Completely Disagree) - 7 (Completely Agree) & & 4.86 (1.89) & 1985\\
    Norm Violation & Using AI is not how political campaigns should act. &  1 (Completely Disagree) - 7 (Completely Agree) & & 4.9 (1.87) & 1985\\
    \addlinespace
    Rise in Voter Involvement && 1 (Completely Disagree) - 7 (Completely Agree) & 2 items, SB = 0.81\footnotemark & 4.16 (1.52) & 1985\\
     & Using AI can make politics more interesting to voters. & &&  3.91 (1.72) & 1985\\
     & Using AI can increase voter engagement. & &&  4.41 (1.6) & 1985\\
     \footnotetext{Spearman–Brown values were calculated for each case separately.} \\
    \addlinespace
    Fairness of Elections && 1 (Completely Disagree) - 7 (Completely Agree) & 3 items, $\alpha$ = 0.78 & 3.74 (1.47) & 1985\\
     & Campaigns often resort to illegal activities to increase their chances of winning. &&& 4.44 (1.62) & 1985\\
     & Elections in this country are conducted fairly. &&& 4.23 (1.92) & 1985\\
     & Most campaigns compete fairly. &&& 3.42 (1.75) & 1985\\
     \addlinespace
     Favorability Political Parties & Parties in general & 1 (Very unfavorable opinion) - 7 (Very favorable opinion) && 3.23 (1.31) & 1985\\
    Favorability Republican Party & Republicans & 1 (Very unfavorable opinion) - 7 (Very favorable opinion) && 3.2 (2.01) & 1985\\
    Favorability Democratic Party & Democrats & 1 (Very unfavorable opinion) - 7 (Very favorable opinion) && 3.68 (1.93) & 1985\\
    \addlinespace
    Stricter Oversight of AI Use in Elections &  & 1 (Completely Disagree) - 7 (Completely Agree) & 3 items, $\alpha$ = 0.92 & 4.8 (1.84) & 1985\\
     & State regulators should limit political parties' and candidates' use of AI, even if this reduces their ability to engage with voters. &&& 4.82 (1.94) & 1985\\
     & Digital platforms like Facebook, Google, Instagram, TikTok, and YouTube should restrict AI use in political content and ads, even if this reduces parties' ability to engage with voters. &&& 4.91 (1.95) & 1985\\
     & Parties and candidates should be banned from digital platforms like Facebook, Google, Instagram, TikTok, or YouTube if they repeatedly publish or share content produced or manipulated with AI. &&& 4.66 (2.06) & 1985\\
     \addlinespace
     AI Contributes to Personal Loss of Control && 1 (Completely Disagree) - 7 (Completely Agree) & 2 items, SB = 0.84\footnotemark & 4.48 (1.66) & 1985\\
     \footnotetext{Spearman–Brown values were calculated for each case separately.}\\
     & As AI increasingly takes over communication, it becomes harder to make well-informed decisions. &&& 4.55 (1.76) & 1985\\
     & As AI increasingly takes over decision-making, we risk losing control over our lives. &&& 4.41 (1.82) & 1985\\
     \addlinespace
     Prioritize Innovation in AI Regulation && 1 - Strong focus on safety, 7 - Strong focus on innovation && 3.24 (1.66) & 1985\\
    \addlinespace
    Stricter Regulation of AI in general && 1 (Completely Disagree) - 7 (Completely Agree) &3 items, $\alpha$ = 0.8& 3.59 (1.66) & 1985\\
     & The development, training, and use of powerful AI systems should be immediately stopped and forbidden for the time being &&& 3.16 (2.08) & 1985\\
     & The development, training, and use of powerful AI systems should only be possible under strict government supervision and control. &&& 4.08 (1.92) & 1985\\
     & Societies are better off  not allowing the development, training, and use of powerful AI systems. &&& 3.52 (1.9) & 1985\\
     \addlinespace
     Support for AI Moratorium & Do you agree that the development, training, and use of powerful AI systems should be stopped immediately and forbidden for the time being? & 1=Yes & & 32.49\% & 1985\\
    \addlinespace
    Gender (Male) && 1 = male && 48.72\% & 1985\\
    Education (High) && 1 = Master's degree or higher && 18.34\% & 1985\\

\end{longtable}
\end{landscape}

We used a factual manipulation check for the treatment groups by asking
respondents a knowledge question with three answer options and a ``not
sure'' option. The majority of participants selected the correct answer
(Deception: 75.25 = ``Creation of Fictional AI-Generated Videos,''
Operations: 94.77\% = ``AI-Assisted Campaign Management,'' Outreach:
72.47\% = ``Personalized Voter Outreach Using AI''). Respondents with
incorrect answers in the deception and outreach conditions primarily
selected ``AI-Assisted Campaign Management,'' likely because it was the
most general answer option.

\subsection{Study 3}\label{study-3}

In Study 3, we test whether parties face a penalty for deceptive AI uses
attributed to them and whether partisans' group-protective cognitions
lead to heterogeneous effects of being informed about deceptive uses.

For this preregistered study (Preregistration:
\url{https://osf.io/vugp8/?view_only=5e4387422dc94458bb355e6e2e5fba3d},
we recruited three samples containing only respondents identifying as
partisans for (1) Democrats (n=1,489), (2) Republicans (n=1,485), or as
(3) Independent (n=1,477). \emph{Prolific} prescreened partisans. No
attempt to be representative was made. The survey was fielded between
June 25 and June 30, 2024.

For an overview of variables, question wordings, operationalizations,
and key diagnostics of item measurements for Study 3, see Tables
\ref{tab:study3_measurement_dems}, \ref{tab:study3_measurement_ind}, and
\ref{tab:study3_measurement_reps}. For the complete questionnaire and
answer options, see preregistration
(\url{https://osf.io/vugp8/?view_only=5e4387422dc94458bb355e6e2e5fba3d}.

\begin{landscape}
    
    \begin{longtable}{p{3cm}p{6cm}p{5cm}p{3cm}p{2cm}c}
    \caption{Table for measurements Study 3: Democrat Sample}
    \label{tab:study3_measurement_dems} \\
    \toprule
    Variable & Question Wording  & Operationalization & $\alpha$ / Spearman-Brown & M (SD) & n\\
    \midrule
    \endfirsthead
    
    \toprule
    Variable & Question Wording  & Answer Options & $\alpha$ / Spearman-Brown & M (SD) & n\\
    \midrule
    \endhead
    
    \multicolumn{6}{r}{{ }} \\
    \endfoot
    
    \bottomrule
    \endlastfoot

Worry & The use of AI in campaigns worries me a lot. & 1 (Completely Disagree) - 7 (Completely Agree) && 5.55 (1.71) & 1489\\
Norm Violation & Using AI is not how political campaigns should act. & 1 (Completely Disagree) - 7 (Completely Agree) && 5.65 (1.67) & 1489\\
\addlinespace
Rise in Voter Involvement && 1 (Completely Disagree) - 7 (Completely Agree) & 2 items, SB = 0.8\footnotemark & 3.97 (1.56) & 1489\\
 & Using AI can make politics more interesting to voters. &&& 3.7 (1.79) & 1489\\
 & Using AI can increase voter engagement. &&& 4.25 (1.63) & 1489\\
 \footnotetext{Spearman–Brown values were calculated for each case separately.} \\
 \addlinespace
Fairness of Elections && 1 (Completely Disagree) - 7 (Completely Agree) & 3 items, $\alpha$ = 0.78 & 4.07 (1.41) & 1489\\
 & Campaigns often resort to illegal activities to increase their chances of winning. &&& 4.18 (1.6) & 1489\\
 & Elections in this country are conducted fairly. &&& 4.74 (1.85) & 1489\\
 & Most campaigns compete fairly. &&& 3.65 (1.72) & 1489\\
\addlinespace
Favorability Republican Party & Republicans & 1 (Very unfavorable opinion) -  7 (Very favorable opinion) && 1.95 (1.22) & 1489\\
Favorability Democratic Party & Democrats & 1 (Very unfavorable opinion) -  7 (Very favorable opinion) && 5.29 (1.32) & 1489\\
\addlinespace
Prioritize Innovation in AI Regulation & & 1 (Strong focus on safety) - 7 (Strong focus on innovation) && 2.84 (1.56) & 1489\\
Support for AI Moratorium & Do you agree that the development, training, and use of powerful AI systems should be stopped immediately and forbidden for the time being? & 1 = Yes &&  34.45\% & 1489\\
Gender (Male) & & 1 = male && 49.29\% & 1489\\
Education (High) && 1 = Master's degree or higher && 22.63\% & 1489\\

\end{longtable}

    \begin{longtable}{p{3cm}p{6cm}p{5cm}p{3cm}p{2cm}c}
    \caption{Table for measurements Study 3: Independent Sample.}
    \label{tab:study3_measurement_ind} \\
    \toprule
    Variable & Question Wording  & Operationalization & $\alpha$ / Spearman-Brown & M (SD) & n\\
    \midrule
    \endfirsthead
    
    \toprule
    Variable & Question Wording  & Answer Options & $\alpha$ / Spearman-Brown & M (SD) & n\\
    \midrule
    \endhead
    
    \multicolumn{6}{r}{{ }} \\
    \endfoot
    
    \bottomrule
    \endlastfoot

Worry & The use of AI in campaigns worries me a lot. & 1 (Completely Disagree) - 7 (Completely Agree) && 5.47 (1.69) & 1485\\
Norm Violation & Using AI is not how political campaigns should act. &1 (Completely Disagree) - 7 (Completely Agree) && 5.64 (1.67) & 1485\\
\addlinespace
Rise in Voter Involvement & &1 (Completely Disagree) - 7 (Completely Agree) & 2 items, SB = 0.8\footnotemark & 3.99 (1.56) & 1485\\
 & Using AI can make politics more interesting to voters. &&& 3.73 (1.8) & 1485\\
 & Using AI can increase voter engagement. &&& 4.24 (1.65) & 1485\\
 \footnotetext{Spearman–Brown values were calculated for each case separately.} \\
 \addlinespace
Fairness of Elections && 1 (Completely Disagree) - 7 (Completely Agree) & 3 items, $\alpha$ = 0.78& 3.35 (1.42) & 1485\\
 & Campaigns often resort to illegal activities to increase their chances of winning. &&& 4.62 (1.61) & 1485\\
 & Elections in this country are conducted fairly. &&& 3.68 (1.87) & 1485\\
 & Most campaigns compete fairly. &&& 2.99 (1.67) & 1485\\
\addlinespace
Favorability Republican Party & Republicans & 1 (Very unfavorable opinion) -  7 (Very favorable opinion) && 2.77 (1.58) & 1485\\
Favorability Democratic Party & Democrats & 1 (Very unfavorable opinion) -  7 (Very favorable opinion) && 3.33 (1.58) & 1485\\
\addlinespace
Prioritize Innovation in AI Regulation & & 1 (Strong focus on safety) - 7 (Strong focus on innovation) && 2.93 (1.63) & 1485\\
Support for AI Moratorium & Do you agree that the development, training, and use of powerful AI systems should be stopped immediately and forbidden for the time being?& 1 = Yes &&  36.77\% & 1485\\
Gender (Male) && 1 = male& & 49.23\% & 1485\\
Education (High) && 1 = Master's degree or higher && 13.94\% & 1485\\

\end{longtable}

    \begin{longtable}{p{3cm}p{6cm}p{5cm}p{3cm}p{2cm}c}
    \caption{Table for measurements Study 3: Republican Sample.}
    \label{tab:study3_measurement_reps} \\
    \toprule
    Variable & Question Wording  & Operationalization & $\alpha$ / Spearman-Brown & M (SD) & n\\
    \midrule
    \endfirsthead
    
    \toprule
    Variable & Question Wording  & Answer Options & $\alpha$ / Spearman-Brown & M (SD) & n\\
    \midrule
    \endhead
    
    \multicolumn{6}{r}{{ }} \\
    \endfoot
    
    \bottomrule
    \endlastfoot

Worry & The use of AI in campaigns worries me a lot.& 1 (Completely Disagree) - 7 (Completely Agree) && 5.11 (1.86) & 1477\\
Norm Violation & Using AI is not how political campaigns should act.& 1 (Completely Disagree) - 7 (Completely Agree) && 5.37 (1.84) & 1477\\
\addlinespace
Rise in Voter Involvement && 1 (Completely Disagree) - 7 (Completely Agree) & 2 items, SB = 0.8\footnotemark & 4.04 (1.6) & 1477\\
 & Using AI can make politics more interesting to voters. &&& 3.8 (1.81) & 1477\\
 & Using AI can increase voter engagement. &&& 4.27 (1.69) & 1477\\
\footnotetext{Spearman–Brown values were calculated for each case separately.} \\
\addlinespace
Fairness of Elections && 1 (Completely Disagree) - 7 (Completely Agree) & 3 items, $\alpha$ = 0.78 & 3.39 (1.39) & 1477\\
 & Campaigns often resort to illegal activities to increase their chances of winning. &&& 4.66 (1.62) & 1477\\
 & Elections in this country are conducted fairly. &&& 3.61 (1.76) & 1477\\
 & Most campaigns compete fairly. &&& 3.23 (1.68) & 1477\\
\addlinespace
Favorability Republican Party & Republicans & 1 (Very unfavorable opinion) -  7 (Very favorable opinion) && 5.4 (1.38) & 1477\\
Favorability Democratic Party & Democrats & 1 (Very unfavorable opinion) -  7 (Very favorable opinion) && 2.45 (1.36) & 1477\\
\addlinespace
Prioritize Innovation in AI Regulation & & 1 (Strong focus on safety) - 7 (Strong focus on innovation) && 3.14 (1.76) & 1477\\
Support for AI Moratorium & Do you agree that the development, training, and use of powerful AI systems should be stopped immediately and forbidden for the time being?& 1 = Yes &&  40.69\% & 1477\\
Gender (Male) & & 1 = male & & 49.42\% & 1477\\
Education (High) && 1 = Master's degree or higher && 18.75\% & 1477\\

\end{longtable}
\end{landscape}

Respondents in these three samples were exposed to either of two
treatments or were assigned a pure control group that did not receive
any information. Treatments contained information about deceptive uses
of AI by candidates from the Democratic Party (T1) or the Republican
Party (T2). This allows us to identify whether group-protective
cognition leads partisans to discount information about uses of AI by
parties they support, compared to adjusting related attitudes when being
informed about deceptive uses by parties they oppose and how this
compares to reactions by Independents. We registered our research
design, analysis plan, and hypotheses about outcomes before the survey.
We did not deviate from the registered procedure.

\begin{tcolorbox}[enhanced jigsaw, rightrule=.15mm, left=2mm, colbacktitle=quarto-callout-tip-color!10!white, breakable, bottomrule=.15mm, leftrule=.75mm, opacitybacktitle=0.6, bottomtitle=1mm, title=\textcolor{quarto-callout-tip-color}{\faLightbulb}\hspace{0.5em}{Treatment 1: Democrat Deception}, toprule=.15mm, coltitle=black, toptitle=1mm, colframe=quarto-callout-tip-color-frame, titlerule=0mm, arc=.35mm, opacityback=0, colback=white]

It was recently reported that candidates from the Democratic Party use
AI in their campaigns.

Democrats use AI technology to produce videos depicting fictional
scenarios involving their opposing candidates. Picture a scenario where
a video portrays an opposing candidate making controversial statements
or engaging in questionable conduct -- all generated using AI.

These resulting videos are frequently captivating and occasionally gain
substantial traction, especially among demographics typically tricky to
engage with for political parties. However, it's essential to note that
these videos are pure fiction and do not reflect actual events or
actions.

\end{tcolorbox}

\begin{tcolorbox}[enhanced jigsaw, rightrule=.15mm, left=2mm, colbacktitle=quarto-callout-tip-color!10!white, breakable, bottomrule=.15mm, leftrule=.75mm, opacitybacktitle=0.6, bottomtitle=1mm, title=\textcolor{quarto-callout-tip-color}{\faLightbulb}\hspace{0.5em}{Treatment 2: Republican Deception}, toprule=.15mm, coltitle=black, toptitle=1mm, colframe=quarto-callout-tip-color-frame, titlerule=0mm, arc=.35mm, opacityback=0, colback=white]

It was recently reported that candidates from the Republican Party use
AI in their campaigns.

Republicans use AI technology to produce videos depicting fictional
scenarios involving their opposing candidates. Picture a scenario where
a video portrays an opposing candidate making controversial statements
or engaging in questionable conduct -- all generated using AI.

These resulting videos are frequently captivating and occasionally gain
substantial traction, especially among demographics typically tricky to
engage with for political parties. However, it's essential to note that
these videos are pure fiction and do not reflect actual events or
actions.

\end{tcolorbox}

We used a factual manipulation check for the treatment groups by asking
respondents two knowledge questions with three answer options and a
``not sure'' option. Most participants in all three samples selected the
correct answers for both questions.

In the Democratic sample, a majority correctly identified both the
described case (Republican Deception: 91.58\% = ``Creation of Fictional
AI-Generated Videos''; Democrat Deception: 89.14\% = ``Creation of
Fictional AI-Generated Videos'') and the party cue (Republican Deception
= 96.59\% correct; Democrat Deception=97.13\% correct).

In the Independents sample, the majority also selected the correct case
(Republican Deception = 90.84\% correct; Democrat Deception = 90.69\%
correct) and party cue option (Republican Deception = 95.11\% correct;
Democrat Deception = 95.34\% correct).

The same pattern could be observed in the Republican sample for the case
(Republican Deception = 89.23\% correct; Democrat Deception = 85.31\%
correct) and party cue (Republican Deception =91.30\% correct; Democrat
Deception = 94.08\% correct).

\section{Preregistrations}\label{appendix_pre_reg_hyps}

We preregistered our research, design, analytical procedure, and
hypotheses of outcomes for our studies:

\begin{itemize}
\tightlist
\item
  Attitudes toward uses of AI in elections. Preregistration:
  \url{https://osf.io/3nrb4/?view_only=1d82e100d6084edd81d9c4af46f31a30}.
\item
  Effects of being informed about specific uses of AI in elections.
  Preregistration
  \url{https://osf.io/wsrkv/?view_only=6d55d846ae8d4ba886c3e3ce8076d845}.
\item
  Impact of partisanship on effects of being informed about specific
  uses of AI in elections:
  \url{https://osf.io/vugp8/?view_only=5e4387422dc94458bb355e6e2e5fba3d}.
\end{itemize}

We did not deviate from the pregistrations in our research design and
analyses.

In Tables Table~\ref{tbl-pre_reg_hyps_camp},
Table~\ref{tbl-pre_reg_hyps_effects}, Table~\ref{tbl-pre_reg_hyps_ind},
Table~\ref{tbl-pre_reg_hyps_reps}, and
Table~\ref{tbl-pre_reg_hyps_dems}, we list each hypothesis and report
whether it was supported by our analysis or not.

\subsection{Study 1: Preregistered
Hypotheses}\label{study-1-preregistered-hypotheses}

\begin{longtable}[]{@{}
  >{\raggedright\arraybackslash}p{(\columnwidth - 10\tabcolsep) * \real{0.0541}}
  >{\raggedright\arraybackslash}p{(\columnwidth - 10\tabcolsep) * \real{0.5946}}
  >{\raggedright\arraybackslash}p{(\columnwidth - 10\tabcolsep) * \real{0.1081}}
  >{\raggedright\arraybackslash}p{(\columnwidth - 10\tabcolsep) * \real{0.1081}}
  >{\raggedright\arraybackslash}p{(\columnwidth - 10\tabcolsep) * \real{0.1081}}
  >{\raggedright\arraybackslash}p{(\columnwidth - 10\tabcolsep) * \real{0.0270}}@{}}
\caption{Preregistered hypotheses on attitudes toward use of AI in
elections (Study 1). See Table
\ref{tab:study1_imputed}.}\label{tbl-pre_reg_hyps_camp}\tabularnewline
\toprule\noalign{}
\begin{minipage}[b]{\linewidth}\raggedright
\end{minipage} & \begin{minipage}[b]{\linewidth}\raggedright
Hypotheses
\end{minipage} & \begin{minipage}[b]{\linewidth}\raggedright
Support
\end{minipage} & \begin{minipage}[b]{\linewidth}\raggedright
Est.
\end{minipage} & \begin{minipage}[b]{\linewidth}\raggedright
CI
\end{minipage} & \begin{minipage}[b]{\linewidth}\raggedright
p
\end{minipage} \\
\midrule\noalign{}
\endfirsthead
\toprule\noalign{}
\begin{minipage}[b]{\linewidth}\raggedright
\end{minipage} & \begin{minipage}[b]{\linewidth}\raggedright
Hypotheses
\end{minipage} & \begin{minipage}[b]{\linewidth}\raggedright
Support
\end{minipage} & \begin{minipage}[b]{\linewidth}\raggedright
Est.
\end{minipage} & \begin{minipage}[b]{\linewidth}\raggedright
CI
\end{minipage} & \begin{minipage}[b]{\linewidth}\raggedright
p
\end{minipage} \\
\midrule\noalign{}
\endhead
\bottomrule\noalign{}
\endlastfoot
H1 & Deceptive AI use will be perceived as more worrisome than AI use
for \ldots{} & & & & \\
a) & \ldots{} voter outreach. & Yes & -0.50 & -0.72 -- -0.28 &
\textless0.001 \\
b) & \ldots{} improving internal operations. & Yes & -0.57 & -0.79 --
-0.35 & \textless0.001 \\
H2 & Deceptive AI use will be perceived as a stronger norm violation
than AI use for \ldots{} & & & & \\
a) & \ldots{} voter outreach. & Yes & -0.43 & -0.63 -- -0.23 &
\textless0.001 \\
b) & \ldots{} improving internal operations. & Yes & -0.46 & -0.66 --
-0.26 & \textless0.001 \\
H3 & The expected benefits for the political process of deceptive AI use
will be lower than for AI use for \ldots{} & & & & \\
a) & \ldots{} voter outreach. & Yes & 0.23 & 0.13 -- 0.33 &
\textless0.001 \\
b) & \ldots{} improving internal operations. & Yes & 0.26 & 0.16 -- 0.36
& \textless0.001 \\
H4 & AI applications involving deception are more likely to be
associated with risks specifically mentioning deception, compared to
\ldots{} & & & & \\
a) & \ldots{} AI applications focused on improving a campaign's
operations. (See Table \ref{tab:open_answer_deception}). & Yes & Odds
ratio: 0.32 & 0.21 -- 0.51 & \textless0.001 \\
b) & \ldots{} AI applications that improve a campaign's voter outreach.
(See Table \ref{tab:open_answer_deception}). & Yes & Odds ratio: 0.28 &
0.18 -- 0.44 & \textless0.001 \\
H5 & AI applications that improve a campaign's voter outreach are more
likely to be associated with risks specifically referencing the reduced
agency of voters compared to \ldots{} & & & & \\
a) & \ldots{} AI applications focused on improving a campaign's
operations. (See Table \ref{tab:open_answer_agency}). & Yes & Odds
ratio: 0.62 & 0.45 -- 0.86 & 0.004 \\
b) & \ldots{} AI applications involving deception. (See Table
\ref{tab:open_answer_agency}). & Yes & Odds ratio: 0.57 & 0.41 -- 0.79 &
0.001 \\
H6 & The stronger the belief in AI's benefits to society, the lower the
level of worry regarding its use in campaigns. & Yes & -0.16 & -0.22 --
-0.1 & \textless0.001 \\
H7 & \ldots{} the less likely AI use in campaigns is to be perceived as
a norm violation. & Yes & -0.09 & -0.15 -- -0.03 & \textless0.001 \\
H8 & \ldots{} the stronger the expectation of AI's positive impact on
politics when used in political campaigns. & Yes & 0.52 & 0.46 -- 0.58 &
\textless0.001 \\
H9 & The stronger the belief in AI's risks to society, the higher the
level of worry regarding its use in campaigns. & Yes & 0.51 & 0.45 --
0.57 & \textless0.001 \\
H10 & \ldots{} the more likely AI use in campaigns is to be perceived as
a norm violation. & Yes & 0.38 & 0.32 -- 0.44 & \textless0.001 \\
H11 & \ldots{} the weaker the expectation of AI's positive impact on
politics when used in political campaigns. & Yes & -0.07 & -0.13 --
-0.01 & \textless0.001 \\
\end{longtable}

\subsection{Study 2: Preregistered
Hypotheses}\label{study-2-preregistered-hypotheses}

\begin{longtable}[]{@{}
  >{\raggedright\arraybackslash}p{(\columnwidth - 10\tabcolsep) * \real{0.0541}}
  >{\raggedright\arraybackslash}p{(\columnwidth - 10\tabcolsep) * \real{0.5946}}
  >{\raggedright\arraybackslash}p{(\columnwidth - 10\tabcolsep) * \real{0.1081}}
  >{\raggedright\arraybackslash}p{(\columnwidth - 10\tabcolsep) * \real{0.1081}}
  >{\raggedright\arraybackslash}p{(\columnwidth - 10\tabcolsep) * \real{0.1081}}
  >{\raggedright\arraybackslash}p{(\columnwidth - 10\tabcolsep) * \real{0.0270}}@{}}
\caption{Preregistered hypotheses on the effects of different AI uses in
elections (Study 2).}\label{tbl-pre_reg_hyps_effects}\tabularnewline
\toprule\noalign{}
\begin{minipage}[b]{\linewidth}\raggedright
\end{minipage} & \begin{minipage}[b]{\linewidth}\raggedright
Hypotheses
\end{minipage} & \begin{minipage}[b]{\linewidth}\raggedright
Support
\end{minipage} & \begin{minipage}[b]{\linewidth}\raggedright
Est.
\end{minipage} & \begin{minipage}[b]{\linewidth}\raggedright
CI
\end{minipage} & \begin{minipage}[b]{\linewidth}\raggedright
p
\end{minipage} \\
\midrule\noalign{}
\endfirsthead
\toprule\noalign{}
\begin{minipage}[b]{\linewidth}\raggedright
\end{minipage} & \begin{minipage}[b]{\linewidth}\raggedright
Hypotheses
\end{minipage} & \begin{minipage}[b]{\linewidth}\raggedright
Support
\end{minipage} & \begin{minipage}[b]{\linewidth}\raggedright
Est.
\end{minipage} & \begin{minipage}[b]{\linewidth}\raggedright
CI
\end{minipage} & \begin{minipage}[b]{\linewidth}\raggedright
p
\end{minipage} \\
\midrule\noalign{}
\endhead
\bottomrule\noalign{}
\endlastfoot
H1 & Individuals informed about the deceptive use of AI in election
campaigns will express greater concern about the use of AI in elections
compared to those informed about \ldots{} (See Table
\ref{tab:study2_worry}) & & & & \\
a) & \ldots{} AI use for voter outreach. & Yes & -1.29 & -1.50 - -1.07 &
\textless0.001 \\
b) & \ldots{} AI use for campaign operations. & Yes & -1.35 & -1.57 -
-1.13 & \textless0.001 \\
c) & \ldots{} those not informed about AI uses in elections (control
group). & Yes & -0.74 & -0.95 - -0.52 & \textless0.001 \\
H2 & Individuals informed about the deceptive use of AI in election
campaigns will perceive a greater sense of norm violation by campaigns
using AI compared to those informed about \ldots{} (See Table
\ref{tab:study2_norm}) & & & & \\
a) & \ldots{} AI use for voter outreach. & Yes & -1.62 & -1.83 - -1.41 &
\textless0.001 \\
b) & \ldots{} AI use for campaign operations. & Yes & -1.41 & -1.63 -
-1.19 & \textless0.001 \\
c) & \ldots{} those not informed about AI uses in elections (control
group). & Yes & -0.65 & -0.85 - -0.44 & \textless0.001 \\
H3 & Individuals informed about the deceptive use of AI in election
campaigns will perceive less potential for a rise in voter involvement
in the use of AI for politics compared to those informed about \ldots{}
(See Table \ref{tab:study2_rise}) & & & & \\
a) & \ldots{} AI use for voter outreach. & Yes & 0.65 & 0.47 - 0.83 &
\textless0.001 \\
b) & \ldots{} AI use for campaign operations. & No & -0.08 & -0.27 -
0.10 & 0.389 \\
c) & \ldots{} those not informed about AI uses in elections (control
group). & No & -0.29 & -0.48 - -0.09 & 0.004 \\
H4 & Individuals informed about the deceptive use of AI in election
campaigns will perceive elections as less fair compared to those
informed about \ldots{} (See Table \ref{tab:study2_fairness}) & & & & \\
a) & \ldots{} AI use for voter outreach. & No & 0.08 & -0.10 - 0.26 &
0.398 \\
b) & \ldots{} AI use for campaign operations. & No & 0.12 & -0.06 - 0.31
& 0.195 \\
c) & \ldots{} those not informed about AI uses in elections (control
group). & No & 0.10 & -0.08 - 0.28 & 0.294 \\
H5 & Individuals informed about the deceptive use of AI in election
campaigns will have less favorable opinions of i) specific parties and
ii) political parties in general compared to those informed about
\ldots{} (See Tables \ref{tab:study2_parties}, \ref{tab:study2_reps},
and \ref{tab:study2_dems}) & & & & \\
a) & \ldots{} AI use for voter outreach. & & & & \\
i) & specific parties (Rep) & No & 0.03 & -0.13 - 0.19 & 0.703 \\
i) & specific parties (Dem) & No & 0.08 & -0.09 - 0.26 & 0.348 \\
ii) & parties in general & No & -0.07 & -0.23 - 0.10 & 0.434 \\
b) & \ldots{} AI use for campaign operations. & & & & \\
i) & specific parties (Rep) & No & 0.01 & -0.16 - 0.18 & 0.879 \\
i) & specific parties (Dem) & No & -0.07 & -0.25 - 0.10 & 0.416 \\
ii) & parties in general & No & -0.02 & -0.19 - 0.14 & 0.777 \\
c) & \ldots{} those not informed about AI uses in elections (control
group). & & & & \\
i) & specific parties (Rep) & No & 0.11 & -0.06 - 0.29 & 0.201 \\
i) & specific parties (Dem) & No & 0.11 & -0.06 - 0.29 & 0.198 \\
ii) & parties in general & No & 0.05 & -0.11 - 0.22 & 0.517 \\
H6 & Individuals informed about the deceptive use of AI in election
campaigns will express stronger support for governmental regulation of
election campaigns compared to those informed about \ldots{} (See Table
\ref{tab:study2_reg_elections}) & & & & \\
a) & \ldots{} AI use for voter outreach. & Yes & -1.14 & -1.36- -0.93 &
\textless0.001 \\
b) & \ldots{} AI use for campaign operations. & Yes & -1.12 & -1.34 -
-0.90 & \textless0.001 \\
c) & \ldots{} those not informed about AI uses in elections (control
group). & Yes & -0.61 & -0.82 - -0.39 & \textless0.001 \\
H7 & Individuals informed about the deceptive use of AI in election
campaigns will experience a greater sense of personal loss of control
compared to those informed about \ldots{} (See Table
\ref{tab:study2_loss}) & & & & \\
a) & \ldots{} AI use for voter outreach. & No & -0.20 & -0.41 - 0.00 &
0.051 \\
b) & \ldots{} AI use for campaign operations. & No & -0.11 & -0.31 -
0.09 & 0.289 \\
c) & \ldots{} those not informed about AI uses in elections (control
group). & Yes & -0.28 & -0.48 - -0.07 & 0.008 \\
H8 & Individuals informed about the deceptive use of AI in election
campaigns will more strongly prioritize safety in AI regulation compared
to those informed about \ldots{} (See Table \ref{tab:study2_prios}) & &
& & \\
a) & \ldots{} AI use for voter outreach. & No & 0.15 & -0.05 - 0.35 &
0.142 \\
b) & \ldots{} AI use for campaign operations. & Yes & 0.26 & 0.06 - 0.46
& 0.011 \\
c) & \ldots{} those not informed about AI uses in elections (control
group). & Yes & 0.23 & 0.02 - 0.43 & 0.031 \\
H9 & Individuals informed about the deceptive use of AI in election
campaigns will express greater support for an AI moratorium compared to
those informed about \ldots{} (See Table \ref{tab:study2_mora}) & & &
& \\
a) & \ldots{} AI use for voter outreach. & Yes & -0.07 & -0.13 - -0.01 &
0.024 \\
b) & \ldots{} AI use for campaign operations. & No & -0.06 & -0.12 -
0.00 & 0.062 \\
c) & \ldots{} those not informed about AI uses in elections (control
group). & Yes & -0.09 & -0.15 - -0.03 & 0.002 \\
H10 & Individuals informed about the deceptive use of AI in election
campaigns will express stronger support for stricter measures of AI
regulation compared to those informed about \ldots{} (See Table
\ref{tab:study2_reg_ai}) & & & & \\
a) & \ldots{} AI use for voter outreach. & No & -0.16 & -0.37 - 0.05 &
0.138 \\
b) & \ldots{} AI use for campaign operations. & No & -0.18 & -0.39 -
0.02 & 0.083 \\
c) & \ldots{} those not informed about AI uses in elections (control
group). & No & -0.20 & -0.41 - 0.01 & 0.058 \\
\end{longtable}

\subsection{Study 3: Preregistered
Hypotheses}\label{study-3-preregistered-hypotheses}

\begin{longtable}[]{@{}
  >{\raggedright\arraybackslash}p{(\columnwidth - 10\tabcolsep) * \real{0.0541}}
  >{\raggedright\arraybackslash}p{(\columnwidth - 10\tabcolsep) * \real{0.5946}}
  >{\raggedright\arraybackslash}p{(\columnwidth - 10\tabcolsep) * \real{0.1081}}
  >{\raggedright\arraybackslash}p{(\columnwidth - 10\tabcolsep) * \real{0.1081}}
  >{\raggedright\arraybackslash}p{(\columnwidth - 10\tabcolsep) * \real{0.1081}}
  >{\raggedright\arraybackslash}p{(\columnwidth - 10\tabcolsep) * \real{0.0270}}@{}}
\caption{Independents Sample, Preregistered hypotheses on the role of
partisanship in effects of different AI uses in elections (Study
3)}\label{tbl-pre_reg_hyps_ind}\tabularnewline
\toprule\noalign{}
\begin{minipage}[b]{\linewidth}\raggedright
\end{minipage} & \begin{minipage}[b]{\linewidth}\raggedright
Hypotheses
\end{minipage} & \begin{minipage}[b]{\linewidth}\raggedright
Support
\end{minipage} & \begin{minipage}[b]{\linewidth}\raggedright
Est.
\end{minipage} & \begin{minipage}[b]{\linewidth}\raggedright
CI
\end{minipage} & \begin{minipage}[b]{\linewidth}\raggedright
p
\end{minipage} \\
\midrule\noalign{}
\endfirsthead
\toprule\noalign{}
\begin{minipage}[b]{\linewidth}\raggedright
\end{minipage} & \begin{minipage}[b]{\linewidth}\raggedright
Hypotheses
\end{minipage} & \begin{minipage}[b]{\linewidth}\raggedright
Support
\end{minipage} & \begin{minipage}[b]{\linewidth}\raggedright
Est.
\end{minipage} & \begin{minipage}[b]{\linewidth}\raggedright
CI
\end{minipage} & \begin{minipage}[b]{\linewidth}\raggedright
p
\end{minipage} \\
\midrule\noalign{}
\endhead
\bottomrule\noalign{}
\endlastfoot
H1 & Independents informed about the deceptive use of AI attributed to
a) the Republican Party or b) the Democratic Party will express greater
concern about AI use in elections than those not given that information
(See Table \ref{tab:study3_i_worry}). & & & & \\
a) & Republican Deception & Yes & 0.70 & 0.49 - 0.91 & \textless0.001 \\
b) & Democratic Deception & Yes & 0.59 & 0.39 - 0.80 & \textless0.001 \\
H2 & Independents informed about the deceptive use of AI attributed to
a) the Republican Party or b) the Democratic Party will perceive a
greater sense of norm violation about AI use in elections compared to
those not given that information (See Table \ref{tab:study3_i_norm}). &
& & & \\
a) & Republican Deception & Yes & 0.62 & 0.41 - 0.83 & \textless0.001 \\
b) & Democratic Deception & Yes & 0.50 & 0.29 - 0.71 & \textless0.001 \\
H3 & Independents informed about the deceptive use of AI attributed to
a) the Republican Party or b) the Democratic Party will perceive less
beneficial potential about AI use in elections than those not given that
information (See Table \ref{tab:study3_i_rise}). & & & & \\
a) & Republican Deception & No & 0.22 & 0.02 - 0.41 & 0.031 \\
b) & Democratic Deception & No & 0.37 & 0.19 - 0.56 & \textless0.001 \\
H4 & Independents informed about the deceptive use of AI attributed to
a) the Republican Party or b) the Democratic Party will more strongly
prioritize safety in AI regulation compared to those not given that
information (See Table \ref{tab:study3_i_prios}). & & & & \\
a) & Republican Deception & No & -0.06 & -0.26 - 0.14 & 0.564 \\
b) & Democratic Deception & No & -0.09 & -0.28 - 0.11 & 0.385 \\
H5 & Independents informed about the deceptive use of AI attributed to
a) the Republican Party or b) the Democratic Party will express greater
support for an AI moratorium compared to those not given that
information (See Table \ref{tab:study3_i_mora}). & & & & \\
a) & Republican Deception & No & 0.04 & -0.01 - 0.10 & 0.142 \\
b) & Democratic Deception & No & 0.03 & -0.03 - 0.09 & 0.265 \\
H6 & Independents informed about the deceptive use of AI attributed to
a) the Republican Party or b) the Democratic Party will perceive
elections as less fair compared to those not given that information (See
Table \ref{tab:study3_i_fair}). & & & & \\
a) & Republican Deception & No & -0.15 & -0.32 - 0.03 & 0.106 \\
b) & Democratic Deception & Yes & -0.23 & -0.40 - -0.05 & 0.012 \\
H7 & Independents informed about the deceptive use of AI attributed to
a) the Republican Party or b) the Democratic Party will assess i) the
Democratic Party and ii) the Republican Party less favorably compared to
those not given that information (See Tables \ref{tab:study3_i_reps} and
\ref{tab:study3_i_dems}). & & & & \\
a) i) & Republican Deception / Democratic Assessment & No & -0.09 &
-0.29 - 0.10 & 0.342 \\
a) ii) & Republican Deception / Republican Assessment & No & 0.11 &
-0.09 - 0.31 & 0.277 \\
b) i) & Democratic Deception / Democratic Assessment & No & -0.09 &
-0.28 - 0.11 & 0.377 \\
b) ii) & Democratic Deception / Republican Assessment & No & 0.06 &
-0.14 - 0.25 & 0.561 \\
\end{longtable}

\begin{longtable}[]{@{}
  >{\raggedright\arraybackslash}p{(\columnwidth - 10\tabcolsep) * \real{0.0541}}
  >{\raggedright\arraybackslash}p{(\columnwidth - 10\tabcolsep) * \real{0.5946}}
  >{\raggedright\arraybackslash}p{(\columnwidth - 10\tabcolsep) * \real{0.1081}}
  >{\raggedright\arraybackslash}p{(\columnwidth - 10\tabcolsep) * \real{0.1081}}
  >{\raggedright\arraybackslash}p{(\columnwidth - 10\tabcolsep) * \real{0.1081}}
  >{\raggedright\arraybackslash}p{(\columnwidth - 10\tabcolsep) * \real{0.0270}}@{}}
\caption{Republicans Sample, Preregistered hypotheses on the role of
partisanship in effects of different AI uses in elections (Study
3)}\label{tbl-pre_reg_hyps_reps}\tabularnewline
\toprule\noalign{}
\begin{minipage}[b]{\linewidth}\raggedright
\end{minipage} & \begin{minipage}[b]{\linewidth}\raggedright
Hypotheses
\end{minipage} & \begin{minipage}[b]{\linewidth}\raggedright
Support
\end{minipage} & \begin{minipage}[b]{\linewidth}\raggedright
Est.
\end{minipage} & \begin{minipage}[b]{\linewidth}\raggedright
CI
\end{minipage} & \begin{minipage}[b]{\linewidth}\raggedright
p
\end{minipage} \\
\midrule\noalign{}
\endfirsthead
\toprule\noalign{}
\begin{minipage}[b]{\linewidth}\raggedright
\end{minipage} & \begin{minipage}[b]{\linewidth}\raggedright
Hypotheses
\end{minipage} & \begin{minipage}[b]{\linewidth}\raggedright
Support
\end{minipage} & \begin{minipage}[b]{\linewidth}\raggedright
Est.
\end{minipage} & \begin{minipage}[b]{\linewidth}\raggedright
CI
\end{minipage} & \begin{minipage}[b]{\linewidth}\raggedright
p
\end{minipage} \\
\midrule\noalign{}
\endhead
\bottomrule\noalign{}
\endlastfoot
H1 & Republicans informed about the deceptive use of AI attributed to a)
the Republican Party or b) the Democratic Party will express greater
concern about AI use in elections than those not given that information
(See Table \ref{tab:study3_r_worry}). & & & & \\
a) & Republican Deception & No & 0.06 & -0.18 - 0.30 & 0.616 \\
b) & Democratic Deception & Yes & 0.43 & 0.21 - 0.66 & \textless0.001 \\
H2 & Republicans informed about the deceptive use of AI attributed to a)
the Republican Party or b) the Democratic Party will perceive a greater
sense of norm violation about AI use in elections compared to those not
given that information (See Table \ref{tab:study3_r_norm}). & & & & \\
a) & Republican Deception & Yes & 0.23 & 0.00 - 0.47 & 0.048 \\
b) & Democratic Deception & Yes & 0.54 & 0.31 - 0.76 & \textless0.001 \\
H3 & Republicans informed about the deceptive use of AI attributed to a)
the Republican Party or b) the Democratic Party will perceive less
potential about AI use in elections than those not given that
information (See Table \ref{tab:study3_r_rise}). & & & & \\
a) & Republican Deception & No & 0.28 & 0.08 - 0.48 & 0.005 \\
b) & Democratic Deception & No & 0.42 & 0.23 - 0.62 & \textless0.001 \\
H4 & Republicans informed about the deceptive use of AI attributed to a)
the Republican Party or b) the Democratic Party will more strongly
prioritize safety in AI regulation compared to those not given that
information (See Table \ref{tab:study3_r_prios}). & & & & \\
a) & Republican Deception & Yes & 0.27 & 0.06 - 0.48 & 0.012 \\
b) & Democratic Deception & No & 0.14 & -0.07 - 0.35 & 0.205 \\
H5 & Republicans informed about the deceptive use of AI attributed to a)
the Republican Party or b) the Democratic Party will express greater
support for an AI moratorium compared to those not given that
information (See Table \ref{tab:study3_r_mora}). & & & & \\
a) & Republican Deception & No & 0.00 & -0.06 - 0.06 & 0.961 \\
b) & Democratic Deception & No & 0.05 & -0.01 - 0.11 & 0.094 \\
H6 & Republicans informed about the deceptive use of AI attributed to a)
the Republican Party or b) the Democratic Party will perceive elections
as less fair compared to those not given that information (See Table
\ref{tab:study3_r_fair}). & & & & \\
a) & Republican Deception & No & 0.03 & -0.13 - 0.20 & 0.691 \\
b) & Democratic Deception & No & 0.09 & -0.08 - 0.26 & 0.315 \\
H7 & Republicans informed about the deceptive use of AI attributed to a)
the Republican Party or b) the Democratic Party will assess i) the
Democratic Party and ii) the Republican Party less favorably compared to
those not given that information (See Tables \ref{tab:study3_r_reps} and
\ref{tab:study3_r_dems}). & & & & \\
a) i) & Republican Deception / Democratic Assessment & No & 0.14 & -0.03
- 0.31 & 0.105 \\
a) ii) & Republican Deception / Republican Assessment & No & 0.06 &
-0.11 - 0.24 & 0.465 \\
b) i) & Democratic Deception / Democratic Assessment & No & -0.19 &
-0.36 - -0.03 & 0.023 \\
b) ii) & Democratic Deception / Republican Assessment & No & -0.02 &
-0.19 - 0.15 & 0.825 \\
\end{longtable}

\begin{longtable}[]{@{}
  >{\raggedright\arraybackslash}p{(\columnwidth - 10\tabcolsep) * \real{0.0541}}
  >{\raggedright\arraybackslash}p{(\columnwidth - 10\tabcolsep) * \real{0.5946}}
  >{\raggedright\arraybackslash}p{(\columnwidth - 10\tabcolsep) * \real{0.1081}}
  >{\raggedright\arraybackslash}p{(\columnwidth - 10\tabcolsep) * \real{0.1081}}
  >{\raggedright\arraybackslash}p{(\columnwidth - 10\tabcolsep) * \real{0.1081}}
  >{\raggedright\arraybackslash}p{(\columnwidth - 10\tabcolsep) * \real{0.0270}}@{}}
\caption{Democrat Sample, Preregistered hypotheses on the role of
partisanship in effects of different AI uses in elections (Study
3)}\label{tbl-pre_reg_hyps_dems}\tabularnewline
\toprule\noalign{}
\begin{minipage}[b]{\linewidth}\raggedright
\end{minipage} & \begin{minipage}[b]{\linewidth}\raggedright
Hypotheses
\end{minipage} & \begin{minipage}[b]{\linewidth}\raggedright
Support
\end{minipage} & \begin{minipage}[b]{\linewidth}\raggedright
Est.
\end{minipage} & \begin{minipage}[b]{\linewidth}\raggedright
CI
\end{minipage} & \begin{minipage}[b]{\linewidth}\raggedright
p
\end{minipage} \\
\midrule\noalign{}
\endfirsthead
\toprule\noalign{}
\begin{minipage}[b]{\linewidth}\raggedright
\end{minipage} & \begin{minipage}[b]{\linewidth}\raggedright
Hypotheses
\end{minipage} & \begin{minipage}[b]{\linewidth}\raggedright
Support
\end{minipage} & \begin{minipage}[b]{\linewidth}\raggedright
Est.
\end{minipage} & \begin{minipage}[b]{\linewidth}\raggedright
CI
\end{minipage} & \begin{minipage}[b]{\linewidth}\raggedright
p
\end{minipage} \\
\midrule\noalign{}
\endhead
\bottomrule\noalign{}
\endlastfoot
H1 & Democrats informed about the deceptive use of AI attributed to a)
the Republican Party or b) the Democratic Party will express greater
concern about AI use in elections than those not given that information
(See Table \ref{tab:study3_d_worry}). & & & & \\
a) & Republican Deception & Yes & 1.02 & 0.81 - 1.22 & \textless0.001 \\
b) & Democratic Deception & Yes & 0.68 & 0.47 - 0.90 & \textless0.001 \\
H2 & Democrats informed about the deceptive use of AI attributed to a)
the Republican Party or b) the Democratic Party will perceive a greater
sense of norm violation about AI use in elections compared to those not
given that information (See Table \ref{tab:study3_d_norm}). & & & & \\
a) & Republican Deception & Yes & 0.84 & 0.64 - 1.04 & \textless0.001 \\
b) & Democratic Deception & Yes & 0.56 & 0.36 - 0.77 & \textless0.001 \\
H3 & Democrats informed about the deceptive use of AI attributed to a)
the Republican Party or b) the Democratic Party will perceive less
beneficial potential about AI use in elections than those not given that
information (See Table \ref{tab:study3_d_rise}). & & & & \\
a) & Republican Deception & No & 0.16 & -0.03 - 0.36 & 0.096 \\
b) & Democratic Deception & No & 0.29 & 0.10 - 0.47 & 0.003 \\
H4 & Democrats informed about the deceptive use of AI attributed to a)
the Republican Party or b) the Democratic Party will more strongly
prioritize safety in AI regulation compared to those not given that
information (See Table \ref{tab:study3_d_prios}). & & & & \\
a) & Republican Deception & Yes & -0.28 & -0.47 - -0.09 & 0.003 \\
b) & Democratic Deception & No & -0.10 & -0.28 - 0.09 & 0.323 \\
H5 & Democrats informed about the deceptive use of AI attributed to a)
the Republican Party or b) the Democratic Party will express greater
support for an AI moratorium compared to those not given that
information (See Table \ref{tab:study3_d_mora}). & & & & \\
a) & Republican Deception & Yes & 0.12 & 0.06 - 0.18 & \textless0.001 \\
b) & Democratic Deception & Yes & 0.07 & 0.01 - 0.12 & 0.024 \\
H6 & Democrats informed about the deceptive use of AI attributed to a)
the Republican Party or b) the Democratic Party will perceive elections
as less fair compared to those not given that information (See Table
\ref{tab:study3_d_fair}). & & & & \\
a) & Republican Deception & Yes & -0.21 & -0.38 - -0.04 & 0.015 \\
b) & Democratic Deception & No & -0.16 & -0.33 - 0.01 & 0.063 \\
H7 & Democrats informed about the deceptive use of AI attributed to a)
the Republican Party or b) the Democratic Party will assess i) the
Democratic Party and ii) the Republican Party less favorably compared to
those not given that information (See Tables \ref{tab:study3_d_reps} and
\ref{tab:study3_d_dems}). & & & & \\
a) i) & Republican Deception / Democratic Assessment & No & -0.08 &
-0.23 - 0.06 & 0.265 \\
a) ii) & Republican Deception / Republican Assessment & No & -0.05 &
-0.20 - 0.11 & 0.559 \\
b) i) & Democratic Deception / Democratic Assessment & No & -0.03 &
-0.19 - 0.13 & 0.735 \\
b) ii) & Democratic Deception / Republican Assessment & No & -0.12 &
-0.28 - 0.05 & 0.168 \\
\end{longtable}

\section{Detailed Responses to Specific Uses of AI in Elections (Study
1)}\label{detailed-responses-to-specific-uses-of-ai-in-elections-study-1}

Table~\ref{tbl-response_shares} shows the shares of responses agreeing
with the statements \emph{This use of AI worries me a lot} (Worry),
\emph{This AI use is not how political campaigns should act} (Norm
Violation), and \emph{This use of AI makes politics more interesting to
voters} and \emph{This use of AI increases voter engagement} (Rise Voter
Involvement). Share is calculated as share of all responsens over the
value of 4 (on a scale of 1-7) of all responsens excluding NA.

\begin{longtable}[]{@{}
  >{\raggedright\arraybackslash}p{(\columnwidth - 6\tabcolsep) * \real{0.4286}}
  >{\raggedleft\arraybackslash}p{(\columnwidth - 6\tabcolsep) * \real{0.2000}}
  >{\raggedleft\arraybackslash}p{(\columnwidth - 6\tabcolsep) * \real{0.1905}}
  >{\raggedleft\arraybackslash}p{(\columnwidth - 6\tabcolsep) * \real{0.1810}}@{}}
\caption{Share responses that agree with assessment (Study
1)}\label{tbl-response_shares}\tabularnewline
\toprule\noalign{}
\begin{minipage}[b]{\linewidth}\raggedright
Campaign Task
\end{minipage} & \begin{minipage}[b]{\linewidth}\raggedleft
Worry (in \%)
\end{minipage} & \begin{minipage}[b]{\linewidth}\raggedleft
Norm Violation (in \%)
\end{minipage} & \begin{minipage}[b]{\linewidth}\raggedleft
Rise Voter Involvement (in \%)
\end{minipage} \\
\midrule\noalign{}
\endfirsthead
\toprule\noalign{}
\begin{minipage}[b]{\linewidth}\raggedright
Campaign Task
\end{minipage} & \begin{minipage}[b]{\linewidth}\raggedleft
Worry (in \%)
\end{minipage} & \begin{minipage}[b]{\linewidth}\raggedleft
Norm Violation (in \%)
\end{minipage} & \begin{minipage}[b]{\linewidth}\raggedleft
Rise Voter Involvement (in \%)
\end{minipage} \\
\midrule\noalign{}
\endhead
\bottomrule\noalign{}
\endlastfoot
Deception: Astroturfing (interactive) & 68.57 & 64.49 & 33.88 \\
Deception: Astroturfing (social media) & 76.37 & 69.86 & 32.18 \\
Deception: Deceptive Robocalls & 68.78 & 64.14 & 28.69 \\
Deception: Deepfakes (negative campaigning) & 71.58 & 68.26 & 29.46 \\
Deception: Deepfakes (self-promotional) & 71.80 & 68.76 & 36.23 \\
Operations: Automating Interactions & 67.72 & 63.56 & 35.84 \\
Operations: Deepfakes (benign) & 56.43 & 57.11 & 40.18 \\
Operations: Resource Allocation & 49.80 & 47.83 & 41.70 \\
Operations: Transcription & 53.83 & 53.42 & 33.13 \\
Operations: Writing & 62.91 & 55.10 & 29.50 \\
Outreach: Ad Optimization & 63.54 & 57.78 & 39.02 \\
Outreach: Data Driven Targeting & 56.11 & 55.90 & 34.16 \\
Outreach: Fundraising & 58.41 & 55.09 & 39.38 \\
Outreach: Message Testing \& Opinion Research & 60.83 & 54.92 & 35.63 \\
Outreach: Outreach Optimization & 60.91 & 58.85 & 32.30 \\
\end{longtable}

\section{Content Analysis Open Answer Fields (Study
1)}\label{content-analysis-open-answer-fields-study-1}

We were also interested in the risks people associate with specific
campaigning uses of AI and whether these risks correspond systematically
with the usage categories identified by use (i.e., campaign operations,
voter outreach, and deception). After the short description of specific
campaign tasks AI was used for (see Table~\ref{tbl-tasks_items}), we
posted the question: ``What risks for society do you see with this use
of AI in political campaigns?''. Respondents were provided with an open
answer field, where they could answer without having specific risks
prompted by us.

Our preregistered expectations (see preregistration
\url{https://osf.io/3nrb4/?view_only=1d82e100d6084edd81d9c4af46f31a30})
were supported by the analysis. We expected that people were
significantly more likely to mention risks associated with deception to
campaign tasks within our \emph{deception} category than to those in the
categories \emph{campaign operations} and \emph{voter outreach} (see
Table~\ref{tbl-pre_reg_hyps_camp} H4a,b). Correspondingly, we expected
that people were significantly more likely to mention risks associated
with reduced voter agency to campaign tasks within our \emph{voter
outreach} category than to those in the categories \emph{campaign
operations} and \emph{deception} (see Table~\ref{tbl-pre_reg_hyps_camp}
H5a,b).

To classify the responses, we used two preregistered prompts with the
OpenAI model ``gpt-4o-mini-2024-07-18'' (temperature=0) (see
\url{https://openai.com/index/gpt-4o-mini-advancing-cost-efficient-intelligence/}).
The workings of both prompts were validated by manual coding. Both
prompts worked well as the manual validation with 50 randomly sampled
answers for each variable indicated -- deception (Cohen's Kappa=0.87;
Accuracy=0.94) and reduced agency (Cohen's Kappa=1; Accuracy=1).

We used the following preregistered prompts:

\begin{tcolorbox}[enhanced jigsaw, rightrule=.15mm, left=2mm, colbacktitle=quarto-callout-note-color!10!white, breakable, bottomrule=.15mm, leftrule=.75mm, opacitybacktitle=0.6, bottomtitle=1mm, title=\textcolor{quarto-callout-note-color}{\faInfo}\hspace{0.5em}{Prompt risk 1: Deception (associated with AI-enabled deception)}, toprule=.15mm, coltitle=black, toptitle=1mm, colframe=quarto-callout-note-color-frame, titlerule=0mm, arc=.35mm, opacityback=0, colback=white]

Analyze the following response to an open survey question and determine
if it explicitly mentions deceptive uses of AI in politics.

Deception includes all acts and statements that mislead, hide the truth,
or promote a belief, concept, or idea that is not true. Examples include
but are not limited to the use of deep fakes to generate faked text,
video, or audio content. It also includes the automated generation of
social media posts pretending to be from humans. Another form of
deception are automated interactions with journalists, political elites,
or voters in text, audio, or video pretending to come from humans.
Deception does also include the purposeful generation and distribution
of misinformation, disinformation, and lies.

Reply with 1 if it does, and with 2 if it does not. Reply only with a
number.

Here is the response: {[}survey response was added here{]}

\end{tcolorbox}

\begin{tcolorbox}[enhanced jigsaw, rightrule=.15mm, left=2mm, colbacktitle=quarto-callout-note-color!10!white, breakable, bottomrule=.15mm, leftrule=.75mm, opacitybacktitle=0.6, bottomtitle=1mm, title=\textcolor{quarto-callout-note-color}{\faInfo}\hspace{0.5em}{Prompt risk 2: Reduced agency (associated with AI-enabled voter
outreach)}, toprule=.15mm, coltitle=black, toptitle=1mm, colframe=quarto-callout-note-color-frame, titlerule=0mm, arc=.35mm, opacityback=0, colback=white]

Analyze the following response to an open survey question and determine
if it explicitly mentions uses of AI in politics that reduced the agency
of voters.

Here, reduced agency refers to a situation where individuals' ability to
make informed and autonomous choices in the political sphere is
constrained or limited. Examples for reduced agency include, but are not
limited to, presenting people selected true information that supports
the campaign's goals. Another example is profiling people based on their
behavior on- as well as offline and then adapting communicative
approaches and content to better persuade or influence them to support a
campaign, donate money, or turn up to vote and in general to use these
profiles to undermine voters' critical reasoning. Reduced agency does
not include cases where a campaign actively deceives people or lies to
them.

Reply with 1 if it does, and with 2 if it does not. Reply only with a
number.

Here is the response: {[}survey response was added here{]}

\end{tcolorbox}

% Study 1: Open Answers analysis

\begin{table}[!ht]
    \centering
    \caption{Probability that open answer to questions on risks to deceptive uses of AI in campaigns mentions deception.}
    \label{tab:open_answer_deception}
    \begin{tabular}{llll}
    \hline
        Predictors & Odds Ratios & CI & p \\ \hline
        (Intercept) & 0.19 & 0.13 – 0.28 & \textless0.001 \\ 
        Case Dimension (Operations) & 0.32 & 0.21 – 0.51 & \textless0.001 \\ 
        Case Dimension (Outreach) & 0.28 & 0.18 – 0.44 & \textless0.001 \\ 
        Gender (Male) & 0.78 & 0.56 – 1.09 & 0.145 \\ 
        Education (Binary) & 1.31 & 0.82 – 2.11 & 0.257 \\ \hline
        Random Effects & ~ & ~ & ~ \\ \hline
        $\sigma^2$ & 3.29 & ~ & ~ \\ 
        $\tau_{00}$ respondent & 5.23 & ~ & ~ \\ 
        $\tau_{00}$ case & 0.11 & ~ & ~ \\ 
        ICC & 0.62 & ~ & ~ \\ 
        N (case) & 15 & ~ & ~ \\ 
        N (respondent) & 1199 & ~ & ~ \\ \hline
        Observations & 7194 & ~ & ~ \\ \hline
        Marginal R$^2$ / Conditional R$^2$ & 0.039 / 0.633 & ~ & ~ \\ \hline
    \end{tabular}
\end{table}

\begin{landscape}
\begin{table}[!ht]
    \centering
    \caption{Probability that open answer to questions on risks to AI-enabled voter outreach mentions reduced agency.}
    \label{tab:open_answer_agency}
    \begin{tabular}{lllllllllll}
    \hline
        ~ & \multicolumn{3}{c}{Reduced Agency} &  \multicolumn{3}{c}{Reduced Agency} & ~ \\ \hline
        Predictors & Odds Ratios & CI & p & Odds Ratios & CI & p \\ \hline
        (Intercept) & 0.01 & 0.01 – 0.02 & \textless0.001 & 0.01 & 0.01 – 0.02 & \textless0.001 \\ 
        Case Dimension (Deception) & 0.62 & 0.45 – 0.86 & 0.004 & 0.62 & 0.45 – 0.86 & 0.004 \\ 
        Case Dimension (Operations) & 0.57 & 0.41 – 0.79 & 0.001 & 0.57 & 0.41 – 0.79 & 0.001 \\ 
        Gender (Male) & 1.04 & 0.69 – 1.56 & 0.857 & 1.04 & 0.69 – 1.56 & 0.857 \\ 
        Education (Binary) & 1.06 & 0.58 – 1.91 & 0.854 & 1.06 & 0.58 – 1.91 & 0.854 \\ \hline
        Random Effects & ~ & ~ & ~ & ~ & ~ & ~ \\ \hline
        $\sigma^2$ & 3.29 & ~ & ~ & 3.29 & ~ & ~ \\ 
        $\tau_{00}$ respondent & 5.22 & ~ & ~ & 5.22 & ~ & ~ \\ 
        $\tau_{00}$ case & 0.00 & ~ & ~ & ~ & ~ & ~ \\ 
        ICC & ~ & ~ & ~ & 0.61 & ~ & ~ \\ 
        N (case) & 15 & ~ & ~ & ~ & ~ & ~ \\ 
        N (respondent) & 1199 & ~ & ~ & 1199 & ~ & ~ \\ \hline
        Observations & 7194 & ~ & ~ & 7194 & ~ & ~ \\ \hline
        Marginal R$^2$ / Conditional R$^2$ & 0.018 / NA & ~ & ~ & 0.007 / 0.616 & ~ & ~ \\ \hline
        \multicolumn{7}{l}{The version of the model on the right side of the table was fitted without varying intercepts for use cases,}\\
        \multicolumn{7}{l}{as the initial model indicated a singular fit.}\\ 
        \hline
    \end{tabular}
\end{table}
\end{landscape}

Based on these automated analyses, we see the hypotheses H4a,b and H5a,b
as supported.

\section{Equivalence Test, Effects on Party Favorability (Study
3)}\label{equivalence-test-effects-on-party-favorability-study-3}

We also used an equivalence test for the party favorability variables
``to test whether an observed effect is surprisingly small, assuming
that a meaningful effect exists in the population''\textsuperscript{32}.
For all tests, we used Cohen's D of 0.216 from the preregistration as
the smallest effect size of interest for the upper and lower bounds of
the test ($\Delta$L = -0.216, $\Delta$U = 0.216). We used Welch's t-tests for the
equivalence test. All the nonsignificant results for the favorability
scores show a significant equivalence test (two one-sided tests). Thus,
we can assume the effect of deceptive use of AI does not substantially
affect party favorability in all three samples. In Table
\ref{tab:study3_tost}, we report the test for the bound with the smaller
t statistic and, thus, the higher p-value\textsuperscript{32}.''

% Study 3
% Study 3: Equivalence Testing Effects on party favorability

\begin{landscape}
    \begin{table}[h!]
        \centering
        \caption{Equivalence Testing Results for Party Favorability (TOST)}
        \label{tab:study3_tost}
        \begin{tabularx}{\linewidth}{XXXXXXX}
            \toprule
            & \multicolumn{2}{c}{Democrat sample} & \multicolumn{2}{c}{Independent sample} & \multicolumn{2}{c}{Republican sample} \\
            \cmidrule(lr){2-3} \cmidrule(lr){4-5} \cmidrule(lr){6-7}
            Favorability & Dem Deception & Rep Deception & Dem Deception & Rep Deception & Dem Deception & Rep Deception \\
            \midrule
            Democratic Party & $\Delta U, t(987.96)=-2.02, p=.022$ & $\Delta U, t(997.53)=-3.01, p=.001$ & $\Delta U, t(991.15)=-2.57, p=.005$ & $\Delta U, t(988.64)=-2.49, p=.006$ & $\Delta U, t(991.81)=-3.14, p<.001$ & - \\
            Republican Party & $\Delta U, t(852.19)=-2.58, p=.005$ & $\Delta U, t(998.39)=-2.29, p=.011$ & $\Delta L, t(991.47)=2.78, p=.003$ & $\Delta L, t(983.1)=2.32, p=.010$ & $\Delta L, t(991.04)=2.63, p=.004$ & $\Delta L, t(984.54)=1.90, p=.029$ \\
            \bottomrule
        \end{tabularx}
    \end{table}
    \end{landscape}

\section{Regression Tables}\label{regression-tables}

\subsection{Supporting Tables -- Study 1: Regression tables, Figure
3}\label{supporting-tables-study-1-regression-tables-figure-3}

% Study 1
% Study 1: results missing data
\begin{landscape}
    \begin{table}[!ht]
        \caption{Attitudes toward AI uses in elections, regression model - Original data without missing responses (Figure 3)}
        \label{tab:study1_wo_missings}
        \centering
        \resizebox{600pt}{!}{
        \begin{tabular}{llllllllll}
        \hline
        ~ & Worry & ~ & ~ & Norm Violation & ~ & ~ & Rise Voter Involvement & ~ & ~ \\ \hline
        Predictors & Estimates & CI & p & Estimates & CI & p & Estimates & CI & p \\ \hline
        Intercept & 3.34 & 2.90 – 3.79 & \textless0.001 & 3.67 & 3.23 – 4.12 & \textless0.001 & 1.25 & 0.84 – 1.65 & \textless0.001 \\ 
        Campaign Task Operations vs Deception & -0.53 & -0.75 – -0.32 & \textless0.001 & -0.42 & -0.60 – -0.24 & \textless0.001 & 0.23 & 0.14 – 0.32 & \textless0.001 \\ 
        Campaign Task Voter Outreach vs Deception & -0.48 & -0.70 – -0.26 & \textless0.001 & -0.37 & -0.55 – -0.19 & \textless0.001 & 0.21 & 0.11 – 0.30 & \textless0.001 \\ 
        AI Benefits & -0.19 & -0.24 – -0.14 & \textless0.001 & -0.12 & -0.18 – -0.07 & \textless0.001 & 0.61 & 0.56 – 0.66 & \textless0.001 \\ 
        AI Risks & 0.57 & 0.51 – 0.63 & \textless0.001 & 0.43 & 0.37 – 0.49 & \textless0.001 & -0.08 & -0.14 – -0.02 & 0.007 \\ 
        Gender (Male) & -0.09 & -0.26 – 0.07 & 0.283 & -0.19 & -0.36 – -0.02 & 0.027 & 0.01 & -0.15 – 0.17 & 0.868 \\ 
        Education & -0.16 & -0.39 – 0.07 & 0.168 & 0.11 & -0.12 – 0.35 & 0.344 & 0.10 & -0.12 – 0.32 & 0.375 \\ \hline
        Random Effects & ~ & ~ & ~ & ~ & ~ & ~ & ~ & ~ & ~ \\ \hline
        $\sigma^2$ & 1.36 & ~ & ~ & 1.80 & ~ & ~ & 0.85 & ~ & ~ \\ 
        $\tau_{00}$ & 1.31 respondent & ~ & ~ & 1.29 respondent & ~ & ~ & 1.27 respondent & ~ & ~ \\ 
        ~ & 0.03 case & ~ & ~ & 0.02 case & ~ & ~ & 0.00 case & ~ & ~ \\ 
        ICC & 0.50 & ~ & ~ & 0.42 & ~ & ~ & 0.60 & ~ & ~ \\ 
        N & 15 case & ~ & ~ & 15 case & ~ & ~ & 15 case & ~ & ~ \\ 
        ~ & 867 respondents & ~ & ~ & 867 respondents & ~ & ~ & 867 respondents & ~ & ~ \\ \hline
        Observations & 5202 & ~ & ~ & 5202 & ~ & ~ & 5202 & ~ & ~ \\ \hline
        Marginal R2 / Conditional R2 & 0.233 / 0.613 & ~ & ~ & 0.128 / 0.494 & ~ & ~ & 0.301 / 0.719 & ~ & ~ \\ \hline
            \end{tabular}
        }
\end{table}
    
% Study 1 results based on data imputation
\begin{table}[!ht]
\caption{Attitudes toward AI uses in elections, regression model - Imputed data (Figure 3)}
\label{tab:study1_imputed}
\centering
    \resizebox{600pt}{!}{
    \begin{tabular}{llllllllll}
    \hline
        ~ & Worry & ~ & ~ & Norm Violation & ~ & ~ & Political Impact & ~ & ~ \\ \hline
        Predictors & Estimates & CI & p & Estimates & CI & p & Estimates & CI & p \\ \hline
        Intercept & 3.56 & 3.11 – 4.01 & \textless0.001 & 3.81 & 3.38 – 4.24 & \textless0.001 & 1.43 & 1.04 – 1.82 & \textless0.001\\
        Campaign Task Operations vs Deception & -0.57 & -0.79 – -0.35 & \textless0.001 & -0.46 & -0.66 – -0.26 & \textless0.001 & 0.26 & 0.16 – 0.36 & \textless0.001\\
        Campaign Task Voter Outreach vs Deception & -0.50 & -0.72 – -0.28 & \textless0.001 & -0.43 & -0.63 – -0.23 & \textless0.001 & 0.23 & 0.13 – 0.33 & \textless0.001\\
        AI Benefits & -0.16 & -0.22 – -0.1 & \textless0.001 & -0.09 & -0.15 – -0.03 & \textless0.001 & 0.52 & 0.46 – 0.58 & \textless0.001\\
        AI Risks & 0.51 & 0.45 – 0.57 & \textless0.001 & 0.38 & 0.32 – 0.44 & \textless0.001 & -0.07 & -0.13 – -0.01 & \textless0.001\\
        Gender (Male) & -0.10 & -0.26 – 0.06 & 0.2 & -0.14 & -0.3 – 0.02 & 0.07 & 0.10 & -0.04 – 0.24 & 0.17\\
        Education & -0.10 & -0.32 – 0.12 & 0.39 & 0.13 & -0.09 – 0.35 & 0.26 & 0.17 & -0.05 – 0.39 & 0.11 \\ \hline
        N & 15 case & ~ & ~ & 15 case & ~ & ~ & 15 case & ~ & ~ \\ 
        ~ & 1199 respondents & ~ & ~ & 1199 respondents & ~ & ~ & 1199 respondents & ~ & ~ \\ \hline
        Observations & 7194 & ~ & ~ & 7194 & ~ & ~ & 7194 & ~ & ~ \\ \hline
        \end{tabular}
    }
\end{table}
\end{landscape}

\newpage

\subsection{Supporting Tables -- Study 2: Regression tables, Figure
4}\label{supporting-tables-study-2-regression-tables-figure-4}

We report the full regression models with Lin (2013)\textsuperscript{33}
covariate adjustment underlying Figure 4.

%% Regression with Lin (2013) covariate adjustment. 95\%-CIs are reported. Outcome variable Worry.
\begin{table}[htp]
\caption{Outcome variable: Worry. Regression with Lin (2013) covariate adjustment. 95\%-CIs are reported.}
\label{tab:study2_worry}
\centering
\begin{tabular}[t]{lrrrrl}
\toprule
Predictors & Estimates & SE & LL & UL & p\\
\midrule
Intercept & 5.70 & 0.07 & 5.56 & 5.84 & \textless0.001\\
Campaign Task Voter Outreach vs Deception & -1.29 & 0.11 & -1.50 & -1.07 & \textless0.001\\
Campaign Task Operations vs Deception & -1.35 & 0.11 & -1.57 & -1.13 & \textless0.001\\
Control Group vs Deception & -0.74 & 0.11 & -0.95 & -0.52 & \textless0.001\\
Education & 0.32 & 0.16 & 0.01 & 0.62 & 0.042\\
\addlinespace
Gender (Male) & -0.31 & 0.14 & -0.60 & -0.03 & 0.029\\
\bottomrule
\end{tabular}
\end{table}

%% Regression with Lin (2013) covariate adjustment. 95\%-CIs are reported. Outcome variable Norm Violation
\begin{table}[htp]
\caption{Outcome variable: Norm Violation. Regression with Lin (2013) covariate adjustment. 95\%-CIs are reported.}
\label{tab:study2_norm}
\centering
\begin{tabular}[t]{lrrrrl}
\toprule
Predictors & Estimates & SE & LL & UL & p\\
    \midrule
    Intercept & 5.82 & 0.07 & 5.68 & 5.96 & \textless0.001\\
    Campaign Task Voter Outreach vs Deception & -1.62 & 0.11 & -1.83 & -1.41 & \textless0.001\\
    Campaign Task Operations vs Deception & -1.41 & 0.11 & -1.63 & -1.19 & \textless0.001\\
    Control Group vs Deception & -0.65 & 0.11 & -0.85 & -0.44 & \textless0.001\\
    Education & 0.23 & 0.17 & -0.11 & 0.57 & 0.176\\
    \addlinespace
    Gender (Male) & -0.32 & 0.15 & -0.60 & -0.03 & 0.028\\
\bottomrule
\end{tabular}
\end{table}
    
 %% Regression with Lin (2013) covariate adjustment. 95\%-CIs are reported. Outcome variable Political Impact.
\begin{table}[htp]
    \caption{Outcome variable: Rises Voter Involvement. Regression with Lin (2013) covariate adjustment. 95\%-CIs are reported.}
	\label{tab:study2_rise}
    \centering
    \begin{tabular}[t]{lrrrrl}
    \toprule
    Predictors & Estimates & SE & LL & UL & p\\
    \midrule
    Intercept & 4.09 & 0.07 & 3.95 & 4.23 & \textless0.001\\
    Campaign Task Voter Outreach vs Deception & 0.65 & 0.09 & 0.47 & 0.83 & \textless0.001\\
    Campaign Task Operations vs Deception & -0.08 & 0.09 & -0.27 & 0.10 & 0.389\\
    Control Group vs Deception & -0.29 & 0.10 & -0.48 & -0.09 & 0.004\\
    Education & 0.07 & 0.17 & -0.27 & 0.40 & 0.682\\
    \addlinespace
    Gender (Male) & -0.01 & 0.14 & -0.29 & 0.26 & 0.927\\
    \bottomrule
    \end{tabular}
    \end{table}

%% Regression with Lin (2013) covariate adjustment. 95\%-CIs are reported. Outcome variable Fairness of Elections.
    \begin{table}[htp]
    \caption{Outcome variable: Fairness of Elections. Regression with Lin (2013) covariate adjustment. 95\%-CIs are reported.}
	\label{tab:study2_fairness}
    \centering
    \begin{tabular}[t]{lrrrrl}
    \toprule
    Predictors & Estimates & SE & LL & UL & p\\
    \midrule
    Intercept & 3.67 & 0.07 & 3.54 & 3.80 & \textless0.001\\
    Campaign Task Voter Outreach vs Deception & 0.08 & 0.09 & -0.10 & 0.26 & 0.398\\
    Campaign Task Operations vs Deception & 0.12 & 0.09 & -0.06 & 0.31 & 0.195\\
    Control Group vs Deception & 0.10 & 0.09 & -0.08 & 0.28 & 0.294\\
    Education & 0.39 & 0.17 & 0.05 & 0.73 & 0.024\\
    \addlinespace
    Gender (Male) & 0.42 & 0.13 & 0.16 & 0.68 & 0.002\\
    \bottomrule
    \end{tabular}
    \end{table}
    
%% Regression with Lin (2013) covariate adjustment. 95\%-CIs are reported. Outcome variable Favorability Political Parties.
    \begin{table}[htp]
    \caption{Outcome variable: Favorability Political Parties. Regression with Lin (2013) covariate adjustment. 95\%-CIs are reported.}
	\label{tab:study2_parties}
    \centering
    \begin{tabular}[t]{lrrrrl}
    \toprule
    Predictors & Estimates & SE & LL & UL & p\\
    \midrule
    Intercept & 3.24 & 0.06 & 3.12 & 3.35 & \textless0.001\\
    Campaign Task Voter Outreach vs Deception & -0.07 & 0.08 & -0.23 & 0.10 & 0.434\\
    Campaign Task Operations vs Deception & -0.02 & 0.08 & -0.19 & 0.14 & 0.777\\
    Control Group vs Deception & 0.05 & 0.08 & -0.11 & 0.22 & 0.517\\
    Education & -0.01 & 0.15 & -0.32 & 0.29 & 0.927\\
    \addlinespace
    Gender (Male) & -0.12 & 0.12 & -0.35 & 0.12 & 0.344\\
    \bottomrule
    \end{tabular}
    \end{table}
    
%% Regression with Lin (2013) covariate adjustment. 95\%-CIs are reported. Outcome variable Favorability Republican Party.
    \begin{table}[htp]    
    \caption{Outcome variable: Favorability Republican Party. Regression with Lin (2013) covariate adjustment. 95\%-CIs are reported.}
	\label{tab:study2_reps}
    \centering
    \begin{tabular}[t]{lrrrrl}
    \toprule
    Predictors & Estimates & SE & LL & UL & p\\
    \midrule
    Intercept & 3.16 & 0.06 & 3.04 & 3.28 & \textless0.001\\
    Campaign Task Voter Outreach vs Deception & 0.03 & 0.08 & -0.13 & 0.19 & 0.703\\
    Campaign Task Operations vs Deception & 0.01 & 0.09 & -0.16 & 0.18 & 0.879\\
    Control Group vs Deception & 0.11 & 0.09 & -0.06 & 0.29 & 0.201\\
    Education & -0.17 & 0.16 & -0.47 & 0.14 & 0.28\\
    \addlinespace
    Gender (Male) & -0.16 & 0.12 & -0.39 & 0.08 & 0.189\\
    \bottomrule
    \end{tabular}
    \end{table}
    
%% Regression with Lin (2013) covariate adjustment. 95\%-CIs are reported. Outcome variable Favorability Democratic Party.
    \begin{table}[htp]
    \caption{Outcome variable: Favorability Democratic Party. Regression with Lin (2013) covariate adjustment. 95\%-CIs are reported.}
    \label{tab:study2_dems}
	\centering
    \begin{tabular}[t]{lrrrrl}
    \toprule
    Predictors & Estimates & SE & LL & UL & p\\
    \midrule
    Intercept & 3.65 & 0.06 & 3.53 & 3.77 & \textless0.001\\
    Campaign Task Voter Outreach vs Deception & 0.08 & 0.09 & -0.09 & 0.26 & 0.348\\
    Campaign Task Operations vs Deception & -0.07 & 0.09 & -0.25 & 0.10 & 0.416\\
    Control Group vs Deception & 0.11 & 0.09 & -0.06 & 0.29 & 0.198\\
    Education & 0.13 & 0.16 & -0.18 & 0.44 & 0.425\\
    \addlinespace
    Gender (Male) & -0.23 & 0.13 & -0.47 & 0.02 & 0.071\\
    \bottomrule
    \end{tabular}
    \end{table}

 %% Regression with Lin (2013) covariate adjustment. 95\%-CIs are reported. Outcome variable Stricter Oversight of AI Use in Elections.
    \begin{table}[htp]
    \caption{Outcome variable: Stricter Oversight of AI Use in Elections. Regression with Lin (2013) covariate adjustment. 95\%-CIs are reported.}
	\label{tab:study2_reg_elections}
    \centering
    \begin{tabular}[t]{lrrrrl}
    \toprule
    Predictors & Estimates & SE & LL & UL & p\\
    \midrule
    Intercept & 5.51 & 0.07 & 5.37 & 5.66 & \textless0.001\\
    Campaign Task Voter Outreach vs Deception & -1.14 & 0.11 & -1.36 & -0.93 & \textless0.001\\
    Campaign Task Operations vs Deception & -1.12 & 0.11 & -1.34 & -0.90 & \textless0.001\\
    Control Group vs Deception & -0.61 & 0.11 & -0.82 & -0.39 & \textless0.001\\
    Education & 0.03 & 0.19 & -0.35 & 0.41 & 0.887\\
    \addlinespace
    Gender (Male) & -0.38 & 0.15 & -0.68 & -0.08 & 0.012\\
    \bottomrule
    \end{tabular}
    \end{table}
    
%% Regression with Lin (2013) covariate adjustment. 95\%-CIs are reported. Outcome variable AI Contributes to Personal Loss of Control.
    \begin{table}[htp]
    \caption{Outcome variable: AI Contributes to Personal Loss of Control. Regression with Lin (2013) covariate adjustment. 95\%-CIs are reported.}
	\label{tab:study2_loss}
    \centering
    \begin{tabular}[t]{lrrrrl}
    \toprule
    Predictors & Estimates & SE & LL & UL & p\\
    \midrule
    Intercept & 4.63 & 0.07 & 4.49 & 4.77 & \textless0.001\\
    Campaign Task Voter Outreach vs Deception & -0.20 & 0.10 & -0.41 & 0.00 & 0.051\\
    Campaign Task Operations vs Deception & -0.11 & 0.10 & -0.31 & 0.09 & 0.289\\
    Control Group vs Deception & -0.28 & 0.10 & -0.48 & -0.07 & 0.008\\
    Education & 0.07 & 0.18 & -0.28 & 0.43 & 0.689\\
    \addlinespace
    Gender (Male) & -0.29 & 0.15 & -0.58 & 0.00 & 0.049\\
    \bottomrule
    \end{tabular}
    \end{table}
    
%% Regression with Lin (2013) covariate adjustment. 95\%-CIs are reported. Outcome variable Prioritize Innovation in AI Regulation.
    \begin{table}[htp]    
    \caption{Outcome variable: Prioritize Innovation in AI Regulation. Regression with Lin (2013) covariate adjustment. 95\%-CIs are reported.}
	\label{tab:study2_prios}
    \centering
    \begin{tabular}[t]{lrrrrl}
    \toprule
    Predictors & Estimates & SE & LL & UL & p\\
    \midrule
    Intercept & 3.08 & 0.07 & 2.93 & 3.22 & \textless0.001\\
    Campaign Task Voter Outreach vs Deception & 0.15 & 0.10 & -0.05 & 0.35 & 0.142\\
    Campaign Task Operations vs Deception & 0.26 & 0.10 & 0.06 & 0.46 & 0.011\\
    Control Group vs Deception & 0.23 & 0.10 & 0.02 & 0.43 & 0.031\\
    Education & 0.12 & 0.18 & -0.24 & 0.47 & 0.517\\
    \addlinespace
    Gender (Male) & 0.48 & 0.15 & 0.19 & 0.77 & 0.001\\
    \bottomrule
    \end{tabular}
    \end{table}
    
%% Regression with Lin (2013) covariate adjustment. 95\%-CIs are reported. Outcome variable Support for AI Moratorium.
    \begin{table}[htp] 
    \caption{Outcome variable: Support for AI Moratorium. Regression with Lin (2013) covariate adjustment. 95\%-CIs are reported.}
	\label{tab:study2_mora}
    \centering
    \begin{tabular}[t]{lrrrrl}
    \toprule
    Predictors & Estimates & SE & LL & UL & p\\
    \midrule
    Intercept & 0.38 & 0.02 & 0.34 & 0.42 & \textless0.001\\
    Campaign Task Voter Outreach vs Deception & -0.07 & 0.03 & -0.13 & -0.01 & 0.024\\
    Campaign Task Operations vs Deception & -0.06 & 0.03 & -0.12 & 0.00 & 0.062\\
    Control Group vs Deception & -0.09 & 0.03 & -0.15 & -0.03 & 0.002\\
    Education & -0.11 & 0.06 & -0.21 & 0.00 & 0.056\\
    \addlinespace
    Gender (Male) & -0.08 & 0.04 & -0.16 & 0.01 & 0.081\\
    \bottomrule
    \end{tabular}
    \end{table}
    
%% Regression with Lin (2013) covariate adjustment. 95\%-CIs are reported. Outcome variable Stricter Regulation of AI in general.
    \begin{table}[htp]  
    \caption{Outcome variable: Stricter Regulation of AI in general. Regression with Lin (2013) covariate adjustment. 95\%-CIs are reported.}
	\label{tab:study2_reg_ai}
    \centering
    \begin{tabular}[t]{lrrrrl}
    \toprule
    Predictors & Estimates & SE & LL & UL & p\\
    \midrule
    Intercept & 3.73 & 0.08 & 3.57 & 3.88 & \textless0.001\\
    Campaign Task Voter Outreach vs Deception & -0.16 & 0.11 & -0.37 & 0.05 & 0.138\\
    Campaign Task Operations vs Deception & -0.18 & 0.11 & -0.39 & 0.02 & 0.083\\
    Control Group vs Deception & -0.20 & 0.11 & -0.41 & 0.01 & 0.058\\
    Education & -0.05 & 0.19 & -0.43 & 0.34 & 0.812\\
    \addlinespace
    Gender (Male) & -0.34 & 0.16 & -0.65 & -0.03 & 0.033\\
    \bottomrule
    \end{tabular}
    \end{table}

\newpage

\subsection{Supporting Tables -- Study 3: Regression tables, Figure
5}\label{supporting-tables-study-3-regression-tables-figure-5}

We report the full regression models with Lin (2013)\textsuperscript{33}
covariate adjustment underlying Figure 5.

\subsubsection{Democrat Sample}\label{democrat-sample}

Regression models calculated on Democrat partisans.

% Study 3: Democrats

%% Regression with Lin (2013) covariate adjustment. 95\%-CIs are reported. Outcome variable Worry.
\begin{table}[htp]
    \caption{Outcome variable: Worry. Regression with Lin (2013) covariate adjustment. 95\%-CIs are reported.}
    \label{tab:study3_d_worry}
    \centering
    \begin{tabular}[t]{lrrrrl}
    \toprule
    Predictors & Estimates & SE & LL & UL & p\\
    \midrule
    Intercept & 4.99 & 0.08 & 4.82 & 5.15 & \textless0.001\\
    Republican Deception vs Control & 1.02 & 0.11 & 0.81 & 1.22 & \textless0.001\\
    Democrat Deception vs Control & 0.68 & 0.11 & 0.47 & 0.90 & \textless0.001\\
    Education & -0.12 & 0.20 & -0.52 & 0.28 & 0.55\\
    Gender (Male) & -0.47 & 0.17 & -0.80 & -0.14 & 0.006\\
    \bottomrule
    \end{tabular}
    \end{table}
   
    %% Regression with Lin (2013) covariate adjustment. 95\%-CIs are reported. Outcome variable Norm Violation
    \begin{table}[htp]
    \caption{Outcome variable: Norm Violation. Regression with Lin (2013) covariate adjustment. 95\%-CIs are reported.}
    \label{tab:study3_d_norm}
    \centering
    \begin{tabular}[t]{lrrrrl}
    \toprule
    Predictors & Estimates & SE & LL & UL & p\\
    \midrule
    Intercept & 5.19 & 0.08 & 5.04 & 5.34 & \textless0.001\\
    Republican Deception vs Control & 0.84 & 0.10 & 0.64 & 1.04 & \textless0.001\\
    Democrat Deception vs Control & 0.56 & 0.11 & 0.36 & 0.77 & \textless0.001\\
    Education & -0.44 & 0.19 & -0.81 & -0.07 & 0.019\\
    Gender (Male) & -0.32 & 0.15 & -0.62 & -0.02 & 0.036\\
    \bottomrule
    \end{tabular}
    \end{table}
    
    %% Regression with Lin (2013) covariate adjustment. 95\%-CIs are reported. Outcome variable Rises Voter Involvement.
    \begin{table}[htp]
    \caption{Outcome variable: Rise Voter Involvement. Regression with Lin (2013) covariate adjustment. 95\%-CIs are reported.}
    \centering
    \label{tab:study3_d_rise}
    \begin{tabular}[t]{lrrrrl}
    \toprule
    Predictors & Estimates & SE & LL & UL & p\\
    \midrule
    Intercept & 3.83 & 0.07 & 3.70 & 3.96 & \textless0.001\\
    Republican Deception vs Control & 0.16 & 0.10 & -0.03 & 0.36 & 0.096\\
    Democrat Deception vs Control & 0.29 & 0.10 & 0.10 & 0.47 & 0.003\\
    Education & 0.15 & 0.15 & -0.15 & 0.45 & 0.331\\
    Gender (Male) & 0.37 & 0.13 & 0.12 & 0.63 & 0.004\\
    \bottomrule
    \end{tabular}
    \end{table}
    
    %% Regression with Lin (2013) covariate adjustment. 95\%-CIs are reported. Outcome variable Fairness of Elections.
    \begin{table}[htp]
    \caption{Outcome variable: Fairness of Elections. Regression with Lin (2013) covariate adjustment. 95\%-CIs are reported.}
    \label{tab:study3_d_fair}
    \centering
    \begin{tabular}[t]{lrrrrl}
    \toprule
    Predictors & Estimates & SE & LL & UL & p\\
    \midrule
    Intercept & 4.19 & 0.06 & 4.07 & 4.32 & \textless0.001\\
    Republican Deception vs Control & -0.21 & 0.09 & -0.38 & -0.04 & 0.015\\
    Democrat Deception vs Control & -0.16 & 0.09 & -0.33 & 0.01 & 0.063\\
    Education & 0.10 & 0.14 & -0.17 & 0.38 & 0.46\\
    Gender (Male) & 0.50 & 0.12 & 0.26 & 0.74 & \textless0.001\\
    \bottomrule
    \end{tabular}
    \end{table}
    
    %% Regression with Lin (2013) covariate adjustment. 95\%-CIs are reported. Outcome variable Favorability Republican Party.
    \begin{table}[htp] 
    \caption{Outcome variable: Favorability Republican Party. Regression with Lin (2013) covariate adjustment. 95\%-CIs are reported.}
    \label{tab:study3_d_reps}
    \centering
    \begin{tabular}[t]{lrrrrl}
    \toprule
    Predictors & Estimates & SE & LL & UL & p\\
    \midrule
    Intercept & 1.99 & 0.05 & 1.89 & 2.10 & \textless0.001\\
    Republican Deception vs Control & -0.08 & 0.08 & -0.23 & 0.06 & 0.265\\
    Democrat Deception vs Control & -0.05 & 0.08 & -0.20 & 0.11 & 0.559\\
    Education & 0.13 & 0.13 & -0.13 & 0.38 & 0.321\\
    Gender (Male) & 0.28 & 0.11 & 0.06 & 0.49 & 0.011\\
    \bottomrule
    \end{tabular}
    \end{table}
    
    %% Regression with Lin (2013) covariate adjustment. 95\%-CIs are reported. Outcome variable Favorability Democratic Party.
    \begin{table}[htp] 
    \caption{Outcome variable: Favorability Democratic Party. Regression with Lin (2013) covariate adjustment. 95\%-CIs are reported.}
    \label{tab:study3_d_dems}
    \centering
    \begin{tabular}[t]{lrrrrl}
    \toprule
    Predictors & Estimates & SE & LL & UL & p\\
    \midrule
    Intercept & 5.34 & 0.06 & 5.22 & 5.46 & \textless0.001\\
    Republican Deception vs Control & -0.03 & 0.08 & -0.19 & 0.13 & 0.735\\
    Democrat Deception vs Control & -0.12 & 0.08 & -0.28 & 0.05 & 0.168\\
    Education & 0.16 & 0.14 & -0.11 & 0.43 & 0.249\\
    Gender (Male) & 0.01 & 0.12 & -0.22 & 0.25 & 0.918\\
    \bottomrule
    \end{tabular}
    \end{table}
    
    %% Regression with Lin (2013) covariate adjustment. 95\%-CIs are reported. Outcome variable Prioritize Innovation in AI Regulation.
    \begin{table}[htp]
    \caption{Outcome variable: Prioritize Innovation in AI Regulation. Regression with Lin (2013) covariate adjustment. 95\%-CIs are reported.}
    \label{tab:study3_d_prios}
    \centering
    \begin{tabular}[t]{lrrrrl}
    \toprule
    Predictors & Estimates & SE & LL & UL & p\\
    \midrule
    Intercept & 2.97 & 0.07 & 2.84 & 3.10 & \textless0.001\\
    Republican Deception vs Control & -0.28 & 0.10 & -0.47 & -0.09 & 0.003\\
    Democrat Deception vs Control & -0.10 & 0.10 & -0.28 & 0.09 & 0.323\\
    Education & 0.00 & 0.16 & -0.31 & 0.31 & 0.989\\
    Gender (Male) & 0.55 & 0.13 & 0.29 & 0.81 & \textless0.001\\
    \bottomrule
    \end{tabular}
    \end{table}
    
    %% Regression with Lin (2013) covariate adjustment. 95\%-CIs are reported. Outcome variable Support for AI Moratorium.
    \begin{table}[htp]
    \caption{Outcome variable: Support for AI Moratorium. Regression with Lin (2013) covariate adjustment. 95\%-CIs are reported.}
    \label{tab:study3_d_mora}
    \centering
    \begin{tabular}[t]{lrrrrl}
    \toprule
    Predictors & Estimates & SE & LL & UL & p\\
    \midrule
    Intercept & 0.28 & 0.02 & 0.24 & 0.32 & \textless0.001\\
    Republican Deception vs Control & 0.12 & 0.03 & 0.06 & 0.18 & \textless0.001\\
    Democrat Deception vs Control & 0.07 & 0.03 & 0.01 & 0.12 & 0.024\\
    Education & -0.07 & 0.05 & -0.16 & 0.02 & 0.142\\
    Gender (Male) & -0.06 & 0.04 & -0.14 & 0.02 & 0.15\\
    \bottomrule
    \end{tabular}
    \end{table}

\newpage

\subsubsection{Indendent Sample}\label{indendent-sample}

Regression models calculated on Independents.

% Study 3: Independents

%% Regression with Lin (2013) covariate adjustment. 95\%-CIs are reported. Outcome variable Worry.
\begin{table}[htp]
\caption{Outcome variable: Worry. Regression with Lin (2013) covariate adjustment. 95\%-CIs are reported.}
\label{tab:study3_i_worry}
\centering
\begin{tabular}[t]{lrrrrl}
\toprule
Predictors & Estimates & SE & LL & UL & p\\
\midrule
Intercept & 5.04 & 0.08 & 4.88 & 5.19 & \textless0.001\\
Republican Deception vs Control & 0.70 & 0.11 & 0.49 & 0.91 & \textless0.001\\
Democrat Deception vs Control & 0.59 & 0.11 & 0.39 & 0.80 & \textless0.001\\
Education & -0.22 & 0.22 & -0.65 & 0.22 & 0.326\\
Gender (Male) & -0.21 & 0.16 & -0.52 & 0.11 & 0.199\\
\bottomrule
\end{tabular}
\end{table}

%% Regression with Lin (2013) covariate adjustment. 95\%-CIs are reported. Outcome variable Norm Violation
\begin{table}[htp]
\caption{Outcome variable: Norm Violation. Regression with Lin (2013) covariate adjustment. 95\%-CIs are reported.}
\label{tab:study3_i_norm}
\centering
\begin{tabular}[t]{lrrrrl}
\toprule
Predictors & Estimates & SE & LL & UL & p\\
\midrule
Intercept & 5.27 & 0.08 & 5.12 & 5.42 & \textless0.001\\
Republican Deception vs Control & 0.62 & 0.11 & 0.41 & 0.83 & \textless0.001\\
Democrat Deception vs Control & 0.50 & 0.11 & 0.29 & 0.71 & \textless0.001\\
Education & -0.19 & 0.22 & -0.62 & 0.24 & 0.386\\
Gender (Male) & -0.23 & 0.15 & -0.53 & 0.07 & 0.133\\
\bottomrule
\end{tabular}
\end{table}

%% Regression with Lin (2013) covariate adjustment. 95\%-CIs are reported. Outcome variable Rises Voter Involvement.
\begin{table}[htp]
\caption{Outcome variable: Rises Voter Involvement. Regression with Lin (2013) covariate adjustment. 95\%-CIs are reported.}
\label{tab:study3_i_rise}
\centering
\begin{tabular}[t]{lrrrrl}
\toprule
Predictors & Estimates & SE & LL & UL & p\\
\midrule
Intercept & 3.79 & 0.07 & 3.66 & 3.93 & \textless0.001\\
Republican Deception vs Control & 0.22 & 0.10 & 0.02 & 0.41 & 0.031\\
Democrat Deception vs Control & 0.37 & 0.10 & 0.19 & 0.56 & \textless0.001\\
Education & 0.26 & 0.19 & -0.10 & 0.63 & 0.16\\
Gender (Male) & 0.36 & 0.14 & 0.10 & 0.63 & 0.007\\
\bottomrule
\end{tabular}
\end{table}

%% Regression with Lin (2013) covariate adjustment. 95\%-CIs are reported. Outcome variable Fairness of Elections.
\begin{table}[htp]
\caption{Outcome variable: Fairness of Elections. Regression with Lin (2013) covariate adjustment. 95\%-CIs are reported.}
\label{tab:study3_i_fair}
\centering
\begin{tabular}[t]{lrrrrl}
\toprule
Predictors & Estimates & SE & LL & UL & p\\
\midrule
Intercept & 3.48 & 0.07 & 3.35 & 3.60 & \textless0.001\\
Republican Deception vs Control & -0.15 & 0.09 & -0.32 & 0.03 & 0.106\\
Democrat Deception vs Control & -0.23 & 0.09 & -0.40 & -0.05 & 0.012\\
Education & 0.58 & 0.19 & 0.21 & 0.95 & 0.002\\
Gender (Male) & 0.28 & 0.13 & 0.03 & 0.54 & 0.03\\
\bottomrule
\end{tabular}
\end{table}

%% Regression with Lin (2013) covariate adjustment. 95\%-CIs are reported. Outcome variable Favorability Republican Party.
\begin{table}[htp]
\caption{Outcome variable: Favorability Republican Party. Regression with Lin (2013) covariate adjustment. 95\%-CIs are reported.}
\label{tab:study3_i_reps}
\centering
\begin{tabular}[t]{lrrrrl}
\toprule
Predictors & Estimates & SE & LL & UL & p\\
\midrule
Intercept & 2.71 & 0.07 & 2.58 & 2.85 & \textless0.001\\
Republican Deception vs Control & 0.11 & 0.10 & -0.09 & 0.31 & 0.277\\
Democrat Deception vs Control & 0.06 & 0.10 & -0.14 & 0.25 & 0.561\\
Education & -0.02 & 0.19 & -0.38 & 0.34 & 0.915\\
Gender (Male) & -0.05 & 0.14 & -0.32 & 0.22 & 0.701\\
\bottomrule
\end{tabular}
\end{table}

%% Regression with Lin (2013) covariate adjustment. 95\%-CIs are reported. Outcome variable Favorability Democratic Party.
\begin{table}[htp]
\caption{Outcome variable: Favorability Democratic Party. Regression with Lin (2013) covariate adjustment. 95\%-CIs are reported.}
\label{tab:study3_i_dems}
\centering
\begin{tabular}[t]{lrrrrl}
\toprule
Predictors & Estimates & SE & LL & UL & p\\
\midrule
Intercept & 3.39 & 0.07 & 3.25 & 3.53 & \textless0.001\\
Republican Deception vs Control & -0.09 & 0.10 & -0.29 & 0.10 & 0.342\\
Democrat Deception vs Control & -0.09 & 0.10 & -0.28 & 0.11 & 0.377\\
Education & 0.33 & 0.20 & -0.07 & 0.72 & 0.107\\
Gender (Male) & 0.02 & 0.14 & -0.26 & 0.30 & 0.897\\
\bottomrule
\end{tabular}
\end{table}

%% Regression with Lin (2013) covariate adjustment. 95\%-CIs are reported. Outcome variable Prioritize Innovation in AI Regulation.
\begin{table}[htp]
\caption{Outcome variable: Prioritize Innovation in AI Regulation. Regression with Lin (2013) covariate adjustment. 95\%-CIs are reported.}
\label{tab:study3_i_prios}
\centering
\begin{tabular}[t]{lrrrrl}
\toprule
Predictors & Estimates & SE & LL & UL & p\\
\midrule
Intercept & 2.98 & 0.07 & 2.84 & 3.11 & \textless0.001\\
Republican Deception vs Control & -0.06 & 0.10 & -0.26 & 0.14 & 0.564\\
Democrat Deception vs Control & -0.09 & 0.10 & -0.28 & 0.11 & 0.385\\
Education & 0.29 & 0.21 & -0.12 & 0.69 & 0.161\\
Gender (Male) & 0.68 & 0.14 & 0.41 & 0.95 & \textless0.001\\
\bottomrule
\end{tabular}
\end{table}

%% Regression with Lin (2013) covariate adjustment. 95\%-CIs are reported. Outcome variable Support for AI Moratorium.
\begin{table}[htp]
\caption{Outcome variable: Support for AI Moratorium. Regression with Lin (2013) covariate adjustment. 95\%-CIs are reported.}
\label{tab:study3_i_mora}
\centering
\begin{tabular}[t]{lrrrrl}
\toprule
Predictors & Estimates & SE & LL & UL & p\\
\midrule
Intercept & 0.34 & 0.02 & 0.30 & 0.38 & \textless0.001\\
Republican Deception vs Control & 0.04 & 0.03 & -0.01 & 0.10 & 0.142\\
Democrat Deception vs Control & 0.03 & 0.03 & -0.03 & 0.09 & 0.265\\
Education & -0.04 & 0.06 & -0.16 & 0.08 & 0.492\\
Gender (Male) & -0.09 & 0.04 & -0.18 & -0.01 & 0.027\\
\bottomrule
\end{tabular}
\end{table}

\newpage

\subsubsection{Republican Sample}\label{republican-sample}

Regression models calculated on Republican partisans.

% Study 3: Republicans

%% Regression with Lin (2013) covariate adjustment. 95\%-CIs are reported. Outcome variable Worry.
\begin{table}[htp]
    \caption{Outcome variable: Worry. Regression with Lin (2013) covariate adjustment. 95\%-CIs are reported.}
    \label{tab:study3_r_worry}
    \centering
    \begin{tabular}[t]{lrrrrl}
    \toprule
    Predictors & Estimates & SE & LL & UL & p\\
    \midrule
    Intercept & 4.94 & 0.08 & 4.78 & 5.11 & \textless0.001\\
    Republican Deception vs Control & 0.06 & 0.12 & -0.18 & 0.30 & 0.616\\
    Democrat Deception vs Control & 0.43 & 0.11 & 0.21 & 0.66 & \textless0.001\\
    Education & -0.20 & 0.22 & -0.63 & 0.24 & 0.376\\
    Gender (Male) & -0.24 & 0.17 & -0.57 & 0.08 & 0.144\\
    \bottomrule
    \end{tabular}
    \end{table}

    %% Regression with Lin (2013) covariate adjustment. 95\%-CIs are reported. Outcome variable Norm Violation
    \begin{table}[htp]
    \caption{Outcome variable: Norm Violation. Regression with Lin (2013) covariate adjustment. 95\%-CIs are reported.}
    \label{tab:study3_r_norm}
    \centering
    \begin{tabular}[t]{lrrrrl}
    \toprule
    Predictors & Estimates & SE & LL & UL & p\\
    \midrule
    Intercept & 5.11 & 0.08 & 4.95 & 5.28 & \textless0.001\\
    Republican Deception vs Control & 0.23 & 0.12 & 0.00 & 0.47 & 0.048\\
    Democrat Deception vs Control & 0.54 & 0.11 & 0.31 & 0.76 & \textless0.001\\
    Education & -0.43 & 0.22 & -0.86 & 0.00 & 0.051\\
    Gender (Male) & 0.08 & 0.17 & -0.25 & 0.42 & 0.62\\
    \bottomrule
    \end{tabular}
    \end{table}

    %% Regression with Lin (2013) covariate adjustment. 95\%-CIs are reported. Outcome variable Political Impact.
    \begin{table}[htp]
    \caption{Outcome variable: Rise Voter Involvement. Regression with Lin (2013) covariate adjustment. 95\%-CIs are reported.}
    \label{tab:study3_r_rise}
    \centering
    \begin{tabular}[t]{lrrrrl}
    \toprule
    Predictors & Estimates & SE & LL & UL & p\\
    \midrule
    Intercept & 3.80 & 0.07 & 3.66 & 3.94 & \textless0.001\\
    Republican Deception vs Control & 0.28 & 0.10 & 0.08 & 0.48 & 0.005\\
    Democrat Deception vs Control & 0.42 & 0.10 & 0.23 & 0.62 & \textless0.001\\
    Education & 0.28 & 0.17 & -0.06 & 0.62 & 0.104\\
    Gender (Male) & 0.25 & 0.14 & -0.03 & 0.53 & 0.078\\
    \bottomrule
    \end{tabular}
    \end{table}

    %% Regression with Lin (2013) covariate adjustment. 95\%-CIs are reported. Outcome variable Fairness of Elections.
    \begin{table}[htp]
    \caption{Outcome variable: Fairness of Elections. Regression with Lin (2013) covariate adjustment. 95\%-CIs are reported.}
    \label{tab:study3_r_fair}
    \centering
    \begin{tabular}[t]{lrrrrl}
    \toprule
    Predictors & Estimates & SE & LL & UL & p\\
    \midrule
    Intercept & 3.35 & 0.06 & 3.23 & 3.47 & \textless0.001\\
    Republican Deception vs Control & 0.03 & 0.09 & -0.13 & 0.20 & 0.691\\
    Democrat Deception vs Control & 0.09 & 0.09 & -0.08 & 0.26 & 0.315\\
    Education & 0.64 & 0.15 & 0.35 & 0.93 & \textless0.001\\
    Gender (Male) & 0.20 & 0.12 & -0.05 & 0.44 & 0.11\\
    \bottomrule
    \end{tabular}
    \end{table}

    %% Regression with Lin (2013) covariate adjustment. 95\%-CIs are reported. Outcome variable Favorability Republican Party.
    \begin{table}[htp]
    \caption{Outcome variable: Favorability Republican Party. Regression with Lin (2013) covariate adjustment. 95\%-CIs are reported.}
    \label{tab:study3_r_reps}
    \centering
    \begin{tabular}[t]{lrrrrl}
    \toprule
    Predictors & Estimates & SE & LL & UL & p\\
    \midrule
    Intercept & 5.33 & 0.06 & 5.21 & 5.45 & \textless0.001\\
    Republican Deception vs Control & 0.14 & 0.09 & -0.03 & 0.31 & 0.105\\
    Democrat Deception vs Control & 0.06 & 0.09 & -0.11 & 0.24 & 0.465\\
    Education & 0.06 & 0.18 & -0.30 & 0.42 & 0.748\\
    Gender (Male) & -0.15 & 0.13 & -0.41 & 0.10 & 0.231\\
    \bottomrule
    \end{tabular}
    \end{table}
 
    %% Regression with Lin (2013) covariate adjustment. 95\%-CIs are reported. Outcome variable Favorability Democratic Party.
    \begin{table}[htp]
    \caption{Outcome variable: Favorability Democratic Party. Regression with Lin (2013) covariate adjustment. 95\%-CIs are reported.}
    \label{tab:study3_r_dems}
    \centering
    \begin{tabular}[t]{lrrrrl}
    \toprule
    Predictors & Estimates & SE & LL & UL & p\\
    \midrule
    Intercept & 2.52 & 0.06 & 2.40 & 2.65 & \textless0.001\\
    Republican Deception vs Control & -0.19 & 0.09 & -0.36 & -0.03 & 0.023\\
    Democrat Deception vs Control & -0.02 & 0.09 & -0.19 & 0.15 & 0.825\\
    Education & 0.39 & 0.18 & 0.04 & 0.75 & 0.029\\
    Gender (Male) & -0.17 & 0.13 & -0.42 & 0.08 & 0.186\\
    \bottomrule
    \end{tabular}
    \end{table}

    %% Regression with Lin (2013) covariate adjustment. 95\%-CIs are reported. Outcome variable Prioritize Innovation in AI Regulation.
    \begin{table}[htp] 
    \caption{Outcome variable: Prioritize Innovation in AI Regulation. Regression with Lin (2013) covariate adjustment. 95\%-CIs are reported.}
    \label{tab:study3_r_prios}
    \centering
    \begin{tabular}[t]{lrrrrl}
    \toprule
    Predictors & Estimates & SE & LL & UL & p\\
    \midrule
    Intercept & 3.01 & 0.08 & 2.86 & 3.15 & \textless0.001\\
    Republican Deception vs Control & 0.27 & 0.11 & 0.06 & 0.48 & 0.012\\
    Democrat Deception vs Control & 0.14 & 0.11 & -0.07 & 0.35 & 0.205\\
    Education & 0.50 & 0.20 & 0.10 & 0.90 & 0.014\\
    Gender (Male) & 0.70 & 0.15 & 0.39 & 1.00 & \textless0.001\\
    \bottomrule
    \end{tabular}
    \end{table}
    
    %% Regression with Lin (2013) covariate adjustment. 95\%-CIs are reported. Outcome variable Support for AI Moratorium.
    \begin{table}[htp]
    \caption{Outcome variable: Support for AI Moratorium. Regression with Lin (2013) covariate adjustment. 95\%-CIs are reported.}
    \label{tab:study3_r_mora}
    \centering
    \begin{tabular}[t]{lrrrrl}
    \toprule
    Predictors & Estimates & SE & LL & UL & p\\
    \midrule
    Intercept & 0.39 & 0.02 & 0.35 & 0.43 & \textless0.001\\
    Republican Deception vs Control & 0.00 & 0.03 & -0.06 & 0.06 & 0.961\\
    Democrat Deception vs Control & 0.05 & 0.03 & -0.01 & 0.11 & 0.094\\
    Education & -0.08 & 0.05 & -0.18 & 0.02 & 0.136\\
    Gender (Male) & -0.14 & 0.04 & -0.22 & -0.05 & 0.002\\
    \bottomrule
    \end{tabular}
    \end{table}

\end{document}